\documentclass{aa}  

\usepackage{graphicx}
\usepackage{txfonts}
\usepackage{times}
\usepackage{float}
\usepackage[usenames,dvipsnames]{color}
\usepackage{hyperref}
\usepackage{xcolor}
\definecolor{orange}{rgb}{0.93, 0.57, 0.13}
\usepackage{caption}
\usepackage{subcaption}
\usepackage{colortbl}
\usepackage{orcidlink}

\begin{document} 

    \title{Linking simulations and observations of star-forming regions}
    \subtitle{I. Masses and temperatures of ALMAGAL-like cores}

   \author{Birka Zimmermann\inst{1}
          \and
          \'Alvaro S\'anchez-Monge\inst{2,3}\orcidlink{0000-0002-3078-9482}
          \and
          Stefanie Walch\inst{1,4}\orcidlink{0000-0001-6941-7638}
          \and
          Gary A.\ Fuller\inst{5,1}\orcidlink{0000-0001-8509-1818}
          }

   \institute{I.\ Physikalisches Institut, Universität zu Köln, Zülpicher Str. 77, 50937 Köln, Germany \\ \email{zimmermann@ph1.uni-koeln.de}
         \and
              Institut de Ci\`encies de l'Espai (ICE), CSIC, Campus UAB, Carrer de Can Magrans s/n, E-08193, Bellaterra, Barcelona, Spain
         \and
              Institut d'Estudis Espacials de Catalunya (IEEC), E-08860, Castelldefels, Barcelona, Spain
         \and
             Center for Data and Simulation Science, University of Cologne, www.cds.uni-koeln.de
         \and
             Jodrell Bank Centre for Astrophysics, Department of Physics \& Astronomy, The University of Manchester, Oxford Road, Manchester M13 9PL, UK\\
             }

   \date{Received February 18, 2026; accepted June 22, 2026}
   
   \titlerunning{Linking simulations and synthetic observations of high-mass star-forming cores}
   \authorrunning{B.\ Zimmermann et al.}

   \abstract
    {Dense cores and clumps in massive star-forming regions are commonly studied through (sub-)millimetre continuum observations, but inferred masses and temperatures can be biased by limited resolution, interferometric filtering, projection, and uncertain dust temperatures.}
    {By comparing numerical simulations with synthetic observations of massive star-forming clumps, we aim to quantify how reliably core masses and temperatures can be inferred from \textsc{ALMA}-like continuum data.}
    {We analysed radiation–hydrodynamic simulations of collapsing 1000~$M_\odot$ clumps post-processed with \textsc{RADMC-3D} and \textsc{CASA} to generate synthetic ALMAGAL-like images. Cores were defined around sink particles, allowing us to compare masses from the 3D simulations, idealised \textsc{RADMC-3D} maps, and synthetic ALMA images. We quantified how resolution, projection, and viewing angle affect core masses, and tested three dust-temperature estimators (clump-scale $L/M$, a $L/M{+}$flux model, and a supervised machine learning-based estimate).}
    {Projection effects cause line-of-sight integrated core masses within 1500~au to exceed the corresponding spherical masses by factors of two to three. The empirical $L/M$ temperature prescription follows the global thermal evolution but misses the core-to-core scatter, producing individual core-mass errors of a factor of a few. Including local flux information improves temperature and mass estimates, and tests with a gradient-boosting machine learning model suggest further improvement. Synthetic ALMAGAL-like images recover the global core structure, but flux-reconstruction uncertainties introduce mass errors up to a factor of about three and limit reliable core-mass recovery to $\gtrsim 0.3$~$M_\odot$, comparable to the ALMAGAL completeness limit. We also derived an empirical relation between projected core mass and expected sink mass.}
    {Comparing synthetic ALMAGAL-like observations with known simulation properties quantifies the main biases in continuum-derived core masses. Individual core masses are typically uncertain by a factor of around three to five, mainly due to dust-temperature assumptions and flux reconstruction. Reducing these biases requires temperature estimates that include local core information rather than only a single clump-scale $L/M$-based prescription.}


   \maketitle

\section{Introduction}
\label{sec:introduction}

High-mass stars and stellar clusters play a central role in shaping the interstellar medium through their ionising radiation, winds, and supernovae \citep[e.g.][]{2007Zinnecker, 2014Tan, 2018Motte}. Yet the physical processes that govern their formation remain poorly constrained. A key step towards understanding high-mass star formation is to characterise the dense cores embedded in massive collapsing clumps since their masses, temperatures, and evolution set the initial conditions for individual stars and multiple stellar systems \citep[e.g.][]{2003McKeeTan, 2014Andre, 2014Offner}.

Large interferometric surveys such as ALMAGAL \citep[ALMA Evolutionary Study of High-Mass Protocluster Formation in the Galaxy, see][]{2025Molinari, 2025Sanchez} are designed to address this problem observationally. ALMAGAL targets more than a thousand dense clumps across the Milky Way spanning a broad range of environments and evolutionary stages and provides (sub-)millimetre continuum and line data at typical spatial resolutions of order of $10^3$~au. These data enable statistical studies of fragmentation, core growth, and feedback, among others, and are expected to serve as a benchmark for theoretical models of high-mass star and stellar cluster formation. The first set of ALMAGAL publications \citep[see e.g.][]{2025Coletta,2025Mininni,2025Elia, 2025Schisano,2026Wallace} uses the ALMA continuum emission at 1.3~mm to identify dense cores and determine properties such as their mass, size, and density.

Interpreting such continuum observations is, however, non-trivial. For instance, core masses inferred from dust continuum emission at millimetre wavelengths depend on the assumed dust temperature, opacity, and geometry \citep[e.g.][]{1983Hildebrand,Beuther2007, 2014Andre} and are further affected by line-of-sight (LOS) projection effects and interferometric filtering \citep[e.g.][]{2012Offner,2017Seifried, 2017AKoepferl,2023Padoan,2025Nucara}. In practice, the dust temperature of individual cores is rarely measured directly. Within ALMAGAL, so far, an empirical prescription assigns a single temperature to all cores in a region (or clump) based on its luminosity-to-mass ratio ($L/M$), which is taken as a proxy for their evolutionary stage \citep{2025Coletta}. This approach captures the global heating history, but it cannot account for core-to-core temperature variations driven by local accretion, clustering, feedback, and geometry. Such temperature mismatches introduce systematic biases in the inferred core masses \citep[e.g.][]{2011Malinen,2023Traficante}, and
additional uncertainties arise from flux reconstruction in interferometric imaging and from the fact that continuum cores are intrinsically projected entities contaminated by foreground and background material \citep[e.g.][]{2012Offner}.

Numerical simulations provide a complementary way to quantify and interpret these biases on core and clump scales \citep[e.g.][]{2020Rosen,2023Padoan,2025Lebreuilly}. When such simulations are post-processed with radiative transfer tools, the resulting synthetic observations become a powerful tool for interpreting real data and for guiding the development of numerical models \citep[e.g.][]{2018Haworth}. When they are additionally imaged with realistic telescope configurations (e.g. mimicking ALMA observations), these synthetic datasets can be analysed in exactly the same way as real observations, while the underlying ``true'' core properties remain known. The post-processing and comparison of 3D simulation data and 2D synthetic observations offers a controlled laboratory to calibrate temperature prescriptions, assess the impact of projection and flux reconstruction, and link observed continuum cores to their underlying stellar content.

In this paper, we use the radiation–hydrodynamic simulations of \citet[][]{2025Zimmermann} as a testbed to confront simulations and synthetic observations of high-mass star-forming clumps. We focus on two runs that differ only in numerical resolution, and we post-processed each snapshot with the radiative transfer code \textsc{RADMC-3D} \citep{2012Dullemond}. Synthetic ALMA observations that mimic ALMAGAL cores were then calculated with CASA \citep{2022CASATeam}. We defined the cores around individual sink particles with a fixed physical radius, thus allowing for a direct one-to-one comparison between simulated structures and their observational counterparts.

Our aims in this work are twofold. (i) On the theoretical side, we seek to quantify how numerical resolution and projection effects bias core and sink masses as well as the clump-scale core formation efficiency (CFE). (ii) On the observational side, we aim to evaluate several dust-temperature estimators and how together with flux reconstruction they affect continuum-based core masses. The paper is organised as follows. In Sect.~\ref{sec:simulations} we briefly introduce the numerical simulations and describe the process of construction of the synthetic observations. In Sect.~\ref{sec:masses} we analyse core and sink masses, projection effects, and the evolution of the CFE. In Sect.~\ref{sec:temps} we investigate different dust-temperature estimators and their impact on mass estimates. In Sect.~\ref{sec:warnings-guidelines} we provide a compact `recipe of warnings' and quantitative guidelines for interpreting observational dense core catalogues, such as that of ALMAGAL, and in Sect.~\ref{sec:conclusions} we summarise our main findings.

\begin{figure*}[t!]
    \centering
        \includegraphics[width=0.92\textwidth,trim=0 0.7cm 0.0cm 0.8cm, clip]{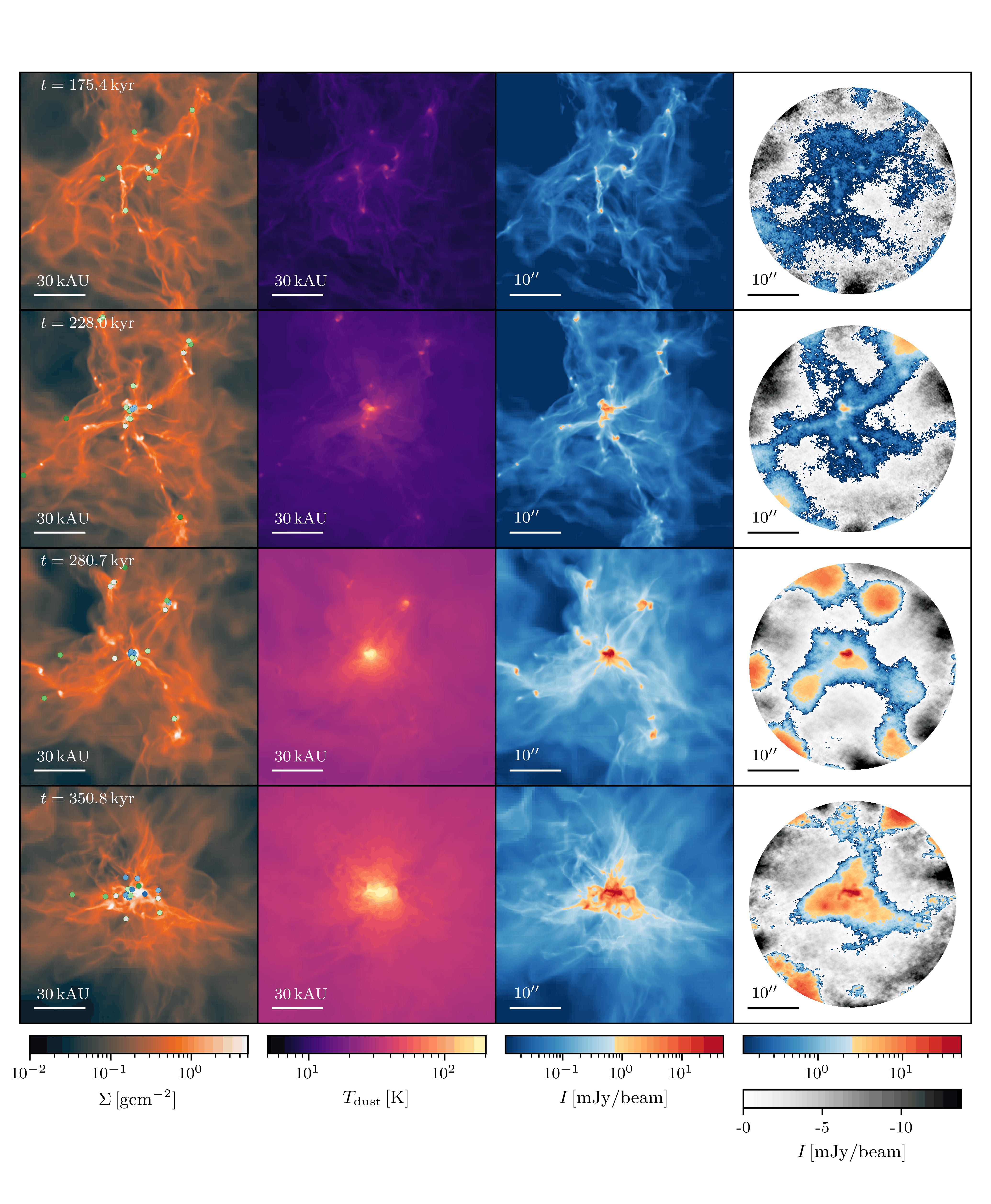}
    \caption{Time evolution of run \textsc{rfl9}. \textit{Left two columns}: Intrinsic simulation outputs: column density and dust temperature. \textit{Third column}: Flux maps obtained through post-processing with \textsc{RADMC-3D} at $1362.7 \, \mu \rm m$ and a distance of $3 \, \rm kpc$. \textit{Right column}: Synthetic \textsc{ALMA} observations generated with CASA while taking into account the interferometric response. }
    \label{fig:introduction_plot}
\end{figure*}

\begin{figure*}[ht!]
    \centering
        \includegraphics[width=0.8\textwidth]{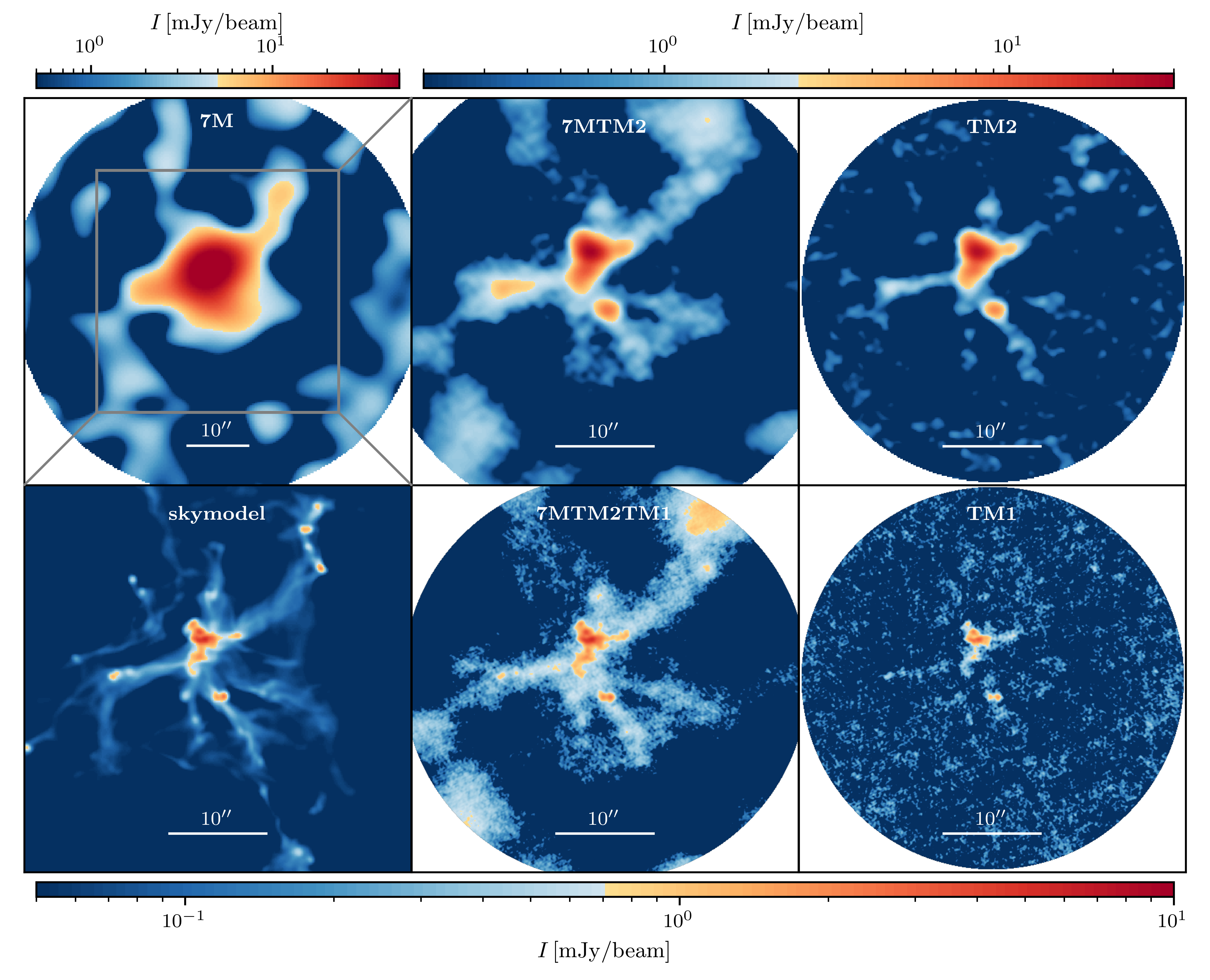}
    \caption{Flux distributions from different \textsc{ALMA} telescope array configurations simulated with CASA at $230 \, \rm kyr$, shown for the LOS along the $z$-direction. \textit{Top row}: The 7~m array (left; geometric mean resolution $6.1^{\prime\prime}$, $18\,360$~au), combined 7~m+TM2 arrays (middle; $1.4^{\prime\prime}$, $4\,340$~au), and TM2 alone (right; $1.3^{\prime\prime}$, $3\,850$~au). \textit{Bottom row}: Sky model (left), combined 7~m+TM2+TM1 configuration (middle; $0.41^{\prime\prime}$, $1\,220$~au), and TM1 alone (right; $0.30^{\prime\prime}$, $890$~au). The combination of short and long baselines is required to recover both compact cores and extended emission.}
    \label{fig:kachel_casa_arrays}
\end{figure*}

\begin{figure*}[t!]
    \centering
    \includegraphics[width=0.9\linewidth,trim=0 0cm 0.6cm 0cm, clip]{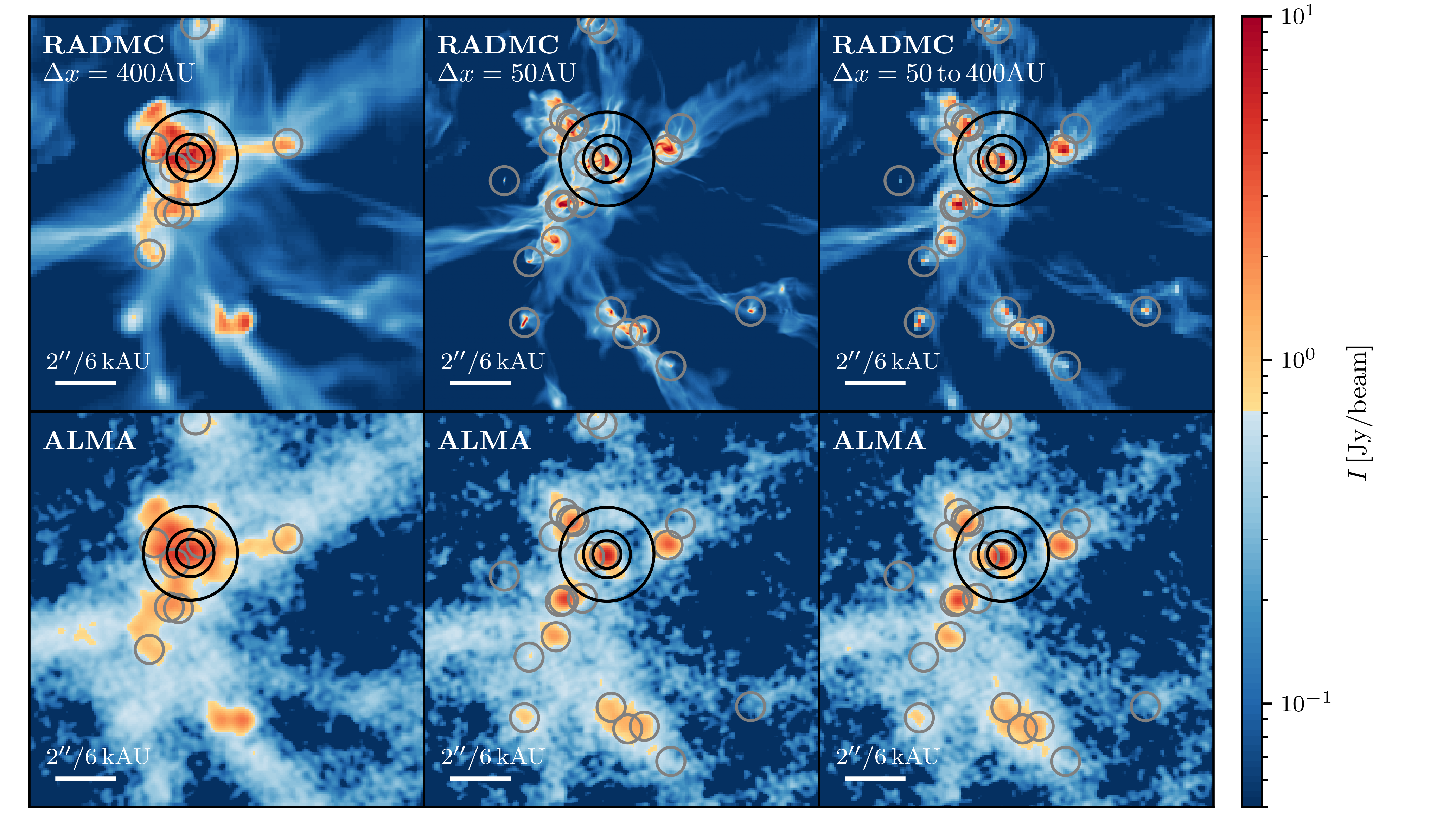}
    \caption{Synthetic \textsc{RADMC-3D} emission (\textit{top panels}) and \textsc{ALMA}-imaged counterparts (\textit{bottom panels}) at $t=230\,\mathrm{kyr}$, shown for the LOS along the $z$-direction. From left to right: Native resolution 400~au (\textsc{rfl9}), native resolution 50~au (\textsc{rfl12}), and initial resolution 50~au. Only for the right-hand panels, the \textsc{RADMC-3D} output was first downsampled to a common grid of 400~au, and the \textsc{ALMA} images were then generated from these downsampled maps. Grey circles indicate the sink particle (core) locations. Black circles mark physical radii of 1500~au, 2500~au, and 5000~au centred on the most massive sink particle. A higher intrinsic simulation resolution resolves more substructure and produces more sharply defined cores. Even after downsampling, the high-resolution run retains fine structure that the lower-resolution simulation cannot reproduce. The \textsc{ALMA} maps are limited by an effective resolution of $\sim1200$~au, but the \textsc{rfl12}-based downsampled \textsc{ALMA} images (bottom right) still show more clearly defined cores than the lower-resolution case. The intrinsic simulation resolution sets which features remain visible and how well the cores are delineated.}
    \label{fig:grid_zoomin_radii}
\end{figure*}

\section{Simulations and synthetic observations}
\label{sec:simulations}

\subsection{Numerical simulations}
\label{sec:NumericalSimulations}

The simulations used in this work are described in detail in \citet{2025Zimmermann}; here we summarise only the main aspects. The simulations follow the collapse of a massive clump with the adaptive mesh refinement  code \textsc{FLASH} \citep{fryxel2000}, using an entropy-stable (magneto-) hydrodynamics scheme \citep{2016Derigs}. Gas and dust are treated as perfectly mixed with a fixed dust-to-gas mass ratio of 0.01. A non-equilibrium chemical network evolves seven species (H$_2$, H, H$^+$, CO, C$^+$, O, e$^-$) with reactions from \citet{Nelson_1997} and updates from \citet{glover_modelling_2010}.

Self-gravity and radiative transfer are handled with the tree-based \textsc{TreeRay} scheme \citep{wunsch2018, 2021Wunsch}. The TreeRay/OpticalDepth module \citep{wunsch2018} accounts for dust shielding as well as the (self-)shielding of molecular hydrogen and CO, while TreeRay/OnTheSpot \citep{2018Haid, 2021Wunsch} models ionising radiation from sink particles using the on-the-spot approximation \citep{Osterbrock_1988}. In these runs, radiation pressure from ionising photons onto gas and dust is calculated, and the corresponding momentum is transferred via TreeRay/RadPressure \citep{2023Kleptiko}. This model also includes the heating of dust from non-ionising photons. The stellar feedback (ionising and non-ionising luminosities) follows the sink accretion histories as described in \citet{2023Kleptiko}.

Sink particles represent individual stars and are created self-consistently following the method of \citet{Dinnbier_2020}. The density threshold for sink formation is tied to the local Jeans length at the maximum refinement level,
\begin{equation}
    \rho_{\rm thresh} = \frac{\pi c_{\rm s}^2}{G \lambda_{\rm J}^2}
    = \frac{\pi c_{\rm s}^2}{G (4\,\Delta x)^2} ,
    \label{eq:rhothresh}
\end{equation}
where $c_{\rm s}$ is the local sound speed, $G$ the gravitational constant, and $\Delta x$ the smallest cell size \citep{truelove1997}. Additional checks ensure that the gas is gravitationally bound and collapsing, that the location is a potential minimum, and that it lies outside the accretion radius ($r_{\mathrm{acc}} = 2.5 \Delta x$) of existing sinks. Each sink is evolved with a fourth-order Hermite integrator, and its radius and effective temperature follow the accretion-dependent protostellar evolution models of \citet{Offner_2009}. Note that at our resolution, a sink particle does not necessarily represent a single star, but rather a protostar together with its disk and inner envelope. This interpretation is appropriate, as the observational data resolve structures on similar spatial scales ($\sim 1200\,\mathrm{au}$).

The initial conditions of the simulations are chosen to be representative of ALMAGAL clumps \citep[see e.g.][]{2025Molinari}. The computational domain is a $(4~\mathrm{pc})^3$ box containing a $1000\, \rm M_\odot$ clump within a radius of 1~pc at solar metallicity. The clump follows a Plummer-like density profile with an outer slope of $-2$ and has a virial parameter of $\alpha_{\rm vir}=0.6$, ensuring global collapse. The initial gas temperature is $20$~K; the dust temperature is set to $10$~K and is rapidly brought into thermal equilibrium. No external interstellar radiation field is included, as the region is deeply embedded. In this work, we use two runs from the resolution study performed in \citet{2025Zimmermann} that share the same global setup but differ in maximum refinement level, $l_{\rm ref}=9$ and 12 (hereafter \textsc{rfl9} and \textsc{rfl12}, respectively), corresponding to minimum cell sizes of 400 and 50~au, respectively. 

In the following analysis, we make use of snapshots that cover the entire evolution of the clump, from the onset of collapse to the stage where stellar feedback disrupts the clump.
Run \textsc{rfl9} reaches $470$\,kyr, while run \textsc{rfl12} reaches $240$\,kyr. In both runs, the time between consecutive snapshots is about 1.8\,kyr.
The temporal evolution of \textsc{rfl9}, showing column density, dust temperature, and the corresponding \textsc{RADMC-3D} (see also Sect. \ref{sec:radmc}) and \textsc{ALMA} (see also Sect. \ref{sec:alma_synth}) post-processed maps (from left to right), is presented in Fig.~\ref{fig:introduction_plot}. Here, the first sink particles form after $\sim 93\,\mathrm{kyr}$. The sink population grows rapidly during the subsequent global collapse phase. Ionising radiation first appears once the most massive sink reaches $8\, \rm M_\odot$, typically at $t \simeq 240\,\mathrm{kyr}$, after which the thermal and density structure of the clump begin to deviate from pure gravitational collapse. As the H\,\textsc{ii} region expands, it eventually disrupts the parent clump at $t \simeq 370\,\mathrm{kyr}$, at which point most of the dense gas is dispersed. These characteristic evolutionary stages (the onset of sink formation, onset of ionisation, eventual clump disruption) provide the temporal context for the synthetic observations and the trends presented in later sections.

\subsection{Radiative transfer and synthetic observations}
\label{sec:radiative transfer}

The numerical simulations described in Sect.~\ref{sec:NumericalSimulations} serve as a testbed to confront simulations with observations. The ultimate goal is to compare with real observational data from the ALMAGAL survey \citep{2025Molinari}. In this work, we compare the simulation outputs with their corresponding synthetic observations. To achieve this, we post-process the simulations using radiative-transfer techniques (see Sect.~\ref{sec:radmc}) and simulate the telescope configurations used in the ALMAGAL project (see Sect.~\ref{sec:alma_synth}), mimicking the conditions of real observations.

\subsubsection{RADMC-3D post-processing}
\label{sec:radmc}

To produce synthetic dust continuum emission maps, we post-process our simulation snapshots with \textsc{RADMC-3D} \citep{2012Dullemond} to compute the intensity by integrating the radiative transfer equation along rays through the three-dimensional grid, accounting for dust absorption and thermal emission and neglecting scattering. For the post-processing, we provide \textsc{RADMC-3D} with the dust temperature computed self-consistently in the simulation, as well as the corresponding gas and dust density distribution. 

With these precomputed fields, we generate synthetic dust continuum emission maps at $1.36\,\mathrm{mm}$ for a clump located at a distance of $3\, \rm kpc$. This corresponds to the wavelength used in the ALMAGAL project to study the continuum emission from cold, dense dust in massive clumps. For the dust opacity we adopt the \citet{weingartner2001} model with $R_V=4.0$, corresponding to an absorption opacity of $\kappa_{1.3\,\mathrm{mm}} = 0.32\,\mathrm{cm^2\,g^{-1}}$ per unit dust mass. Each 3D simulation snapshot is post-processed for three LOS along the $x$, $y$, and $z$ axes, yielding three 2D continuum maps at $1.36\,\mathrm{mm}$.

\subsubsection{ALMA (CASA) post-processing}
\label{sec:alma_synth}

We simulated ALMA band~6 continuum observations using the software \textsc{CASA} \citep[Common Astronomy Software Applications;][]{2022CASATeam}, following the observing and imaging philosophy adopted in the \textsc{ALMAGAL} project \citep[see][]{2025Sanchez}. This step converts the idealised RADMC-3D maps into synthetic interferometric observations.

We initiated the process using noise-free two-dimensional \textsc{RADMC-3D} specific-intensity maps as input sky brightness models. Synthetic interferometric visibilities were generated for these sky models using the task \texttt{simobserve}. We considered three different telescope configurations, corresponding to those used in the ALMAGAL project: 7\,m (ACA or 7M), compact 12\,m (TM2), and extended 12\,m \citep[TM1, see][for more details]{2025Sanchez}. For the spectral setup, we set the central observing frequency at $\nu_0=220$~GHz (i.e. ALMA band~6) and used a total bandwidth $\Delta\nu=3$~GHz, equivalent to the frequency and effective bandwidth of the ALMAGAL continuum observations. We used the same on-source integration times as in the ALMAGAL observations for the different arrays (i.e. between 2 and 4~minutes). The phase centre of the synthetic observations was set at $\alpha_\mathrm{J2000}=16^\mathrm{h}10^\mathrm{m}00^{s}$, $\delta_\mathrm{J2000}=-35\degr30\arcmin00\arcsec$, mimicking the coordinates of a southern-hemisphere source. Finally, thermal noise was added using the \textsc{CASA} atmospheric plus receiver model for band~6, adopting a precipitable water vapour $\mathrm{PWV}=1.7$~mm.

Imaging of the synthetic visibilities was performed using the \texttt{tclean} task. We used a Briggs weighting (\texttt{robust}=0.5) and restricted the imaged region to the 20\% level of the primary-beam area via \texttt{pbmask=pblimit=0.2}. Primary-beam correction was applied to the final images. Deconvolution followed an ALMAGAL-style iterative masking procedure \citep[see][]{2025Sanchez}, in which CLEANing masks were updated based on signal-to-noise emission and residual-noise estimates until convergence. First, we ran a shallow CLEAN with only 100 iterations to obtain an initial partially CLEANed image. Next, we used this first image to build a CLEANing mask based on the signal-to-noise ratio ($\mathrm{S/N}=I/\sigma$, where $I$ is the pixel intensity and $\sigma$ is the rms noise). The mask included those regions with emission above a user-defined low-S/N threshold only when it was spatially connected to a user-defined high-S/N detection. We automatically and iteratively updated this mask (up to five times) and relaxed the S/N thresholds in later iterations. Within each mask iteration, we continued CLEANing in blocks of 100 iterations down to a threshold of $1\sigma$. We stopped when the relative improvement in both the model flux and the residual noise level (estimated from the median absolute deviation) fell below 2\%. Once the iteration process was finished, the final continuum images were generated. As a last step, we applied primary-beam correction with the task \texttt{impbcor} and exported the final images as FITS files. This iterative process was performed to generate images of individual configuration arrays (i.e. 7M, TM2, and TM1) as well as images of combined arrays (i.e. 7M+TM2 and 7M+TM2+TM1). These images correspond to the science products generated within the ALMAGAL project. 

\noindent Figure~\ref{fig:kachel_casa_arrays} shows the flux distributions obtained with the different ALMA array configurations for a representative snapshot of run \textsc{rfl9}. The bottom-left panel displays the sky model, corresponding to the original \textsc{RADMC-3D} input. The 7M image (top-left) has the coarsest resolution ($\sim18{,}360$~au). Its short baselines provide the largest field of view and good sensitivity to extended emission, but blend multiple cores into broad structures. The TM2 configuration (top-right) improves the resolution to $\sim3{,}850$~au, resolving individual features more clearly at the cost of reduced sensitivity to large-scale emission. The 7M+TM2 image (top middle) recovers both extended and intermediate-scale structures. TM1 (bottom-right) reaches the highest resolution ($\sim890$~au), resolving compact cores but filtering out much of the envelope. The fully combined 7M+TM2+TM1 image (bottom-middle, with $\sim1200$~au) incorporates the strengths of all arrays, capturing both compact sources and extended emission. Although some filamentary structure is lost relative to the input map, the main cores and emission peaks remain detectable.

\subsection{Core identification and properties}

In this study, we examine the temperature and mass of individual cores and compare their values from numerical simulations with the corresponding parameters that can be derived from synthetic observations. Throughout this work, we assume that each sink particle in every simulation snapshot represents a core with a radius of 1500~au. As a consequence of this fixed radius, the identified cores may spatially overlap. We note that in this work, our aim is to compare the properties of cores rather than to perform an exhaustive core-identification analysis, as would be required for comparison with observational core-finding strategies, which is left for a later publication. 

The (projected) sink positions therefore define the core locations in the simulations and synthetic observations, ensuring a direct one-to-one comparison between the simulated structures and their observational counterparts. As shown in Fig.~\ref{fig:grid_zoomin_radii}, the sink locations coincide with the peaks of emission in the post-processed RADMC-3D and ALMA images. Before comparing the properties of the cores from synthetic images and numerical observations, we removed objects that do not represent detectable dense cores. For instance, some sinks have very low enclosed masses because ionising radiation clears their inner envelopes. Others may simply form with intrinsically low-mass envelopes. Such objects are not expected to be detectable in post-processing and would introduce non-representative errors.
Since individual cores evolve differently and may be affected by feedback at different times, we do not apply a fixed time cut. Instead, we include only cores with more than $0.1\,M_\odot$ of gas within a 1500~au-radius sphere around the sink particle. This threshold on the spherical mass is guided by the ALMAGAL
core-mass completeness limit of $\sim 0.23\,M_\odot$, which refers to projected continuum-derived masses \citep{2025Coletta}.
Moreover, the smaller field of view of the synthetic ALMA observations (see Sect.~\ref{sec:alma_synth}), compared to the field of view of the numerical and \textsc{RADMC-3D} maps, results in some of the simulated cores being located outside the ALMA field of view. These cores were therefore excluded from the analysis.

Synthetic dust–continuum observations generated with \textsc{RADMC-3D} and CASA allowed us to estimate the masses of dense cores and protostellar envelopes. Under the assumption that the dust emission is optically thin, the total mass can be estimated directly from the observed flux density:
\begin{equation}
\label{eq:mass}
m = \frac{I_\nu D^2}{f_\mathrm{d}\,\kappa(\nu)\,B_\nu(T_\mathrm{d})} \, ,    
\end{equation}
where $I_\nu$ is the measured flux density, $D$ is the distance to the source (set to $3 \, \rm kpc$), $f_\mathrm{d}$ is the dust–to–gas mass ratio (set to 1/100), $\kappa(\nu)$ is the dust opacity at the observing frequency ($\nu = 230 \, \rm GHz$, $\kappa(\nu) =0.32 \, \rm cm^2g^{-1}$, taken from \citealt{weingartner2001}), and $B_\nu(T_\mathrm{d})$ is the Planck function evaluated at the dust temperature $T_\mathrm{d}$ \citep{1983Hildebrand}.

\begin{figure}
    \centering
    \includegraphics[width=1.0\linewidth]{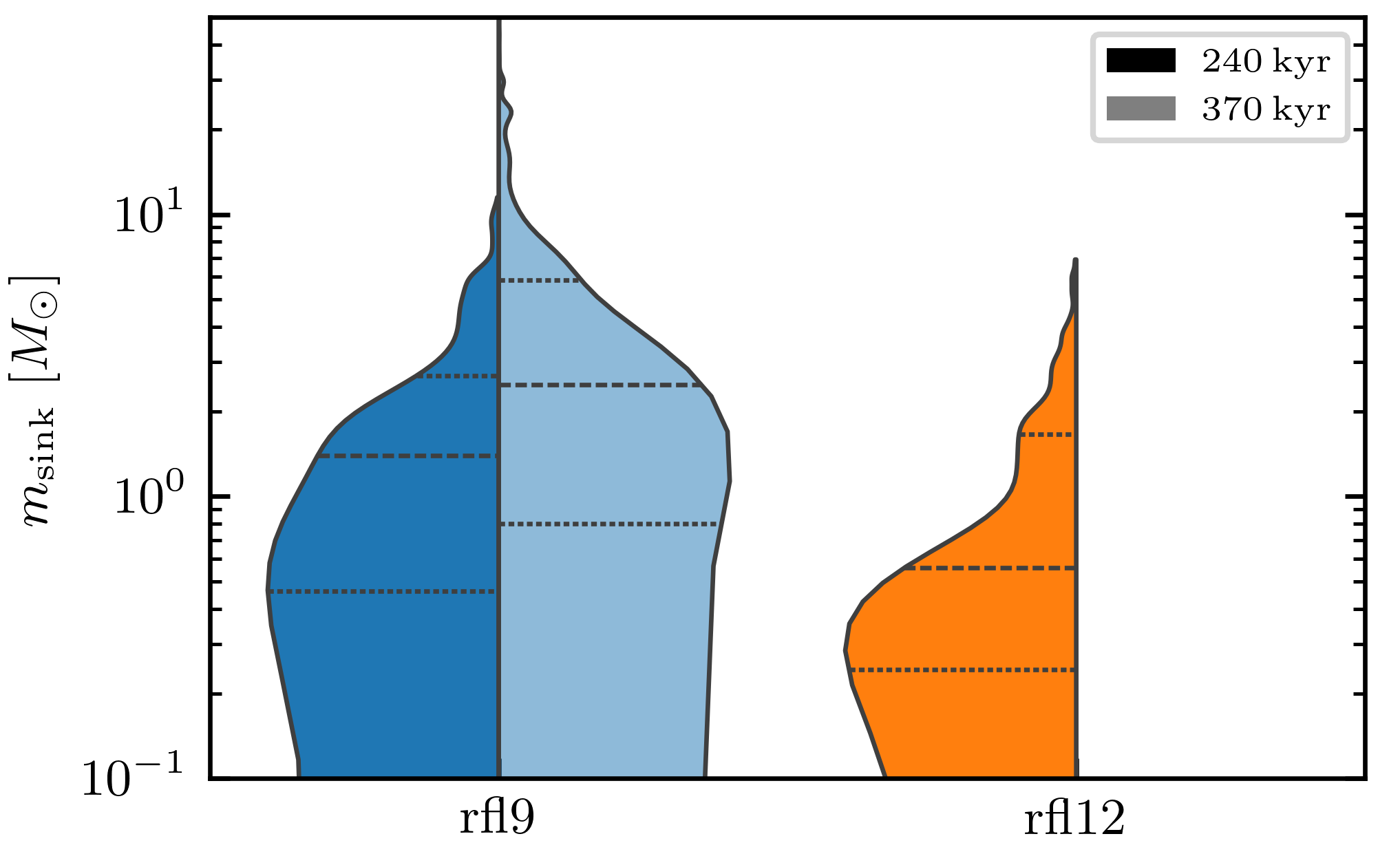}
    \caption{Sink masses for all sinks at each snapshot. Blue and orange colours show \textsc{rfl9} and \textsc{rfl12} up to $240\,\mathrm{kyr}$, respectively. The light blue colour shows \textsc{rfl9} up to $370\,\mathrm{kyr}$. The horizontal lines inside the violins indicate the 25th percentile, median, and 75th percentile. A higher resolution (\textsc{rfl12}) systematically results in lower individual sink masses.}
    \label{fig:violin_flash_Sink_early_vs_all}
\end{figure}

\subsection{Definitions of masses and temperatures}
\label{sec:definitions}

For clarity, we summarise the different mass and temperature parameters used throughout the paper in Table~\ref{tab:mass_temp_definitions}. In the simulations, we distinguish between the sink mass, $m_{\rm sink}$; the spherically enclosed gas mass within radius $r$, $m_{\rm 3D}$; and the projected mass along the LOS, $m_{\rm 2D}$. We additionally considered a purely geometrical cylindrical mass, $m_{\rm cyl}$, defined within a cylinder of radius $r$ and height $2r$. From the \textsc{RADMC-3D} and \textsc{ALMA} emission maps, we derive core masses, $m_{\rm RADMC}$ and $m_{\rm ALMA}$, using Eq.~\ref{eq:mass} with different dust–temperature estimates.

The `true' core dust temperature, $T_{\rm sim,rad}$, is defined as the mass-weighted dust temperature along the LOS within the core radius in projection (2D). The quantity $T_{\rm sim,avg}$ denotes the mass-weighted average of $T_{\rm sim,rad}$ over the core, providing a single representative temperature per core. In addition, we considered several observational temperature estimates (see more in Sect. \ref{sec:LMflux}): (i) $T_{ \rm L/M}$, the empirical temperature assigned from the clump luminosity-to-mass ratio following \citet{2025Coletta}; (ii) a combined $T_{ \rm L/M+F}$ model that depends on both the clump $L/M$ and the flux of each core; and (iii) $T_{\rm ML}$, a supervised machine learning estimate trained on $T_{\rm sim, avg}$. These definitions are used consistently in the following sections.

\begin{table}[t!]
\centering
\caption{Definitions of mass and temperature used in this work.}
\label{tab:mass_temp_definitions}
\begin{tabular}{ll}
\hline \hline \noalign{\smallskip}
Variable  & Description of the variable \\
\hline \noalign{\smallskip}
\multicolumn{2}{c}{Masses} \\
\hline \noalign{\smallskip}
$m_{\rm sink}$      & Sink (stellar) mass from numerical simulation \\
$m_{\rm 3D}$        & Core mass (from simulation) within a sphere \\
                    & \phantom{0} of radius $r_{\rm core}$ (in 3D) \\
$m_{\rm 2D}$        & Projected mass (in 2D) along the LOS \\
                    & \phantom{0} within a circle of radius $r_{\rm core}$ (from simulation) \\
$m_{\rm cyl}$       & Mass within a cylinder of radius $r_{\rm core}$ \\
                    & \phantom{0} and height $2r_{\rm core}$ (from simulation) \\
$m_{\rm RADMC}$     & Mass derived using the flux \\
                    & \phantom{0} of the \textsc{RADMC-3D} continuum map \\
$m_{\rm ALMA}$      & Mass derived using the flux \\
                    & \phantom{0} of the synthetic \textsc{ALMA} continuum image \\
$m_{\rm core}$      & Generic core mass (context-dependent) \\
$m_{\rm clump}$     & Clump mass (from the simulation) \\
                    & \phantom{0} within the adopted clump radius (see Sect.~\ref{sec:clump}) \\
\hline \noalign{\smallskip}
\multicolumn{2}{c}{Temperatures} \\
\hline \noalign{\smallskip}
$T_{\rm sim,rad}$  & Mass-weighted dust temperature \\
                    & \phantom{0} along the LOS (from simulation) \\
$T_{\rm sim,avg}$   & Mass-weighted average of $T_{\rm sim,rad}$ (simulation) \\
$T_{\rm L/M}$           & Temperature from the clump's $L/M$ prescription \\
                    & \phantom{0} (see Sect.~\ref{sec:LM_relation}) \\
$T_{ \rm L/M+F}$         & Temperature from the $L/M{+}$flux model \\
                    & \phantom{0} (see Sect.~\ref{sec:LMflux}) \\
$T_{\rm ML}$        & Temperature from machine learning model \\
                    & \phantom{0} (see Sect.~\ref{sec:MLmodel}) \\
$T_{\rm flux,m2D}$  & Temperature required to reproduce $m_{\rm 2D}$ \\
                    & \phantom{0} from the observed flux \\
\hline
\end{tabular}
\end{table}

\section{Determination and analysis of core masses}\label{sec:masses}

In this section we analyse how the numerical resolution and geometrical projection affect the mass distribution around forming stars (sink particles) in our simulations. We then analyse the core masses derived from the \textsc{RADMC-3D} and ALMA synthetic images in relation to the actual values from the numerical simulations, using the mass and temperature definitions introduced in Sect.~\ref{sec:definitions}.

\subsection{Impact of the numerical resolution}
\label{sec:resolution}

The study of clump fragmentation and sink-particle formation in numerical simulations depends on the simulation's resolution. From \citet{2025Zimmermann}, we know that the number of sink particles increases with increasing resolution, but the total mass in sinks and the corresponding star formation efficiency is comparable across all runs up to $0.45\,t_\mathrm{ff}$ \citep[$\approx 240$~kyr; see][Fig.~13-left panel, and Fig.~14-top panel]{2025Zimmermann}. In this work, we analyse two sets of simulations, with different resolutions: \textsc{rfl9}, with a minimum cell size of 400~au and used for the main analysis in this work, and a higher–resolution run, \textsc{rfl12}, which reaches 50~au. Both achieve spatial resolutions significantly higher than the $\sim 1200$~au resolution of the ALMAGAL observations. 

Figure~\ref{fig:violin_flash_Sink_early_vs_all} shows the distribution of sink masses for the \textsc{rfl9} and \textsc{rfl12} runs at two different evolutionary stages: the onset of ionising radiation at 240~kyr (left) and the point at which the H\,\textsc{ii} region disrupts the clump at 370~kyr (right). Note that simulation \textsc{rfl12} did not reach the later stage due to computational limitations. As indicated in \citet{2025Zimmermann}, higher-resolution runs systematically have lower average sink masses compared to the lower-resolution runs. This trend arises because finer spatial resolution in the numerical simulations enables the gas to reach higher densities, allowing smaller-scale fragmentation to be resolved. As a result, the simulations produce a larger number of sinks with lower individual masses while maintaining a similar total stellar mass.

We now explore the effects of different numerical resolutions in the synthetic images. Figure~\ref{fig:grid_zoomin_radii} shows the synthetic \textsc{RADMC-3D} emission maps (top panels) and the corresponding ALMA synthetic images (bottom panels) for the two simulation runs: \textsc{rfl9} (400~au; left) and \textsc{rfl12} (50~au; middle and right). The top-right panel shows the \textsc{RADMC-3D} image from the high-resolution \textsc{rfl12} run, down-sampled to a 400~au grid, i.e. equivalent to the low-resolution \textsc{rfl9} run. This down-sampled \textsc{RADMC-3D} image is then post-processed to produce a synthetic ALMA image directly comparable to the synthetic ALMA image of the low-resolution \textsc{rfl9} run (see bottom-left panel). Even after downsampling, the higher–resolution simulation \textsc{rfl12} retains a finer substructure level of detail that the lower–resolution run cannot capture, demonstrating that the intrinsic simulation resolution sets the level of detail in the predicted emission, even when post-processed and downsampled to a lower resolution. The synthetic \textsc{ALMA} images shown in the bottom panels of Fig.~\ref{fig:grid_zoomin_radii} have an effective angular resolution set by the synthesised-beam full width at half maximum of $\sim1200$~au. The pixel size of 400~au only specifies the image sampling and corresponds to $\sim3$ pixels across the beam, not to a 400~au resolving power. At this lower resolution, the smallest structures from the numerical simulations are smoothed out. Nevertheless, cores originating from the higher–resolution run (\textsc{rfl12}) remain sharper and more centrally concentrated, whereas those from the lower–resolution run (\textsc{rfl9}) appear more diffuse. This shows that, even after accounting for telescope limitations, the intrinsic simulation resolution still strongly influences the observable appearance of protostellar cores and their immediate environment.

\begin{figure}
     \centering
     \begin{subfigure}[b]{0.45\textwidth}
         \centering
         \includegraphics[width=\columnwidth,trim=0 0 0 0.056cm, clip]{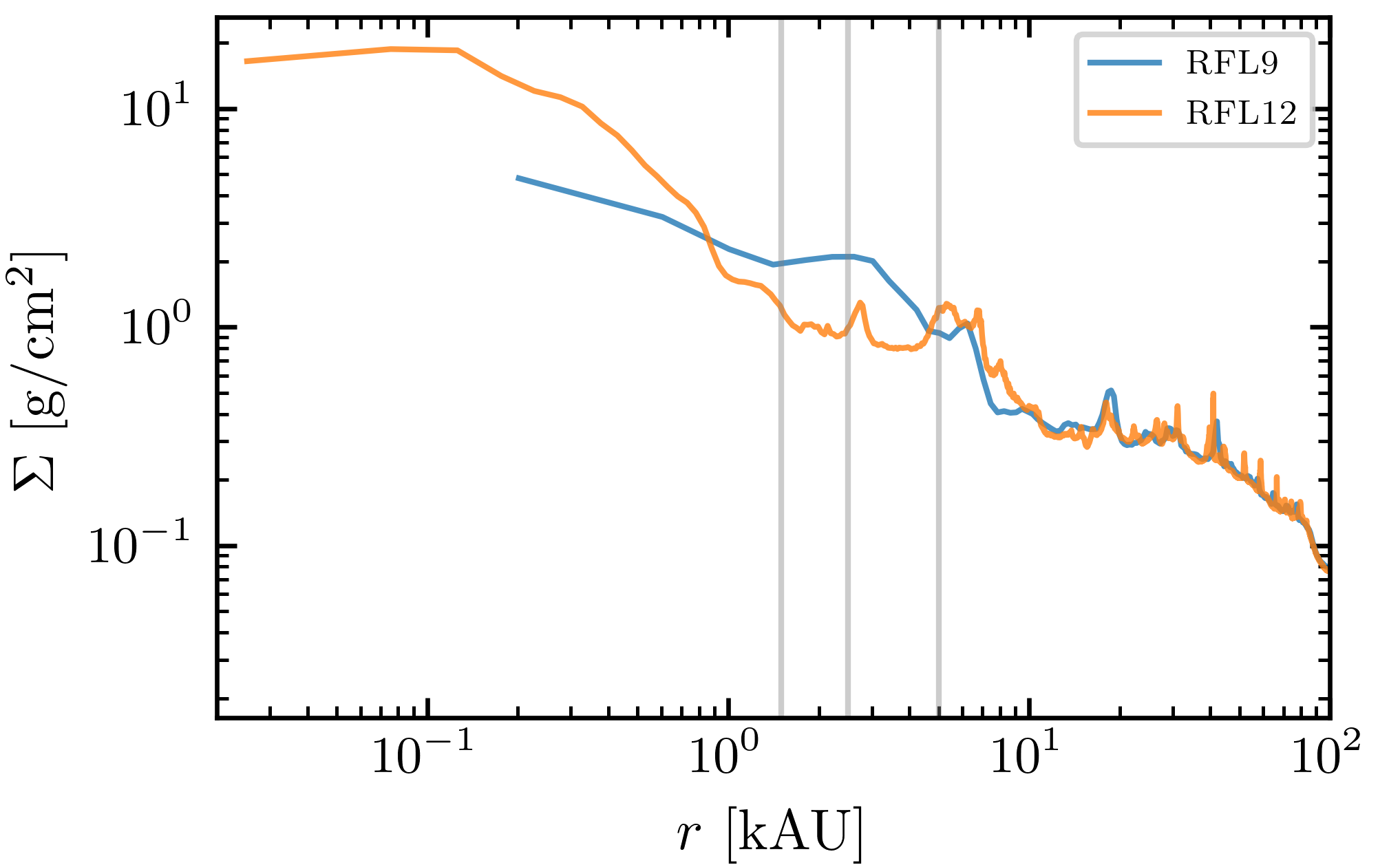}
     \end{subfigure}\\
     \begin{subfigure}[b]{0.45\textwidth}
         \centering
         \vspace{-0.89cm}
         \includegraphics[width=\columnwidth,trim=0 0 0 0.1cm, clip]{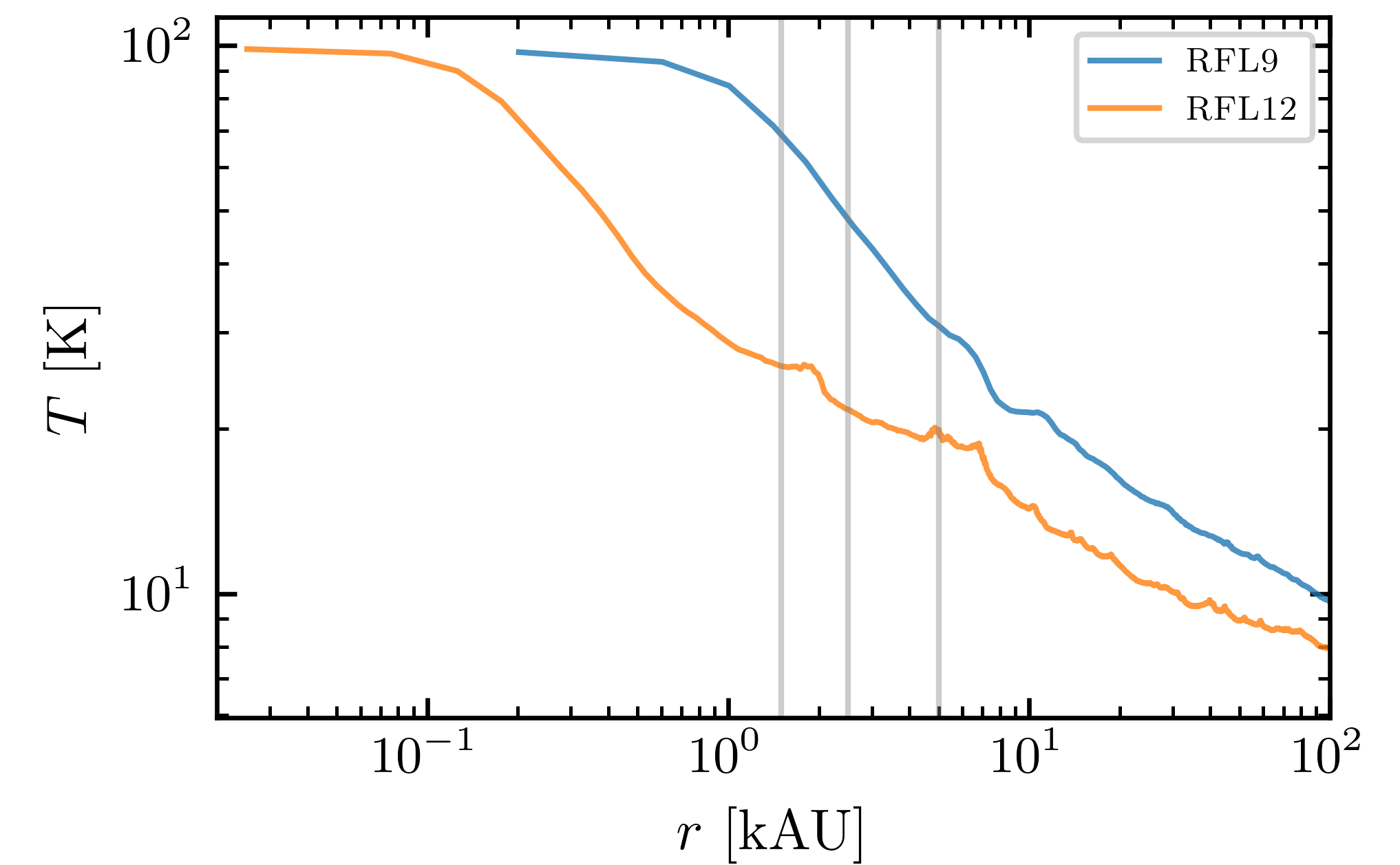}
     \end{subfigure}\\
     \begin{subfigure}[b]{0.45\textwidth}
         \centering
         \vspace{-0.9cm}
         \includegraphics[width=\columnwidth,trim=0 0 0 0.1cm, clip]{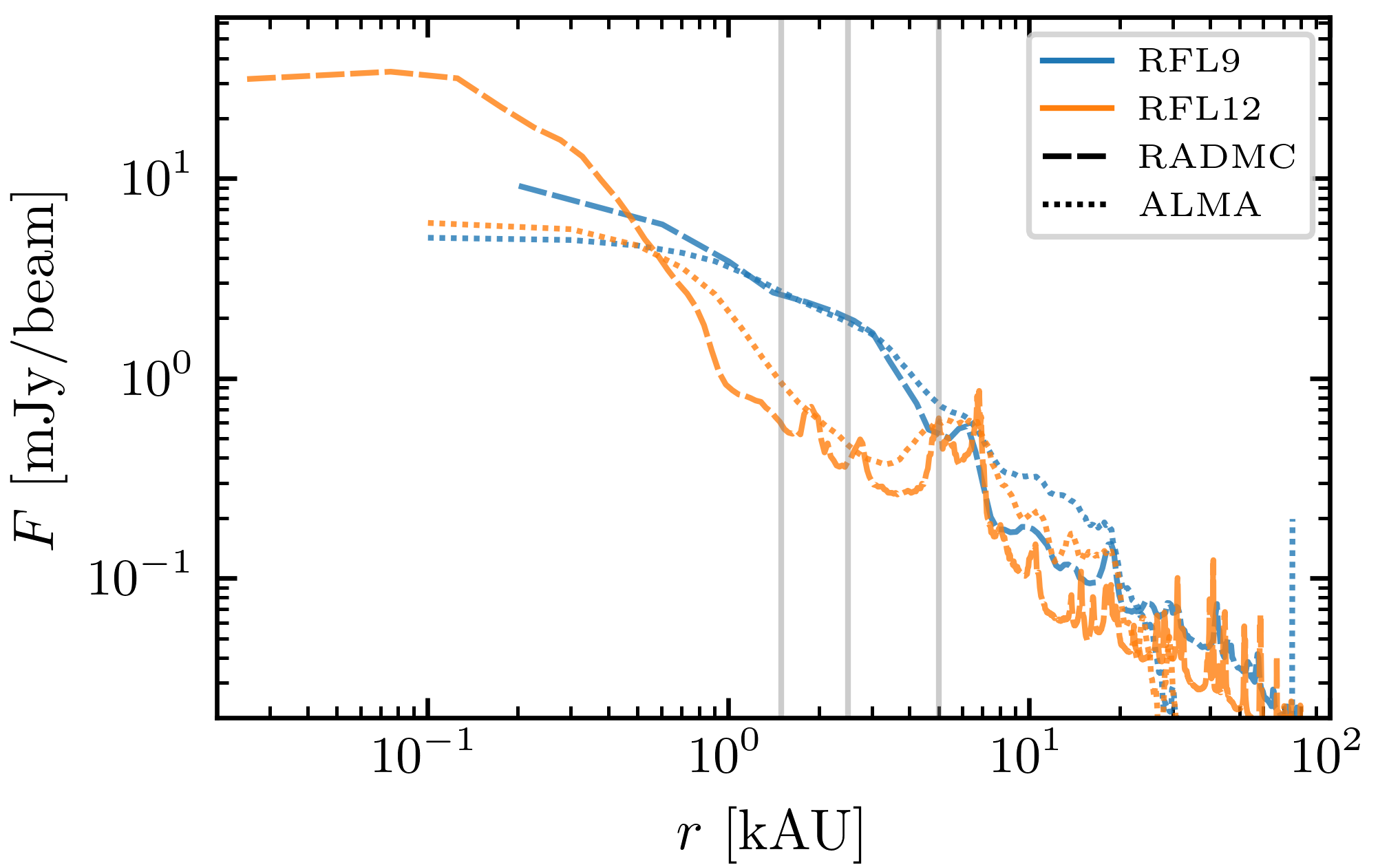}
     \end{subfigure}\\
\caption{Averaged radial profiles of the column density  (\textit{top}), the gas temperature (\textit{middle}), and continuum flux (\textit{bottom}) centred on the most massive sink particle at $t=230,\mathrm{kyr}$. The flux profiles are from both the \textsc{RADMC-3D} images and the corresponding  \textsc{ALMA}  images, while the temperature and column density profiles are directly from the simulation. Results are shown for two spatial resolutions: \textsc{rfl9} ($400\,\mathrm{au}$) and \textsc{rfl12} ($50\,\mathrm{au}$). The vertical lines show the reference radii (from left to right $1.5$, $2.5$ and $5 \, \rm kau$) from Fig. \ref{fig:grid_zoomin_radii}. The higher–resolution run (\textsc{rfl12}) shows a higher central column density and a more rapidly decreasing temperature towards larger radii. \textsc{RADMC-3D} fluxes generally peak higher than the  \textsc{ALMA}  fluxes, particularly at small radii and for the higher–resolution model. In contrast, the lower–resolution run (\textsc{rfl9}) produces a lower central flux peak but a broader outer tail in the radial flux distribution.}
\label{fig:average_radial_flux_profiles}
\end{figure}

These differences are directly reflected in the averaged radial profiles (Fig.~\ref{fig:average_radial_flux_profiles}). The vertical lines in these profiles correspond to the radii highlighted in Fig.~\ref{fig:grid_zoomin_radii}. The profile centres are defined by the sink locations, which are not necessarily located at the centre of a cell. Each simulation and RADMC profile begins at a radius determined by the effective resolution of that run.
The ALMA radial profiles start at smaller radii because the synthetic ALMA images are sampled on a fixed 1024-pixel grid. This reflects the image sampling only and should not be interpreted as a higher physical resolution, which is instead set by the synthesised beam of 1200 au.

The higher-resolution run (\textsc{rfl12}) reaches higher central column densities and shows a steeper radial decline in column density than \textsc{rfl9} (top panel of Fig.~\ref{fig:average_radial_flux_profiles}). At larger radii ($\gtrsim 5~\mathrm{kau}$), the profiles are comparable. 
This difference is a direct consequence of the resolution-dependent sink-formation threshold. Since the threshold density scales with the inverse square of the minimum cell size (Eq.~\ref{eq:rhothresh}), the higher-resolution run allows the gas to collapse to higher densities before sink particles are formed. In contrast, in the coarser-resolution run \textsc{rfl9}, a similar collapsing structure is represented over a larger spatial scale, and the larger accretion radius removes gas from a broader region around the sink. As a result, the central column-density peak is lower and broader than in \textsc{rfl12}.

The peak temperatures of both runs are similar (see the middle panel of  Fig.~\ref{fig:average_radial_flux_profiles}). However, as in the column-density profiles, \textsc{rfl12} shows an earlier decline of temperature with radius. This is mainly caused by the different density structures at the two resolutions. In \textsc{rfl12}, the central density is higher, so the radiation is more strongly shielded and does not heat the gas out to as large radii. In the coarser run \textsc{rfl9}, the density peak is lower and broader, so the temperature decline starts at larger radii. This gives higher average temperatures at intermediate radii in \textsc{rfl9}, although the overall radial slope remains comparable between the two simulations.

The \textsc{RADMC-3D} flux profiles show a pronounced central peak at small radii  (see the bottom panel of Fig.~\ref{fig:average_radial_flux_profiles}). In the high-resolution \textsc{rfl12} run, this peak is approximately a factor of two higher than in \textsc{rfl9}. In contrast, the lower-resolution run \textsc{rfl9} exhibits a broader and more blended envelope structure, leading to a lower central maximum but a more extended flux tail towards larger radii. The corresponding ALMA profiles show systematically reduced peak intensities. Although the ALMA flux peaks are similar for both resolutions, their radial profile generally reflects the underlying \textsc{RADMC-3D} structure: the lower-resolution run exhibits a broader central peak and a more extended outer tail than the high-resolution case.

\subsection{Geometrical effects}
\label{sec:geometry}

\begin{figure}
    \centering
    \includegraphics[width=1.0\linewidth]{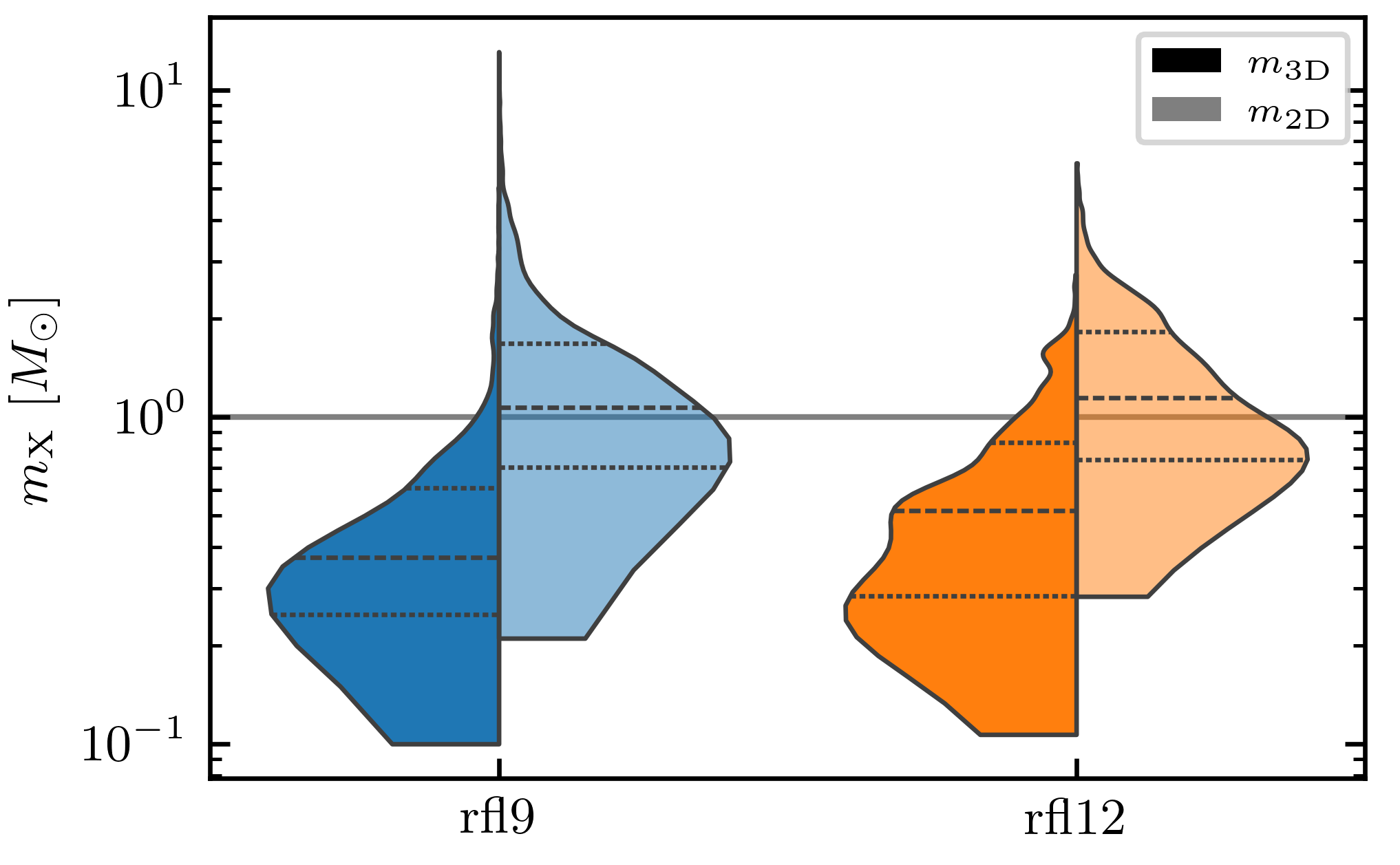}
    \caption{Enclosed mass within a 3D (left halves) or 2D  radius (right halves) of 1500~au up to $240 \, \rm kyr$ for run \textsc{rfl9} (blue) and \textsc{rfl12} (orange). The projected mass, $m_\mathrm{2D}$, is systematically higher than the spherical mass, $m_\mathrm{3D}$, due to the contribution of material in the foreground and background. A higher numerical resolution leads to slightly higher 3D core masses, as higher densities can be resolved.}
    \label{fig:violin_flash_2D_vs_3D}
\end{figure}

From an observational perspective, only the projected mass can be detected through dust continuum emission.  Figure~\ref{fig:violin_flash_2D_vs_3D} presents the enclosed mass within a radius of 1500~au for the different simulation's resolutions. For each run, the right half of the violin represents the projected mass, $m_\mathrm{2D}$, while the left half corresponds to the spherically enclosed mass, $m_\mathrm{3D}$. Across all simulations, $m_{\rm 2D}$ is systematically higher than the true spherical mass, which occurs due to contamination from foreground and background material along the LOS. Although individual sink masses decrease with increasing resolution (see  Fig.~\ref{fig:violin_flash_Sink_early_vs_all}), the corresponding $m_\mathrm{3D}$ increases slightly, representing the opposite effect. This indicates that, at least in these simulations, the spherically enclosed mass is not necessarily a precise proxy for the true protostellar mass, as it primarily reflects the dense gas and dust in the immediate vicinity rather than the stellar component itself (see also Sect.~\ref{sec:sink_core_mass}). Overall, $m_\mathrm{2D}$ shows a similar distribution for both runs, even though the individual sink masses differ.

\begin{figure}
    \centering
    \includegraphics[width=1.0\linewidth]{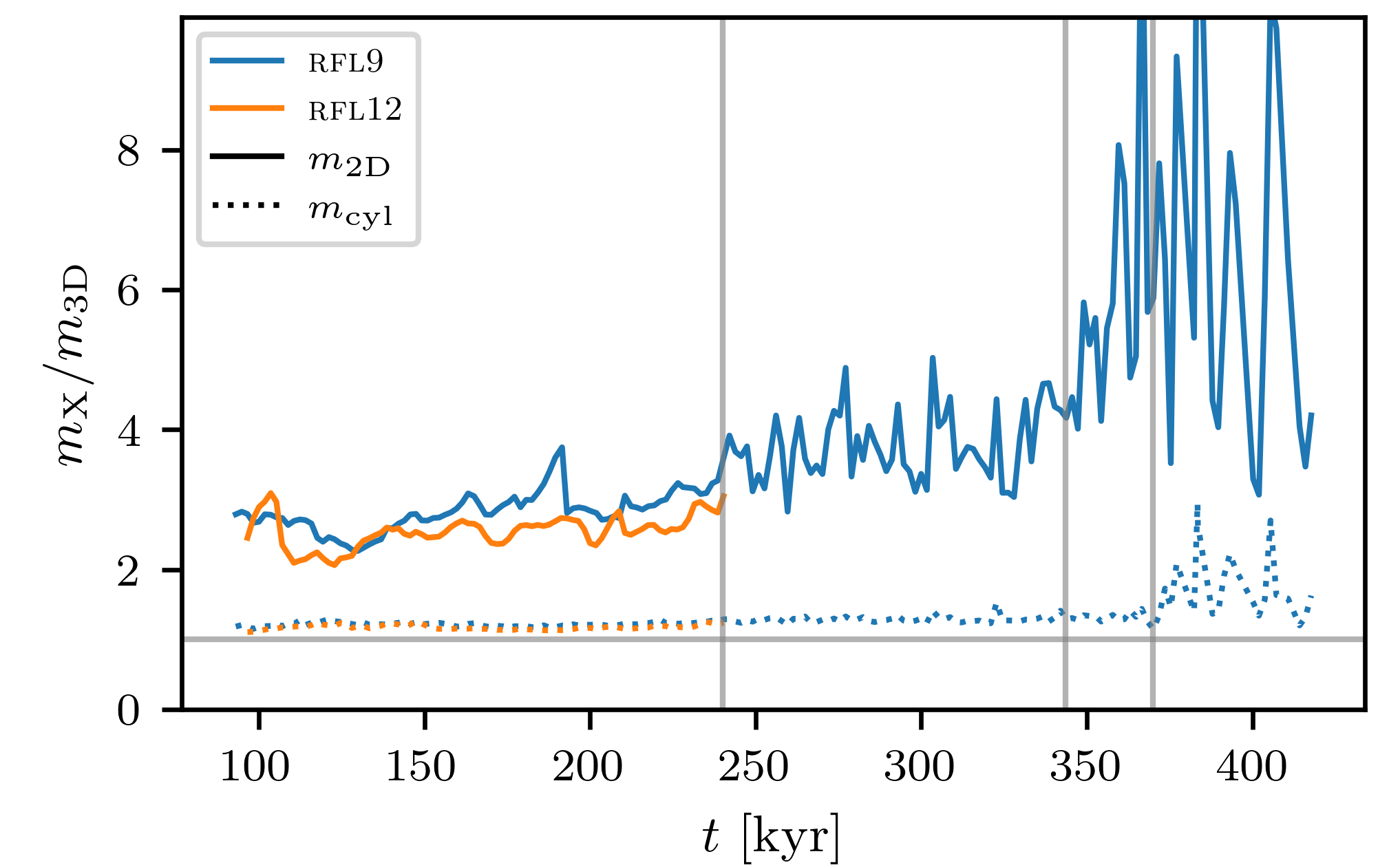}
    \caption{Time evolution of the mean ratio of $m_{\mathrm{2D}}/m_{\mathrm{3D}}$ (solid lines) and $m_{\mathrm{cyl}}/m_{\mathrm{3D}}$ (dotted lines) for different numerical resolutions. In the \textsc{rfl9} run, the $m_{\mathrm{2D}}/m_{\mathrm{3D}}$ ratio remains nearly constant at $\sim 2.9$ until $\sim 240\,\mathrm{kyr}$, after which it increases to $\sim 3.3$ when ionising radiation first appears. Once the expanding H\,\textsc{ii} region begins to disrupt the parental clump at $\sim 370\,\mathrm{kyr}$, both ratios rise sharply. The \textsc{rfl12} run terminates earlier due to computational limitations; in this case, the $m_{\mathrm{2D}}/m_{\mathrm{3D}}$ ratio remains approximately constant at $\sim 2.6$ until $\sim 240\,\mathrm{kyr}$. The cylindrical estimate is closer to the real 3D mass. It systematically overpredicts the enclosed mass by a factor of $\sim 1.2$. }
    \label{fig:resolution_mean_mass_vs_time_all_cnt_flash.png}
\end{figure}

Figure~\ref{fig:resolution_mean_mass_vs_time_all_cnt_flash.png} illustrates the time evolution of the mean ratio between the projected and spherical masses, $m_\mathrm{2D}/m_\mathrm{3D}$, for different numerical resolutions. Up to $\sim 240\,\mathrm{kyr}$, the ratio remains nearly constant at values of 2–3, indicating that the projected mass, $m_\mathrm{2D}$, systematically overestimates the true spherical mass, $m_\mathrm{3D}$, by a factor of two to three during the early stages of collapse. Once ionising radiation begins to affect the clump ($\gtrsim 240\,\mathrm{kyr}$), the ratio increases slightly, and it rises more strongly as the H\,\textsc{ii} region expands ($\sim 340\,\mathrm{kyr}$). This may be due to a decrease in $m_{\rm 3D}$ within the $\sim 1500\,\mathrm{au}$ vicinity as gas is blown out, while $m_{\rm 2D}$ increases because more material disrupted by the H\,\textsc{ii} region falls along the line of sight and is therefore included in the 2D projected mass. By $\sim 370\,\mathrm{kyr}$, the H\,\textsc{ii} region driven by the most massive sink starts to disrupt the clump entirely.

To disentangle the contribution of foreground and background contamination from the purely geometric difference between a cylinder and a sphere, we compute the mass enclosed within a cylindrical volume with a radius of 1500~au and a height of 3000~au (see Fig.~\ref{fig:resolution_mean_mass_vs_time_all_cnt_flash.png}). In this setup, the cylinder is treated purely geometrically, ignoring additional contributions from material along the LOS, so that the comparison isolates the effect of the shape alone. The ratio of cylindrical mass to the spherical mass is nearly constant, with the cylindrical mass exceeding the true spherical mass by approximately 20\,\%. The ratio increases when the parental clump is disrupted at late times. 
For the phases before the strong increase associated with the clump disruption, the projection effect is approximately constant, with $m_{\mathrm{2D}}$ exceeding $m_{\mathrm{3D}}$ by a factor of around two to three.

\subsection{Sink mass versus core mass}
\label{sec:sink_core_mass}

\begin{figure}
    \centering
    \includegraphics[width=1.0\linewidth]{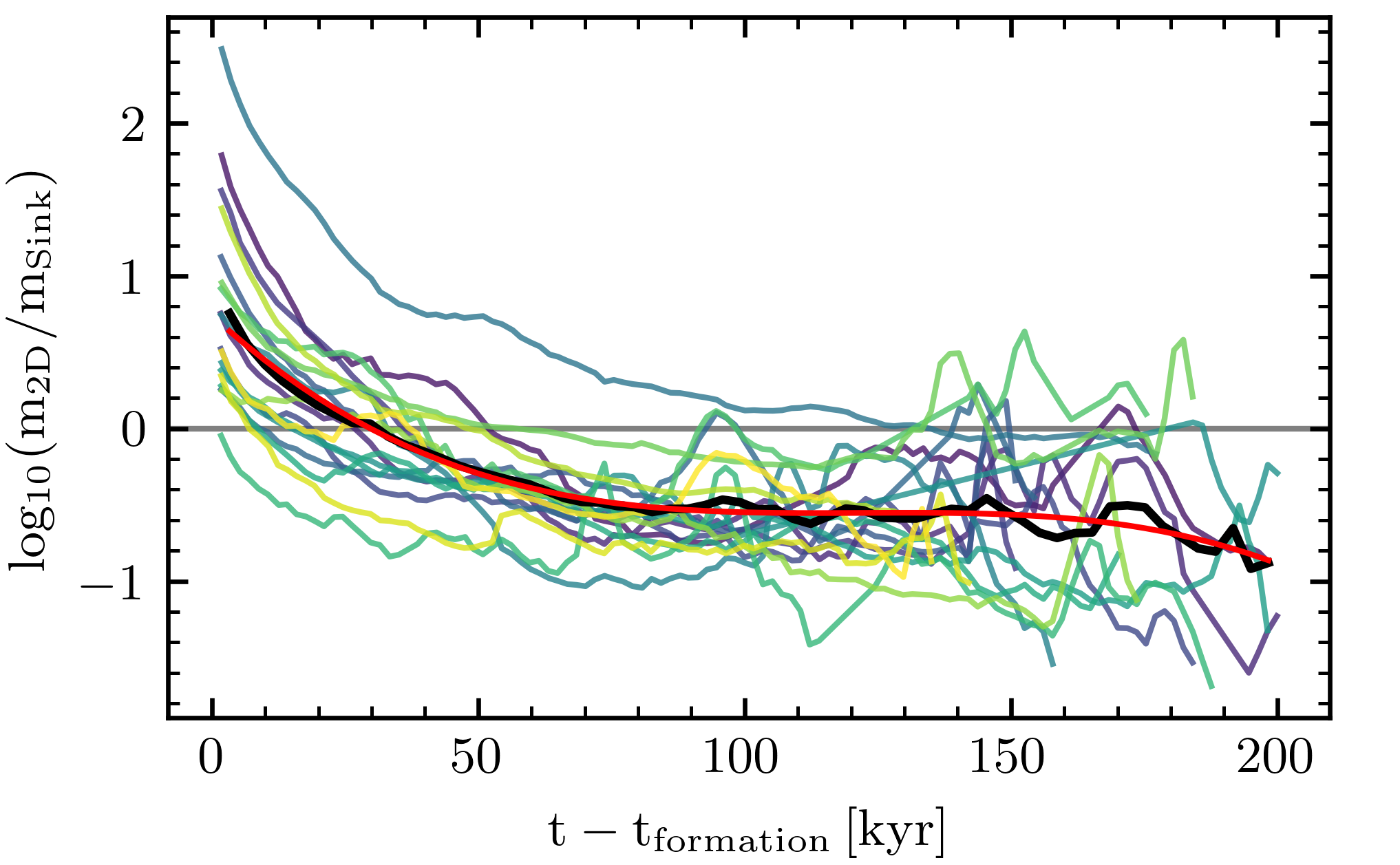}
    \caption{Evolution with time since sink formation of the projected core mass, normalised by the corresponding sink mass. The grey line marks the one-to-one relation where both masses are equal. All sink–core pairs follow a similar trend: At the time of formation, the projected core mass exceeds the sink mass, but this relation rapidly inverts as accretion proceeds. The subsequent fluctuations around the relation likely arise from projection effects and variations in the local density structure.}
    \label{fig:rfl9_update_diffm2D_sink_vs_time}
\end{figure}

In this section, we compare the $m_\mathrm{2D}$ core masses (projected along the z-axis) to the corresponding sink masses. Figure~\ref{fig:rfl9_update_diffm2D_sink_vs_time} shows the time evolution of the ratio between each sink mass and its surrounding core mass, $m_{\rm 2D}$. All sink–core pairs follow a broadly consistent trend. At early times, the inferred core mass is larger than the sink mass for most pairs. As accretion proceeds, the sink mass grows more quickly than the core mass, leading to a crossing point after about $50$~kyr after the formation of the sink, beyond which the sink mass dominates for all pairs.

A similar trend is found when comparing to $m_{\rm 3D}$, which provides the most direct estimate of the intrinsic mass (not shown). From Sect.~\ref{sec:geometry} we know that $m_{\rm 2D}/m_{\rm 3D} \simeq 2-3$. Accordingly, the crossing point occurs earlier, at $\sim 45$~kyr.

\begin{figure}
     \begin{subfigure}[b]{0.45\textwidth}
         \centering
         \includegraphics[width=\columnwidth,trim=0 0 0 0.1cm, clip]{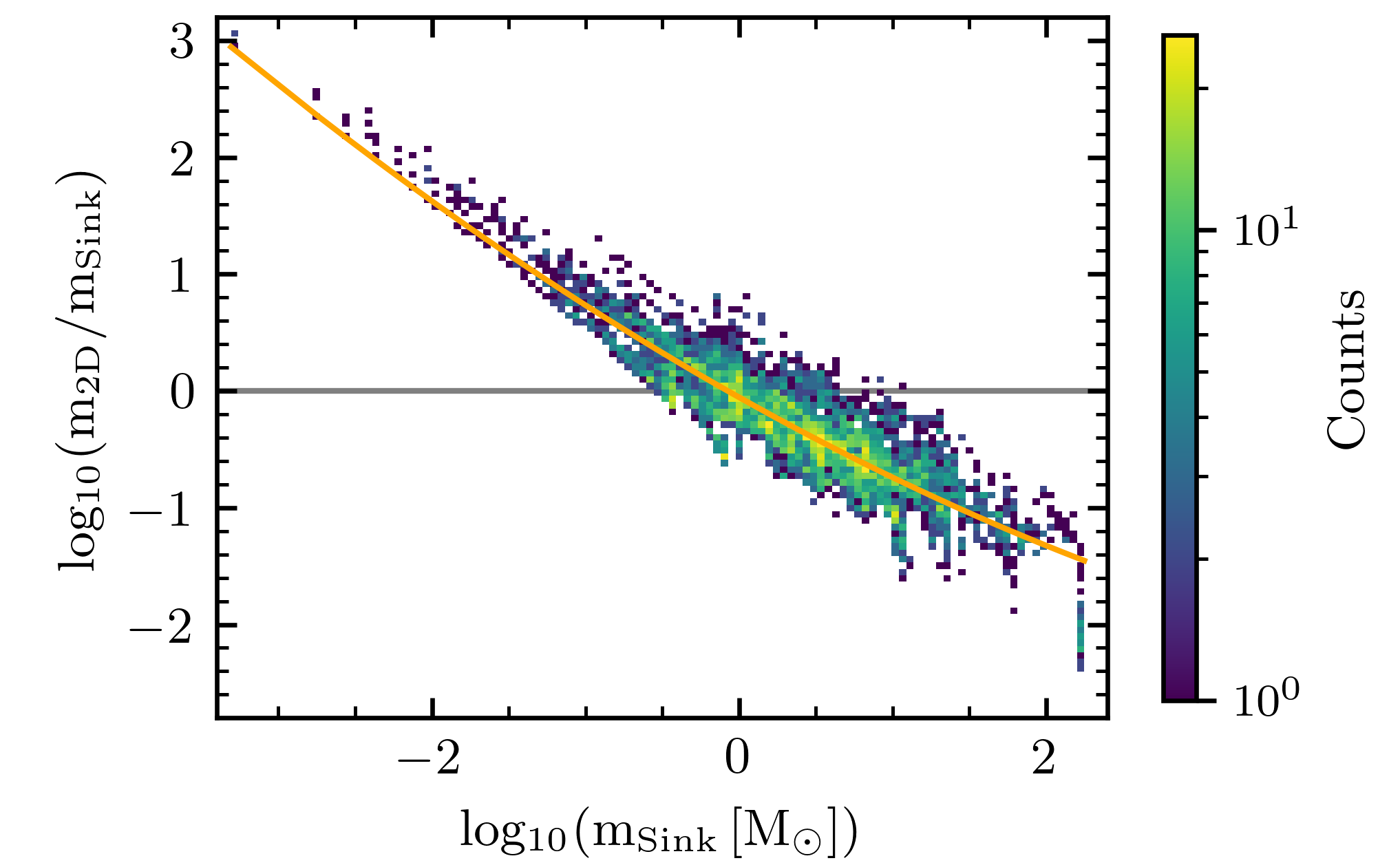}
     \end{subfigure}\\
     \begin{subfigure}[b]{0.45\textwidth}
         \centering
         \vspace{-0.9cm}
         \includegraphics[width=\columnwidth,trim=0 0 0 0.1cm, clip]{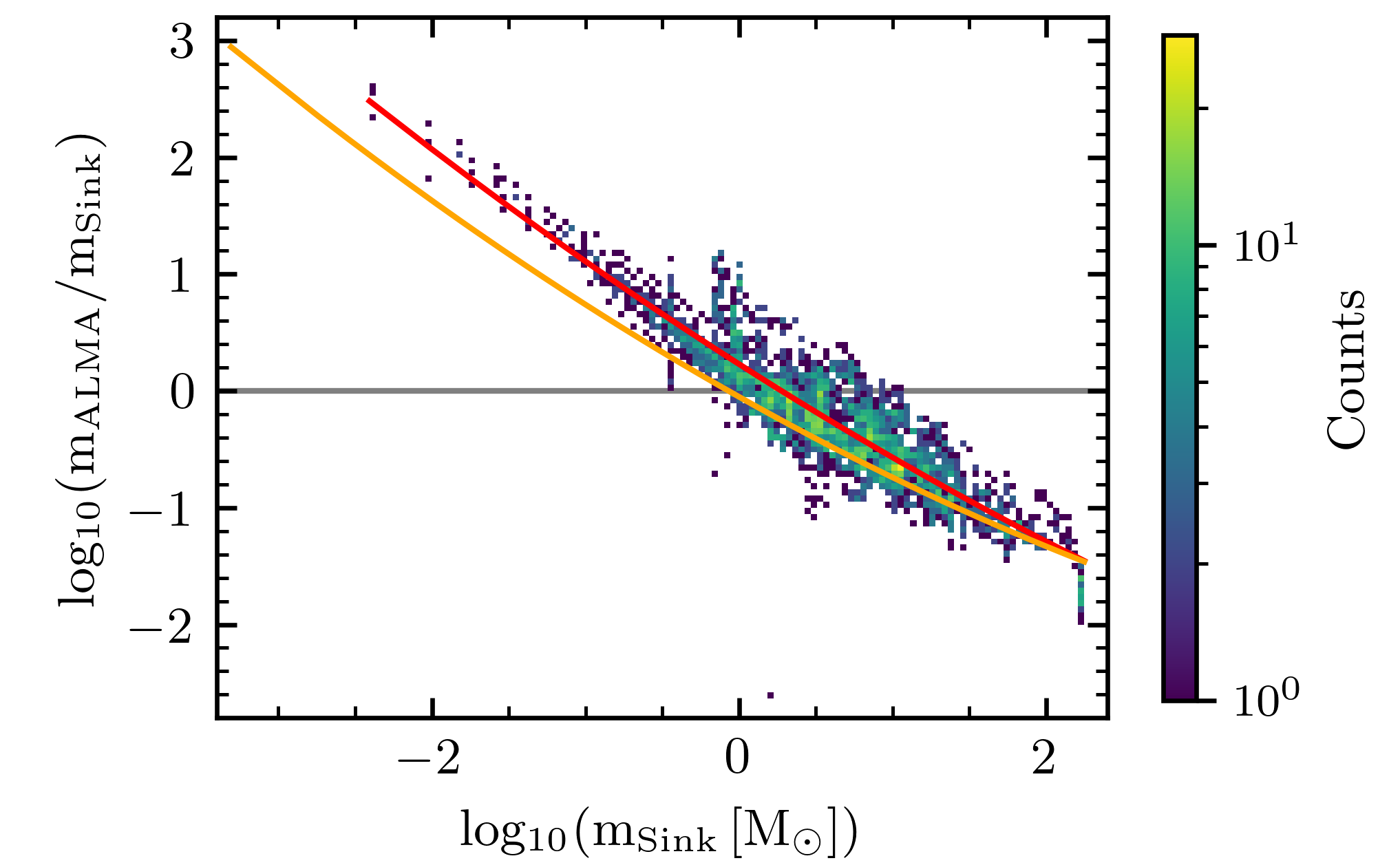}
     \end{subfigure}\\
\caption{\emph{Top:} Projected simulation masses within a radius of 1500~au around each sink plotted against the corresponding sink mass. \emph{Bottom:} ALMA masses within a radius of 1500~au around each sink, normalised by the corresponding sink mass, plotted against the sink mass. In both panels, the horizontal line indicates where the masses are equal. The orange and red
lines show empirical fits to the simulation and ALMA data, respectively. In the early evolutionary stages (when sink masses are still low), the core mass exceeds the sink mass, whereas at later times this trend reverses.}
\label{fig:rfl9_update_hist2D_m2D_vs_msink}
\end{figure}

Figure~\ref{fig:rfl9_update_hist2D_m2D_vs_msink} shows the ratio of core mass to sink mass as a function of sink mass. Here, we consider cores at the three viewing angles for all time steps. $m_{\rm 2D}$s show a general decrease with increasing sink mass (top panel). In the early evolutionary stages, when sink masses are still low, the projected core mass typically exceeds the sink mass. At this stage, the envelope still contains most of the available material, and the sink has only begun to accrete. As evolution progresses, both the envelope and the sink continue to gain mass, but the sink grows more rapidly. Consequently, the ratio of projected core mass to sink mass decreases, and the projected masses begin to underestimate the more evolved sinks. Overall, we find an almost linear anticorrelation between the core and sink masses, with certain scatter. The fact that the core-to-sink mass ratio does not follow a perfect unity line indicates that a given sink mass does not correspond to a unique core mass, reflecting the diversity of accretion histories and core environments.

From our simulations, we derived an empirical fit that provides the expected sink mass for a given projected core mass (see Fig.~\ref{fig:rfl9_update_hist2D_m2D_vs_msink}, top panel):
\begin{equation} 
    \label{eq:msink_m2d}
    \log_{10}(m_{\rm sink}) = -2.6 + \sqrt{\,8 + 20\,\log_{10}(m_{\rm 2D})\,} \, .
\end{equation}
A similar relation of the ALMA-recovered core masses (see also Sect.~\ref{sec:mass_recovery}) is shown in the bottom panel of Fig.~\ref{fig:rfl9_update_hist2D_m2D_vs_msink}, which is empirically fitted by the expression
\begin{equation}
    \label{eq:msink_malma}
    \log_{10}(m_{\rm sink}) = -2 + \sqrt{\,-1.56 + 25\,\log_{10}(m_{\rm ALMA})\,} \, .
\end{equation}
The overall trend is similar to that of the simulation $m_{\rm 2D}$ core masses, though the ALMA cores tend to exhibit systematically higher projected masses (see Sect.~\ref{sec:mass_recovery} for a detailed discussion).

\subsection{Core masses: Simulations versus \textsc{RADMC-3D} versus ALMA}
\label{sec:mass_recovery}

Next, we assess how accurately the simulated core masses can be recovered from the post-processed \textsc{RADMC-3D} and ALMA images, provided the temperature is known (see Sect.~\ref{sec:temps} for the effect of different temperature estimates). In the following, we use either the mass-weighted LOS temperature, $T_\mathrm{sim,rad}$, or the mass-weighted core temperature, $T_\mathrm{sim,avg}$ (see Sect.~\ref{sec:definitions} for their definitions).

\begin{figure}
     \begin{subfigure}[b]{0.45\textwidth}
         \centering
         \includegraphics[width=\columnwidth,trim=0 0 0 0.1cm, clip]{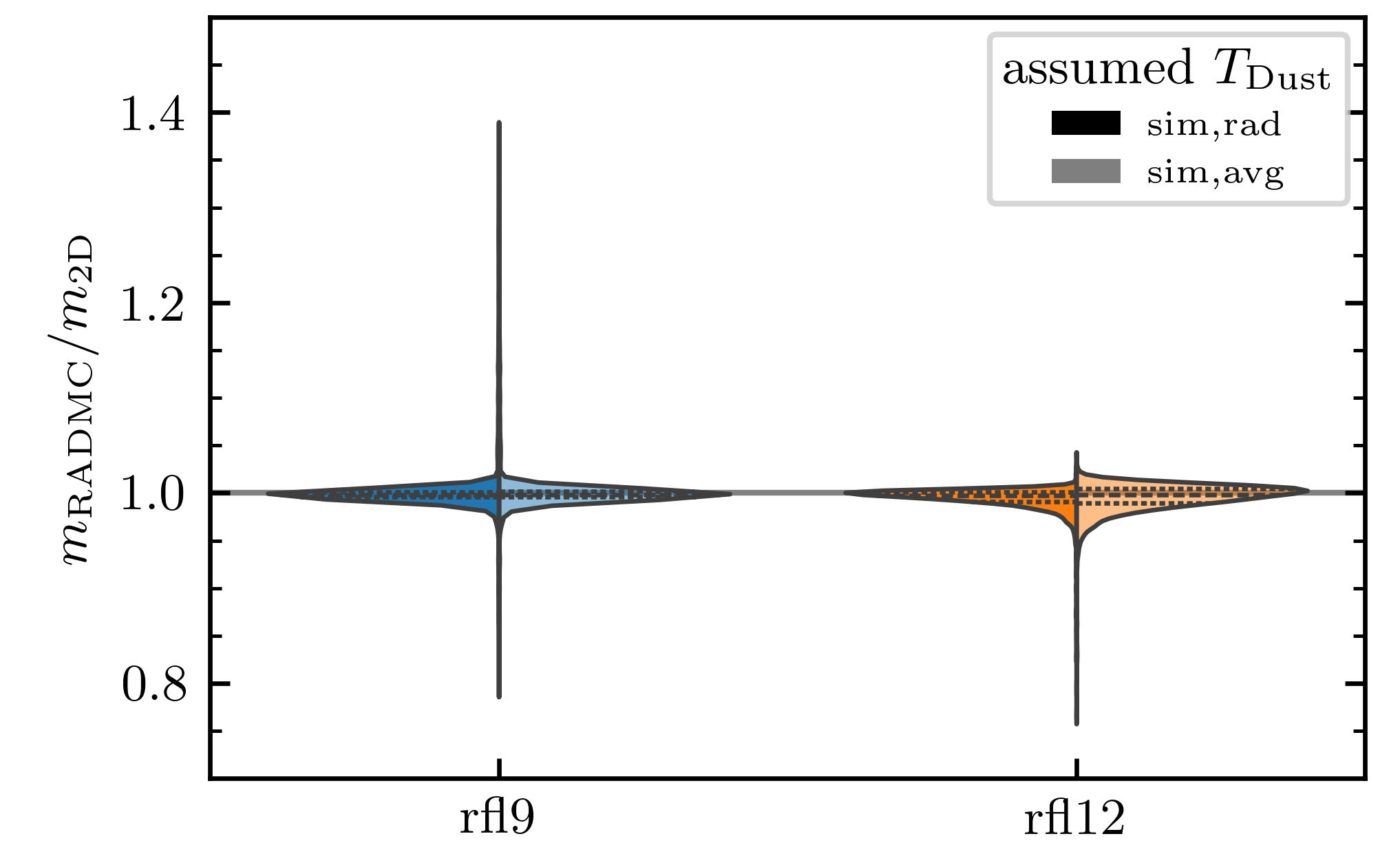}
     \end{subfigure}\\
     \begin{subfigure}[b]{0.45\textwidth}
         \centering
         \vspace{-0.56cm}
         \includegraphics[width=\columnwidth,trim=0 0 0 0.1cm, clip]{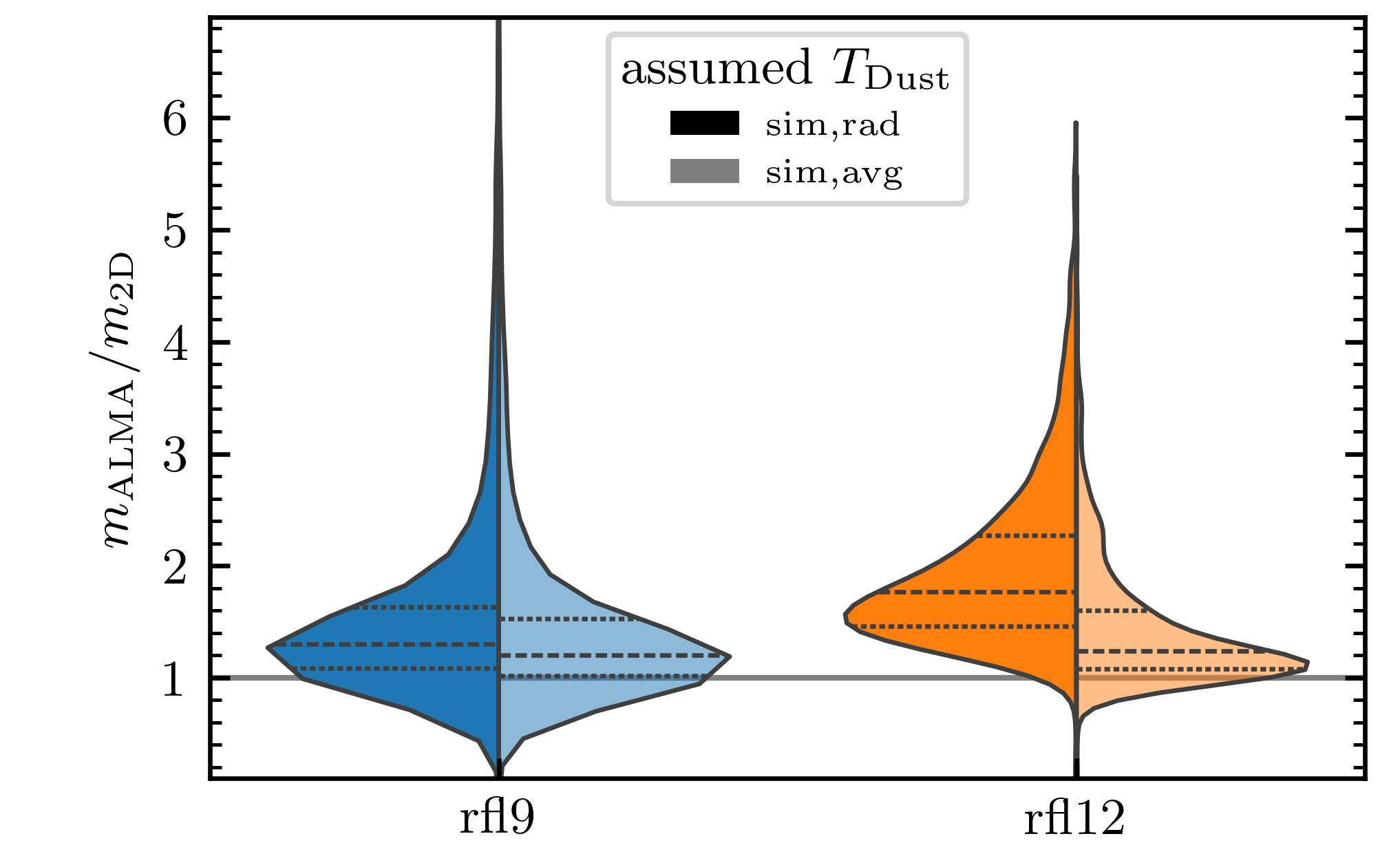}
     \end{subfigure}\\
\caption{Ratio of reconstructed core masses to the true 2D masses for idealised \textsc{RADMC-3D} emission maps (\textit{top}) as well as for synthetic ALMA observations (\textit{bottom}), each for \textsc{rfl9} and \textsc{rfl12}. For both cases, cores were reconstructed using either the mass-weighted LOS temperature ($T_\mathrm{sim,rad}$, dark colours, left half violin) or the mass-weighted average temperature ($T_\mathrm{sim,avg}$, light colours, right half violin). While \textsc{RADMC-3D} maps reproduce the true masses almost perfectly, interferometric filtering in ALMA introduces significant deviations.}
\label{violin_radmc_TD_vs_TDavg}
\end{figure}

Because the \textsc{RADMC-3D} maps provide idealised emission with the same resolution as the numerical simulation, the core masses in run \textsc{rfl9} can be recovered with a high accuracy using either $T_\mathrm{sim,rad}$ or $T_\mathrm{sim,avg}$ (see the top panel of Fig.~\ref{violin_radmc_TD_vs_TDavg}). For this run, using a single averaged temperature per core makes virtually no difference. In contrast, the more centrally concentrated cores in the \textsc{rfl12} run show slightly larger deviations when $T_\mathrm{sim,avg}$ is used instead of the per-pixel temperature, reflecting their steeper temperature gradients (see also the radial profiles in Fig.~\ref{fig:average_radial_flux_profiles}).

The post-processed ALMA images, with a coarser resolution of $\sim 1200$~au, added noise and interferometric artefacts, have less sharply defined cores than in the idealised \textsc{RADMC-3D} maps. As a result, the flux distribution is smoother, and $T_\mathrm{sim,rad}$ becomes a more accurate description of the intrinsic core structure. However, using $T_\mathrm{sim,rad}$ to infer core masses leads to a systematic overestimation (see the bottom panel of Fig.~\ref{violin_radmc_TD_vs_TDavg}), which increases with simulation resolution from a median mass ratio factor of $1.26$ in \textsc{rfl9} to $1.73$ in \textsc{rfl12}. This trend is expected, as the high-resolution simulations resolve structures down to $\sim50$~au, whereas the ALMA synthetic observations are limited to $\sim1200$~au. Using a single representative temperature per core produces better mass estimates: for run \textsc{rfl9} the improvement is modest from $1.26$ to $1.17$, but for the higher resolution run \textsc{rfl12}, the single-temperature approach yields significantly more accurate results, improving the median mass ratio from $1.73$ to $1.2$. As a consequence, using a single temperature rather than a full temperature profile is sufficient as long as the inner structure of the core is not resolved.

\subsection{Reliable mass range of ALMA cores}
\label{sec:alma_mass_reliability}

\begin{figure}
    \centering
    \includegraphics[width=\linewidth]{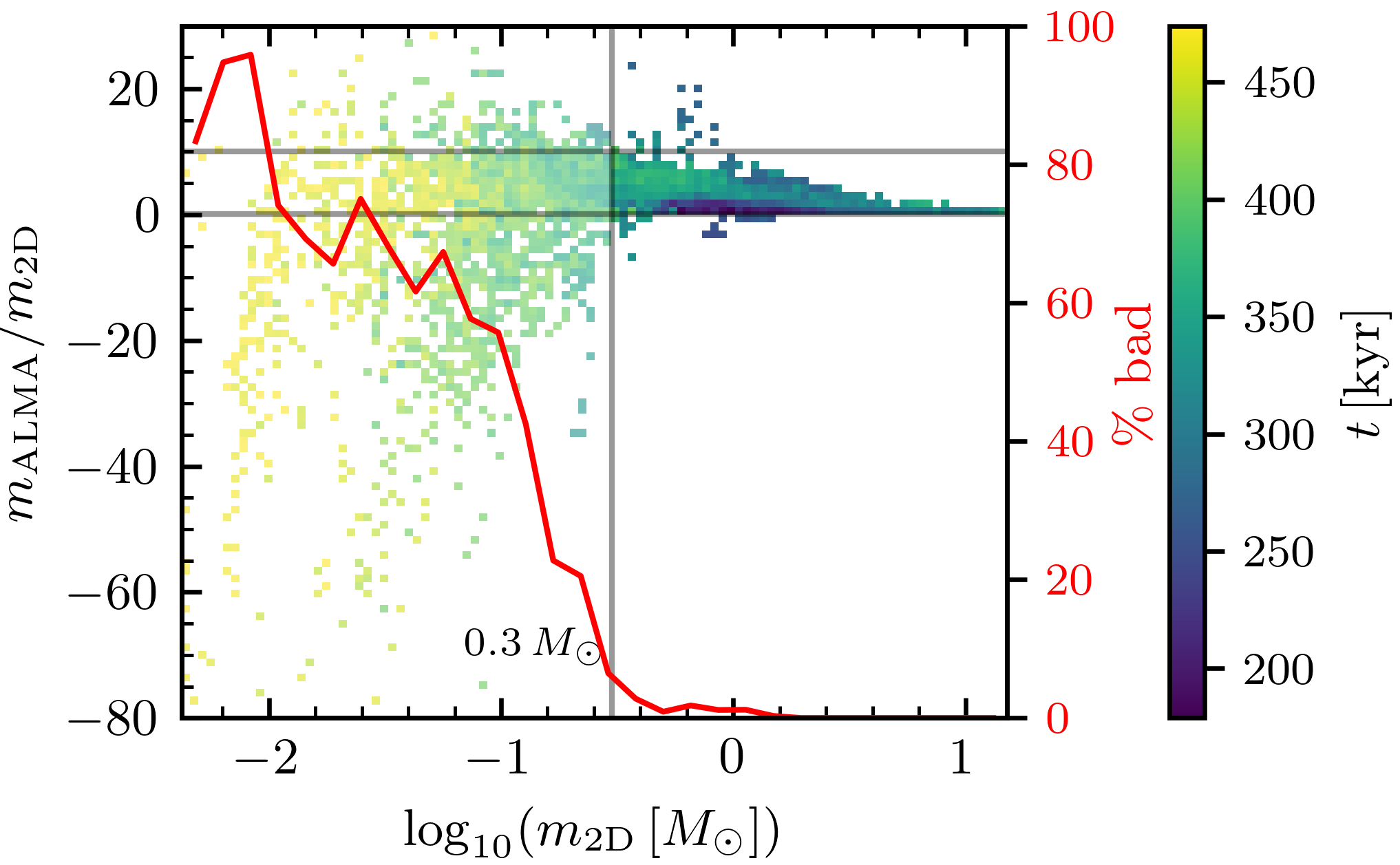}
\caption{Ratio of ALMA-derived core masses to projected simulation masses, $m_{\rm ALMA}/m_{\rm 2D}$, as a function of the intrinsic projected core mass, $m_{\rm 2D}$. The ALMA masses were computed using $T_\mathrm{sim,avg}$, and all cores from all snapshots of run \textsc{rfl9} are included. The colour scale shows the evolutionary time. Horizontal grey lines mark the limits $m_{\rm ALMA}/m_{\rm 2D}=0.1$ and $10$, which define `bad' cores with catastrophic mass under- or overestimates. The red curve (right-hand axis) shows the fraction of such bad cores per mass bin. The vertical grey line at $0.3\,{\rm M}_\odot$ marks the adopted lower limit for reliable core-mass recovery. Note that all the cores are shown without any selection cuts and all time steps are included (see colour bar).}
\label{fig:mass_detection_all}
\end{figure}

To assess the mass range over which ALMA can reliably recover core masses, we compare the ALMA-derived masses, $m_{\rm ALMA}$, of all identified cores in run \textsc{rfl9} with their projected simulation masses, $m_{\rm 2D}$. Figure~\ref{fig:mass_detection_all} shows the ratio $m_{\rm ALMA}/m_{\rm 2D}$ as a function of $m_{\rm 2D}$ together with the fraction of `bad' cores, defined as those with mass errors larger than a factor of ten ($m_{\rm ALMA}/m_{\rm 2D}<0.1$ or $>10$). At low masses, the scatter is extremely large, and the catastrophic-error rate is high, indicating that ALMA cannot reliably measure these cores. In this range, the estimated ALMA core masses can have unphysical, negative values. These negative masses occur when interferometric noise and filtering effects drive the measured core flux below zero. The failure rate decreases rapidly with increasing mass and flattens at values of $\lesssim 5$\,\% above $m_{\rm 2D}\simeq 0.3\,{\rm M}_\odot$. In this regime, the mass ratios cluster much more tightly, and most ALMA-derived core masses deviate from the true values by only factors of a few. This mass threshold is consistent with the core-mass completeness level of $0.23\,M_\odot$ estimated by \citet{2025Coletta} for the ALMAGAL project. We therefore adopted $0.3\,{\rm M}_\odot$ as a conservative lower limit for reliable core-mass estimates in these simulations.

\subsection{Cluster and clump properties}
\label{sec:clump}

\begin{figure}
    \centering
    \includegraphics[width=1.0\linewidth]{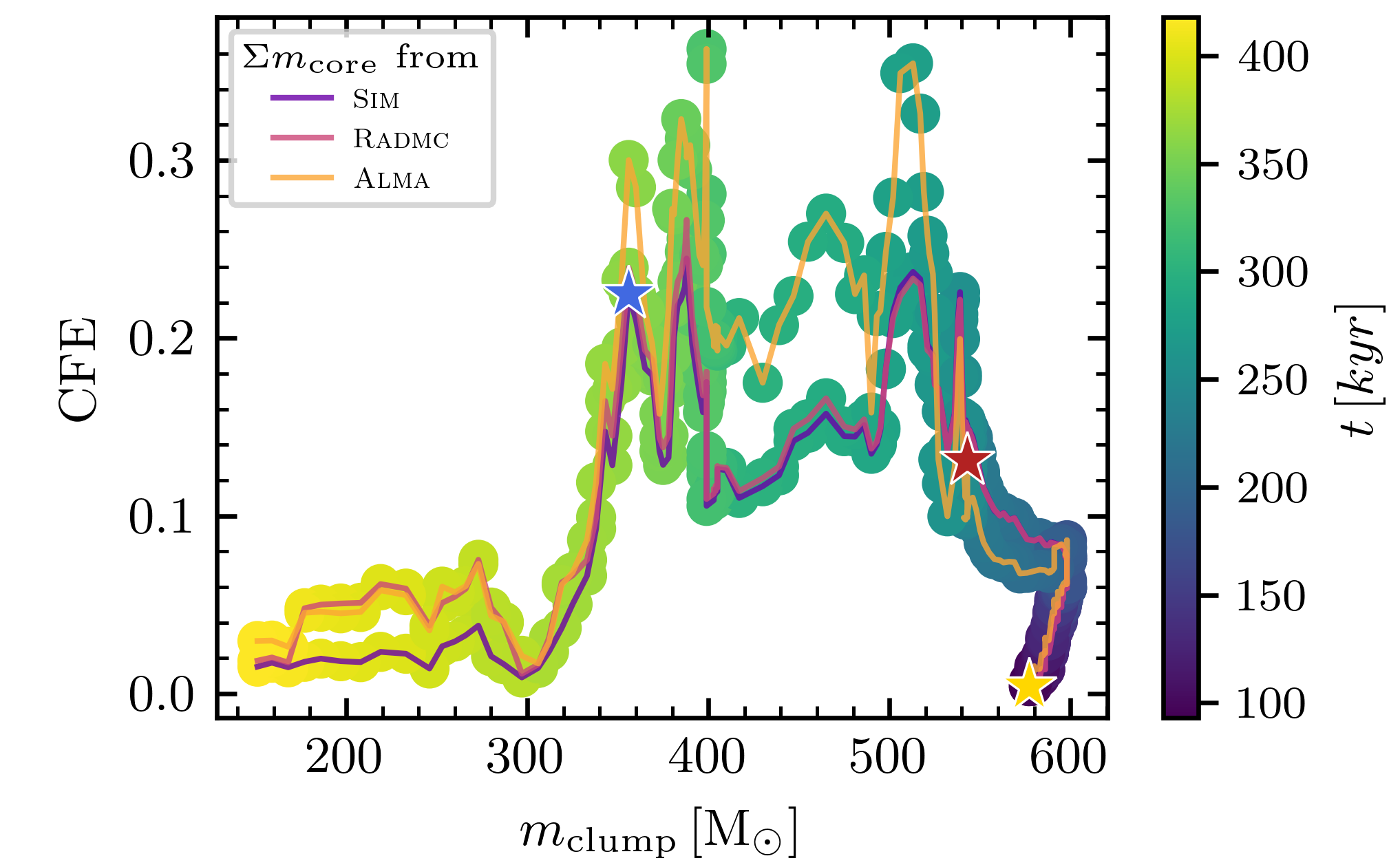}
\caption{Core formation efficiency vs clump mass, colour-coded by time. The curves trace the evolution from global collapse through active core formation to final clump dispersal. The yellow star marks the formation of the first core (93 kyr). The red star indicates the onset of ionising radiation (240 kyr), and the blue star marks clump dispersal (370 kyr).}
\label{fig:rfl9_update_core_mass_vs_cloud_mass}
\end{figure}

Following the ALMAGAL work by \citet{2025Elia}, we examine how the total core population evolves relative to the parental clump mass. For this, we define the clump as a sphere of radius 0.5~pc, comparable to the ALMAGAL primary beam and the Hi-GAL clumps studied in the ALMAGAL project \citep[see][]{2025Elia}, and sum all cores whose centres lie within this region. The clump mass, $m_{\rm clump}$, is taken directly from the simulation. 
The summed core masses, $\Sigma m_\mathrm{core}$, are determined directly from the numerical simulation and from the \textsc{RADMC-3D} and ALMA images using $T_\mathrm{sim,avg}$.
The corresponding core formation efficiency, CFE, is defined as
\begin{equation}
\mathrm{CFE} = \frac{\sum m_{\rm core}}{m_{\rm clump}} \,.
\end{equation}
Figure~\ref{fig:rfl9_update_core_mass_vs_cloud_mass} shows the evolution of the CFE as a function of clump mass. Generally, the CFE evolves from right to left. During the early global collapse, the clump mass within 0.5~pc increases as material flows inwards, while only a few cores are present. Once more sink particles form, and the clump mass decreases as gas is accreted onto sinks and removed from the gas phase, whereas the total mass in cores continues to rise as new cores and sinks form. Ionising radiation begins at $240 \,\rm kyr$. At later times, when ionising radiation disrupts the clump at around 370~kyr, both clump and core masses decline. 
The \textsc{RADMC-3D} core masses closely follow the path described by the core masses calculated in the numerical simulations. The ALMA-derived core masses, despite showing a larger scatter and modest systematic offsets, still reproduce the overall evolutionary trend well.

\begin{figure}
    \centering
    \includegraphics[width=1.0\linewidth]{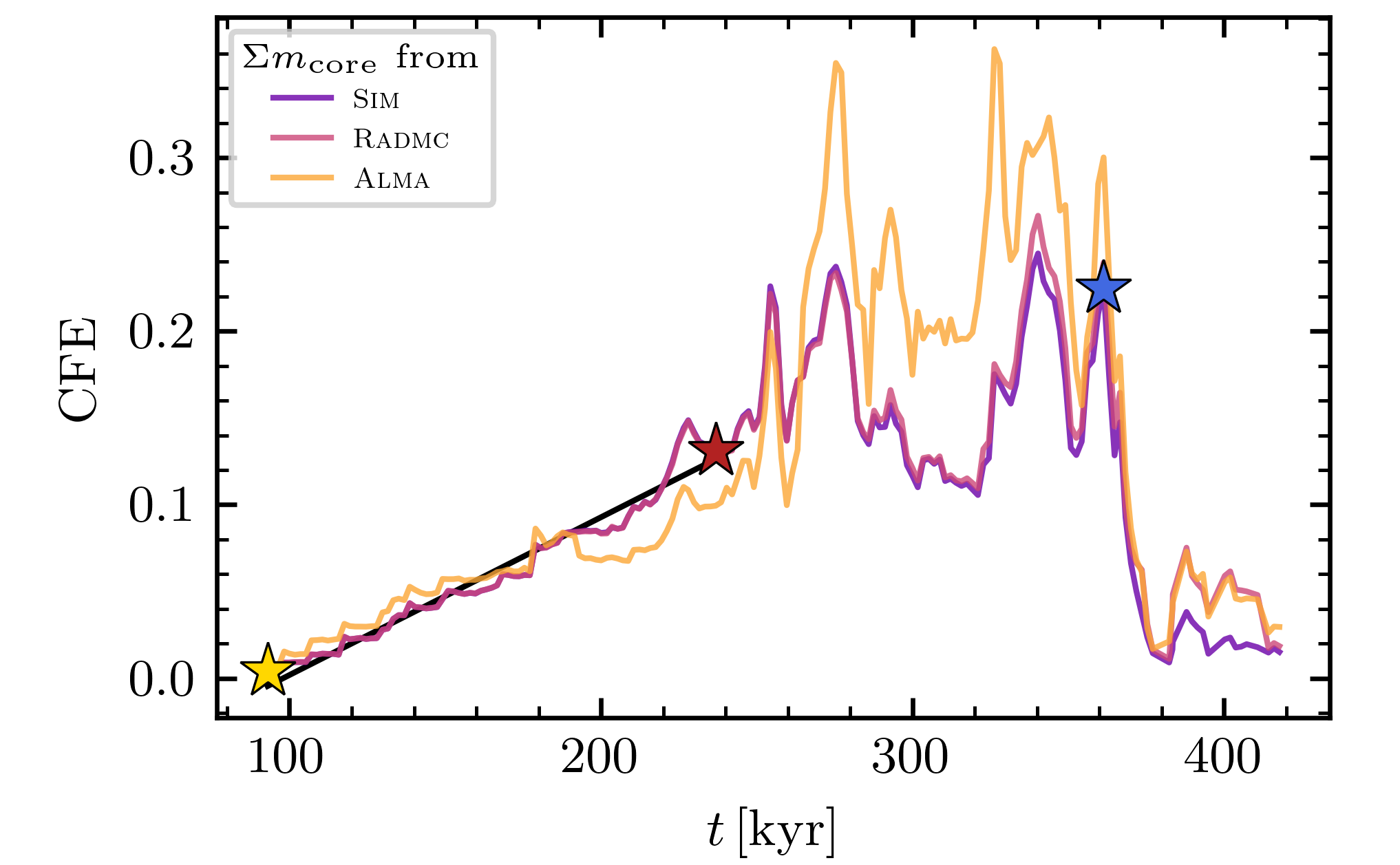}
\caption{Time evolution of the CFE estimated from the simulations, \textsc{RADMC-3D}, and ALMA. The yellow star indicates the first core formation. During the early collapse phase, CFE increases almost linearly. The black line shows a fit to this linear regime with a slope of $8.9 \times 10^{-4}$. The fraction begins to fluctuate once ionising feedback from the first massive stars becomes important (red star) and eventually declines as the expanding H\,\textsc{ii} region disperses the surrounding clump (blue star).}
\label{fig:rfl9_update_core_cloud_mass_vs_time}
\end{figure}

Figure~\ref{fig:rfl9_update_core_cloud_mass_vs_time} shows the time evolution of the CFE. During the early collapse phase, this fraction increases almost linearly; the fitted relation is
\[
\mathrm{CFE}(t) =
\begin{cases}
8.9 \times 10^{-4}\, t - 0.088, & t < 240~\mathrm{kyr}, \\
\text{non-linear evolution}, & t \ge 240~\mathrm{kyr}.
\end{cases}
\]
In this regime, before ionising radiation becomes important, the linear relation implies that the current clump mass can be reconstructed from the total mass in cores. However, once ionising feedback sets in, the evolution becomes highly non-linear, and the CFE declines as the expanding \textsc{H\,ii} region disperses the clump. The SIM and \textsc{RADMC-3D} curves agree very well. The ALMA curve follows the same qualitative trend but shows deviations at the level of a factor of a few, consistent with the mass-recovery uncertainties discussed above.

\section{Interpretation of synthetic observations: Temperatures and masses}
\label{sec:temps}

\subsection{Dust temperature estimates}
\label{sec:temperature_estimates}

The core mass estimate strongly depends on the dust temperature (see Eq.~\ref{eq:mass}). In the simulations, the dust temperature is directly available for every cell, allowing us to compute a mass–weighted temperature distribution along the LOS with perfect spatial coverage. In real observations, however, the dust temperature cannot often be measured directly, and different methods and strategies need to be used. In the following we present and discuss three different methods to estimate the dust temperature.

\subsubsection{Luminosity–mass relation temperature model}
\label{sec:LM_relation}

Within ALMAGAL, \citet{2025Coletta} estimated the core temperature based on the luminosity-to-mass ratio ($L/M$) of the parent clump \citep[following the results by][]{2016Molinari}, which serves as a proxy for its evolutionary stage. The prescription defines three different regimes, summarised in Table~\ref{tab:core_properties}. According to this prescription, a low $L/M$ ratio ($\leq 1\,L_\odot/M_\odot$) is expected to correspond to an early cold phase dominated by accretion and lacking significant internal heating sources, representing pre-stellar clumps with an assumed temperature of 20~K (see Table~\ref{tab:core_properties}). Intermediate values ($1 < L/M \leq 10\,L_\odot/M_\odot$) correspond to clumps in a more evolved stage, where several protostars have already formed and start heating their surroundings, with an assumed temperature of 35~K. Finally, high $L/M$ ratios ($> 10\,L_\odot/M_\odot$) characterise the most advanced evolutionary phase, which is associated with luminous embedded protostars and the emergence of H\,\textsc{ii} regions. The empirical description assumes an increase of temperature with $L/M$.

\begin{figure}
    \centering
    \includegraphics[width=1.0\linewidth]{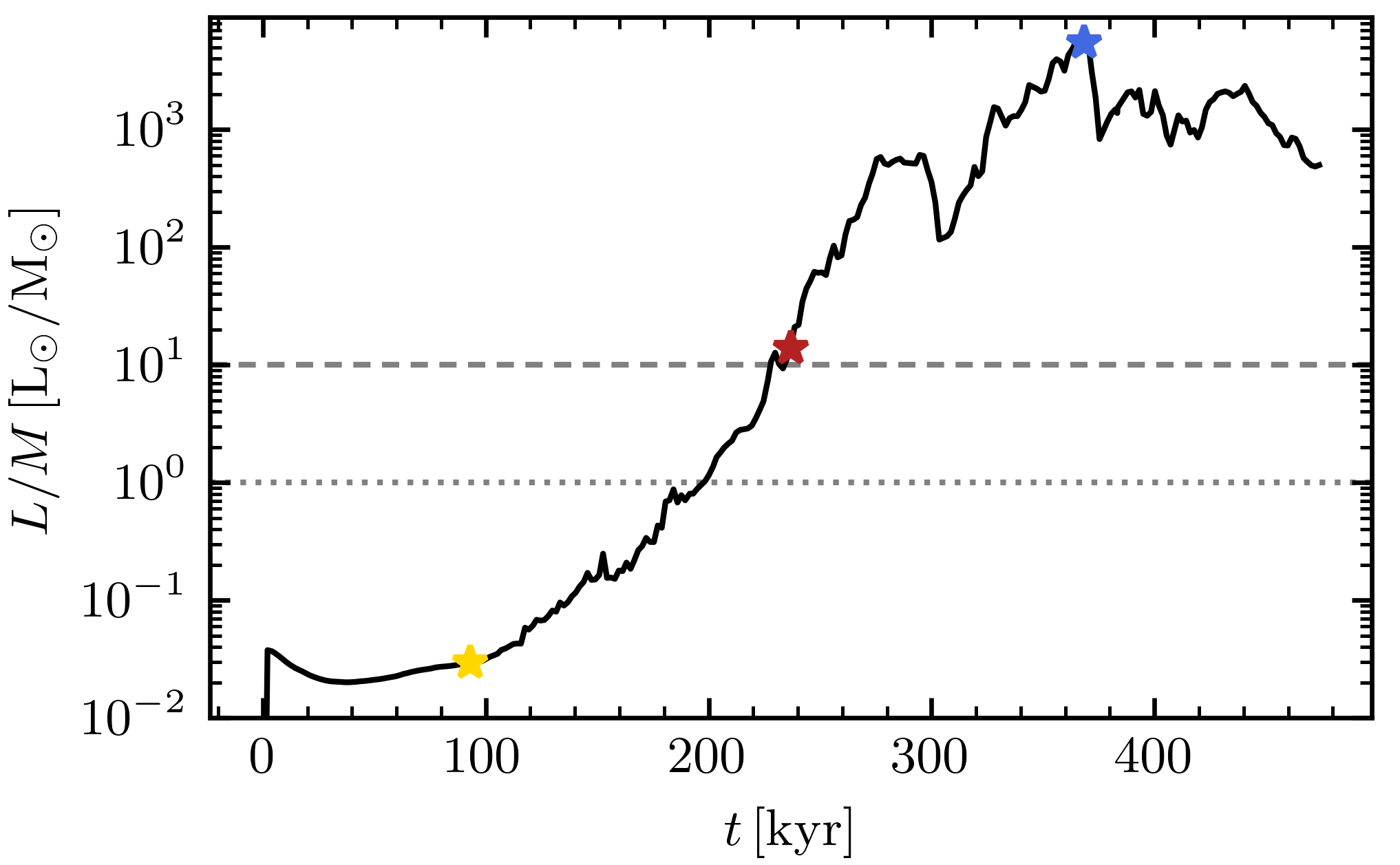}
    \caption{
    Time evolution of the luminosity-to-mass ratio ($L/M$) in the \textsc{rfl9} run.  
    Horizontal lines indicate the transition thresholds for the three temperature regimes: below the dotted line ($L/M=1$), a fixed temperature of 20~K is assumed; between the dotted and dashed lines ($1<L/M\leq10$), 35~K is assumed; and above the dashed line ($L/M=10$), a power-law temperature relation is applied. The yellow star marks the formation of the first sink particle; the red star indicates the onset of ionising radiation, which occurs close to the $L/M=10$ transition; and the blue star denotes the stage when the clump begins to be disrupted.}
    \label{fig:rfl9_update_L_over_M_in_time}
\end{figure}

This empirical prescription can be evaluated in our numerical simulations. The luminosity is obtained by integrating the spectral energy distribution produced with \textsc{RADMC-3D}, while the mass is directly from the simulation. 
Both luminosity and mass are measured within the same region used for the \textsc{RADMC-3D} images, corresponding to a cubic box with a side length of 1~pc centred on the clump.
The evolution of the clump's $L/M$ ratio in the fiducial run is shown in Fig.~\ref{fig:rfl9_update_L_over_M_in_time}. The $L/M$ ratio begins to rise once the first sink particle forms, marking the onset of protostellar activity (i.e. accretion and heating by radiative feedback). The appearance of ionising radiation occurs at $L/M\approx 10\,L_\odot/M_\odot$, consistent with the empirical relation. A pronounced dip at $t \approx 300$~kyr corresponds to a temporary halt in the accretion onto the most massive sink particle in our simulations. Once accretion resumes, $L/M$ increases again. After the clump begins to disperse due to the expanding H\,\textsc{ii} region, the value of $L/M$ decreases. We note that in our simulations, although mass is expelled from the central region in these later evolutionary stages, which would by itself increase $L/M$, the total luminosity within the domain drops as accretion halts and radiation escapes through the disrupted clump. 

\begin{figure}
     \centering
     \begin{subfigure}[b]{0.45\textwidth}
         \centering
         \includegraphics[width=\columnwidth,trim=0 0 0 0.056cm, clip]{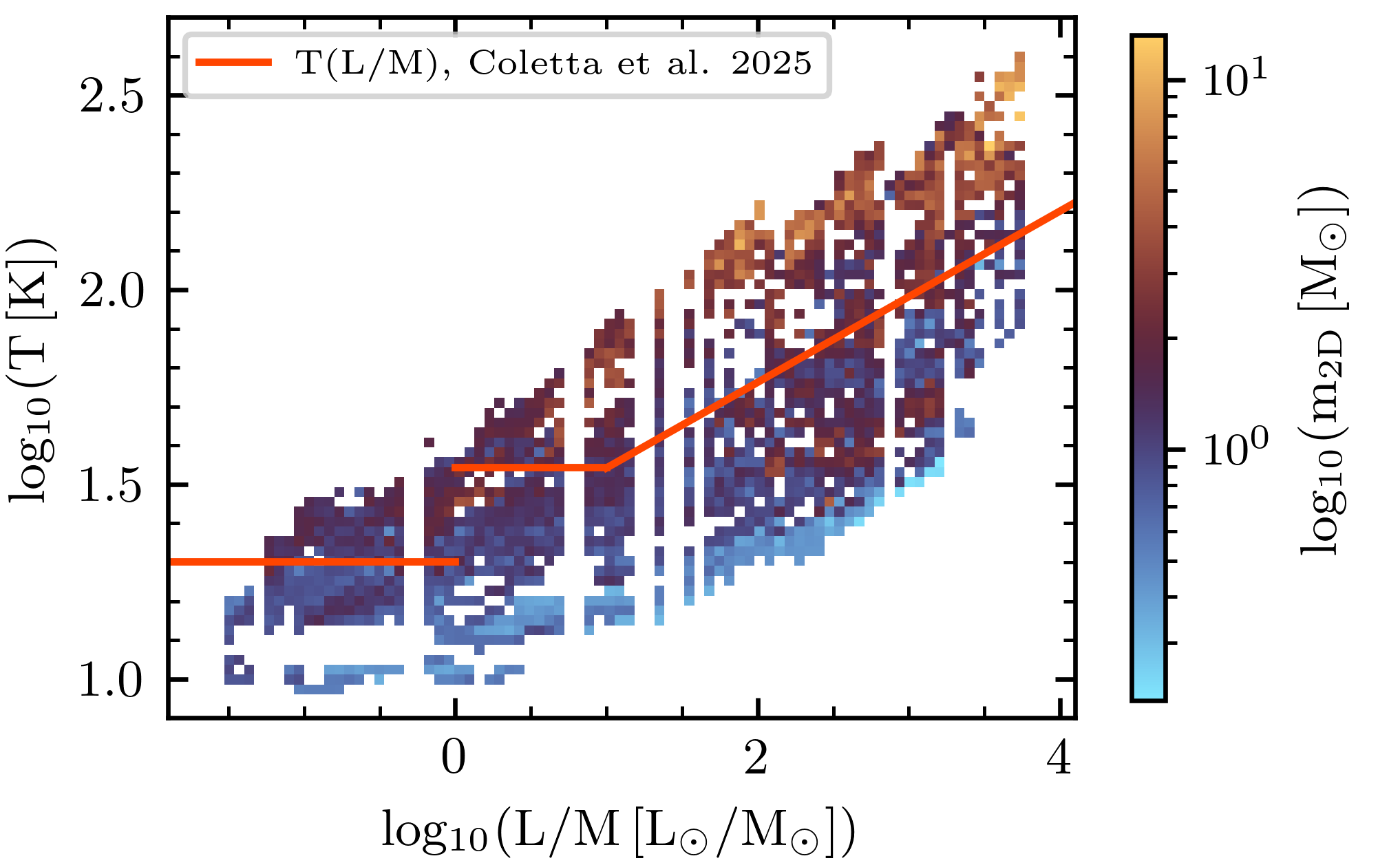}
     \end{subfigure}\\
     \begin{subfigure}[b]{0.45\textwidth}
         \centering
         \vspace{-0.9cm}
         \includegraphics[width=\columnwidth,trim=0 0 0 0.1cm, clip]{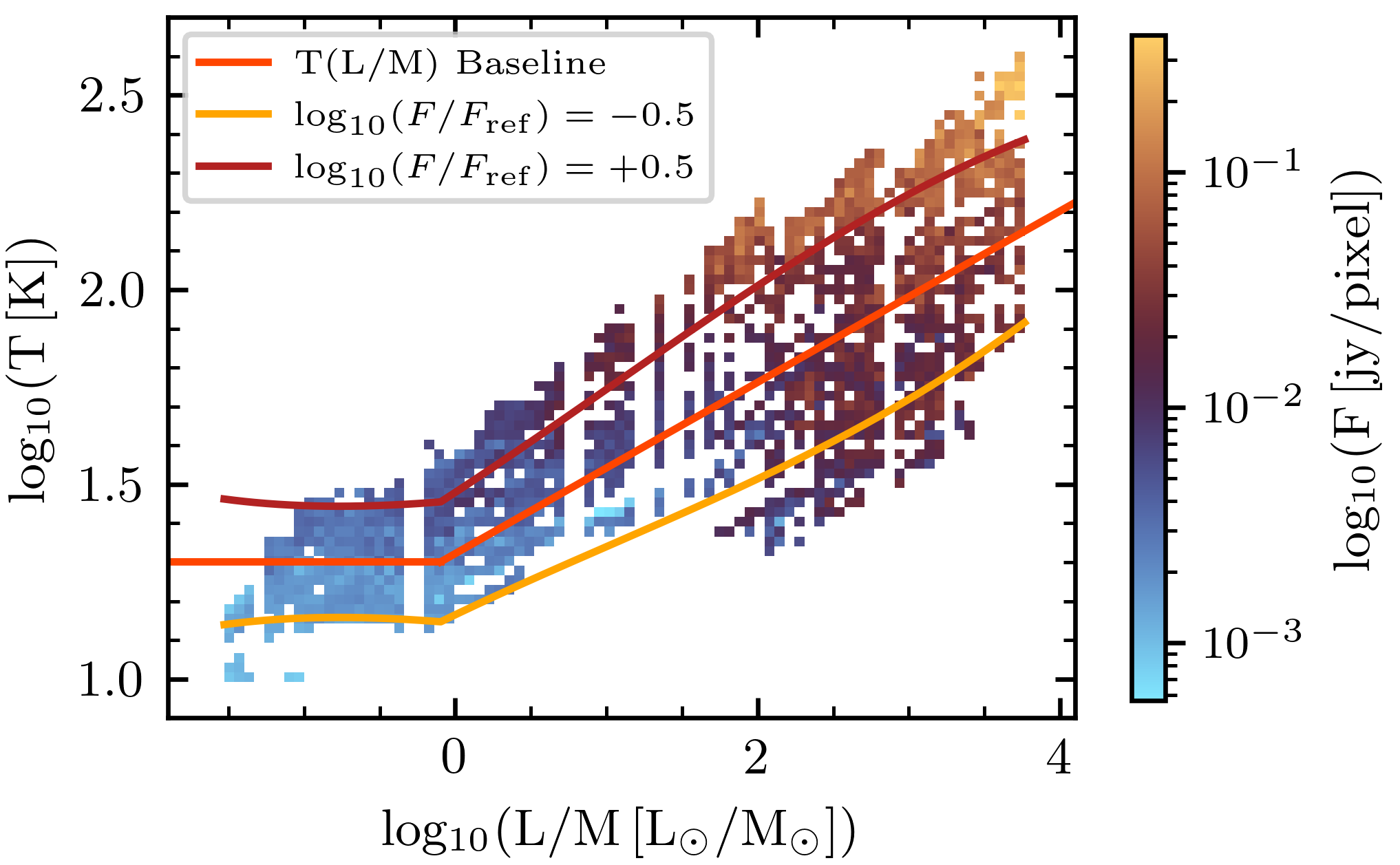}
     \end{subfigure}\\
     \caption{Mass-weighted average core temperature as a function of the luminosity-to-mass ratio ($L/M$), colour-coded by the core mass (\textit{top}) and by the ALMA continuum flux (\textit{bottom}). Not all simulated cores have ALMA counterparts because of the limited interferometric field of view.
     In the top panel, the red line shows the temperature–$L/M$ relation adopted from \citet{2025Coletta}.
     In the bottom panel, the central curve shows the improved baseline temperature–$L/M$ relation, and the two additional curves mark the fitted positive and negative deviations obtained from the flux–temperature response model (Sect.~\ref{sec:LMflux}).}
     \label{fig:rfl9_update_hist2D_tdus_sim_LM_binned_mass}
\end{figure}

Figure~\ref{fig:rfl9_update_hist2D_tdus_sim_LM_binned_mass} (top panel) shows the mass-weighted average core temperature as a function of $L/M$. Overall, the simulations reproduce the global temperature trend used in \citet[see  Fig.~\ref{fig:rfl9_update_hist2D_tdus_sim_LM_binned_mass}, top panel]{2025Coletta}. However, the simulations also reveal a significant intrinsic scatter. For a given $L/M$, individual cores span a wide range of temperatures. This variation arises because the thermal state of a core is determined not only by the global radiative environment of the clump but also by local conditions such as accretion luminosity, density, and optical depth. The results from the simulations also show that more massive cores are systematically warmer, reflecting enhanced accretion heating and stronger internal radiation fields. The bottom panel of Fig.~\ref{fig:rfl9_update_hist2D_tdus_sim_LM_binned_mass} shows that the core temperature is not only related to the clump-wide $L/M$ value but also correlates with the core continuum flux. This motivates a more flexible temperature model that depends on both quantities (see Sect.~\ref{sec:LMflux}).

\begin{table}[t!]
\centering
\caption{Core temperatures estimated from the clump's $L/M$.}
\label{tab:core_properties}
\begin{tabular}{lccc}
\hline
\hline
$(L/M)_\mathrm{clump}$ [$L_\odot/M_\odot$] & $T_\mathrm{core}$ [K] & $N_\mathrm{cores}$ \\
\hline
$\leq 1$ & 20  & 1059 \\
$1 < L/M \leq 10$ & 35 & 570 \\
$> 10$ & $21.1\cdot (L/M)^{0.22}$  & 2081 \\
\hline
\end{tabular}
\tablefoot{Core temperatures assumed in the empirical $L/M$ prescription for different evolutionary stages \citep[see][]{2025Coletta} and the corresponding number of cores in the \textsc{rfl9} run with valid (positive) mass estimates (see Sect.~\ref{sec:alma_mass_reliability}).}
\end{table}

\subsubsection{$L/M$–flux temperature model}
\label{sec:LMflux}

Based on the results shown in the bottom panel of Fig.~\ref{fig:rfl9_update_hist2D_tdus_sim_LM_binned_mass}, we sought to describe the dust temperature ($T$) of each core as a function of the clump luminosity-to-mass ratio $L/M$ (a proxy for the evolutionary stage) and its observed flux ($F$; a proxy for its local radiation intensity or compactness). To that end, we first defined a global dependence of the dust temperature with $L/M$ following previous work \citep[see][]{2016Molinari, 2025Coletta} and the results of our numerical simulations (see Fig.~\ref{fig:rfl9_update_hist2D_tdus_sim_LM_binned_mass}). Based on this trend, we adopted the following as the baseline temperature relation:
\begin{equation}
T_{\rm base}(L/M)=
\begin{cases}
20~{\rm K}, & L/M<0.8,\\[4pt]
21.1\,(L/M)^{0.22}~{\rm K}, & L/M\ge0.8.
\end{cases}
\label{eq:Tbase}
\end{equation}
Similarly, the results shown in the bottom panel of Fig.~\ref{fig:rfl9_update_hist2D_tdus_sim_LM_binned_mass} show that the characteristic flux level associated with a given $L/M$ seems to follow a power-law relation that can be described as
\begin{equation}
F_{\rm base}(L/M)=3.37 \cdot 10^3\,(L/M)^{0.32}.
\label{eq:Fbase}
\end{equation}
We could then express the temperature as a perturbation from the baseline temperature given in Eq.~\ref{eq:Tbase}, where the perturbation depends on the deviation of the observed flux from its expected baseline value (see Eq.~\ref{eq:Fbase}). In its logarithmic form, this expression, including the perturbation factor, reads as 
\begin{equation}
\log_{10}T = \log_{10}T_{\rm base}(L/M) +
\beta\,\bigl[\log_{10}F - \log_{10}F_{\rm base}(L/M)\bigr],
\label{eq:model_log}
\end{equation}
which can be written equivalently in linear form as
\begin{equation}
T(L/M,F)=T_{\rm base}(L/M)\left(\frac{F}{F_{\rm base}(L/M)}\right)^{\beta}.
\label{eq:model}
\end{equation}
The exponent $\beta$ controls how strongly the temperature responds to flux variations at a fixed $L/M$. We adopted $\beta=0.47$ from our fit. Figure~\ref{fig:rfl9_update_hist2D_tdus_sim_LM_binned_mass} shows the baseline $T(L/M)$ relation together with flux-offset tracks at fixed $L/M$. The upper and lower offset tracks correspond to $\log_{10}(F/F_\mathrm{ref})=+0.5$ and $\log_{10}(F/F_\mathrm{ref})=-0.5$, respectively.

This $L/M{+}$flux model captures the intuitive idea that $L/M$ sets the global temperature scale of the clump, while the local flux provides a secondary adjustment reflecting local variations in radiation intensity or compactness. In Sect.~\ref{sec:impact_T} we show that this simple extension already provides a substantial improvement over using $T_{\rm base}(L/M)$ alone.

\subsubsection{Machine learning temperature model}
\label{sec:MLmodel}

As an exploratory step for a new temperature determination approach, we have used a supervised machine learning model to predict dust temperatures from a small set of observable quantities. This exercise is intended as a proof of concept rather than a fully developed method for observational use, and we therefore limit the description to the basic setup and performance.

Our model is trained directly on the simulated core catalogue, where the true dust temperature corresponds to the mass-weighted, LOS averaged value $T_{\rm sim,avg}$ for each core. We adopt a Gradient Boosting Regressor \citep{2001Friedman}, a decision-tree ensemble method that builds a prediction model by adding many small decision trees in sequence, where each tree learns to correct the errors made by the previous ones. By iteratively fitting these residuals, the method captures complex non-linear dependencies and feature interactions without imposing a fixed analytical form.\footnote{We used the \texttt{GradientBoostingRegressor} implementation in \texttt{scikit-learn} \citep{2011Pedregosa}; see \url{https://scikit-learn.org/stable/modules/generated/sklearn.ensemble.GradientBoostingRegressor.html}.}

We combined four physically motivated input quantities:
(i)~the luminosity-to-mass ratio ($L/M$) of the clump,
(ii)~the local (dust) continuum flux at 1.36~mm of the core,
(iii)~the projected distance of the core from the centre of the clump/cluster, and
(iv)~the empirical $L/M$–based temperature estimate $T_{L/M}$.
These quantities jointly encode information about both global and local heating processes: $L/M$ traces the global evolutionary stage, the flux and $T_{L/M}$ characterise the local environment and a physically motivated baseline, and the radial distance of the core to the centre of the clump/cluster captures systematic radial temperature gradients and geometric dilution. The Gradient Boosting model is trained to predict the logarithm of the dust temperature, $\log_{10} T_{\rm dust}$, using 20\% of the simulated cores and evaluated on the remaining 80\%. Even with this limited training fraction, for the test set, the model reproduces approximately 93\,\% of the variance in the simulated temperature distribution and achieves a root mean squared error of $0.092\,\mathrm{dex}$ in $\log_{10} T_{\rm dust}$ (corresponding to a multiplicative error of $\sim 1.24$, i.e.\ $\sim 24\%$).

\subsubsection{Model accuracy in temperature retrieval}
\label{sec:impact_T}

In this section we compare the three temperature estimates ($T_{L/M}$ as described in Sect.~\ref{sec:LM_relation}, the $T_{L/M+F}$ as described in Sect.~\ref{sec:LMflux}, and the machine learning-based $T_{\rm ML}$ as described in Sect.~\ref{sec:MLmodel}) to the mass-weighted average core temperatures from the simulation, $T_{\rm sim,avg}$. We used the coefficient of determination ($R^2$) to quantify how well the predicted temperatures reproduce the variance of the simulated values. Higher $R^2$ values indicate more accurate models, with $R^2 \approx 1$ reflecting near-perfect agreement between model predictions and the simulation data.

\begin{figure}
     \centering
     \begin{subfigure}[b]{0.45\textwidth}
         \centering
         \includegraphics[width=\columnwidth,trim=0 0 0 0.056cm, clip]{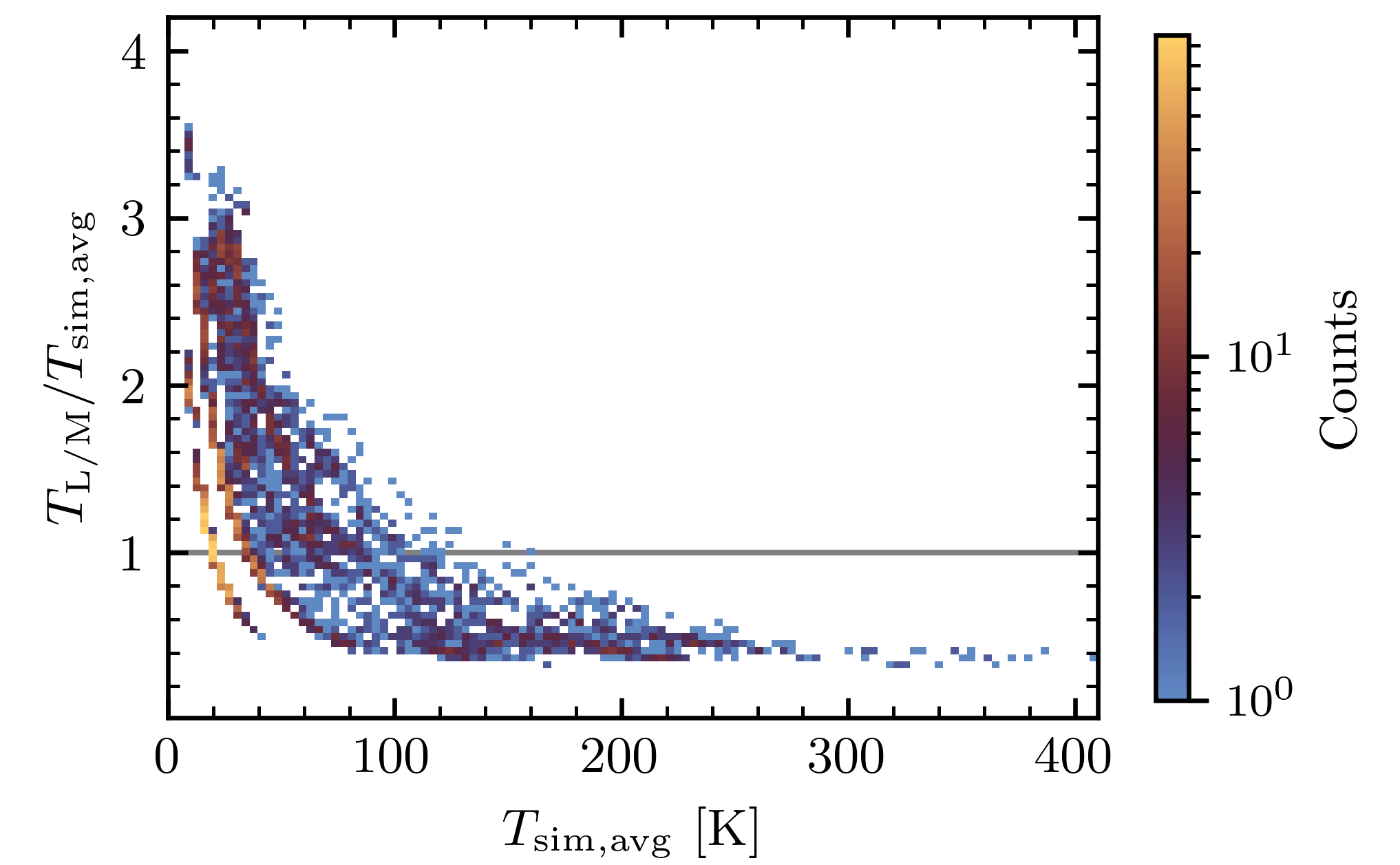}
     \end{subfigure}\\
     \begin{subfigure}[b]{0.45\textwidth}
         \centering
         \vspace{-0.9cm}
         \includegraphics[width=\columnwidth,trim=0 0 0 0.1cm, clip]{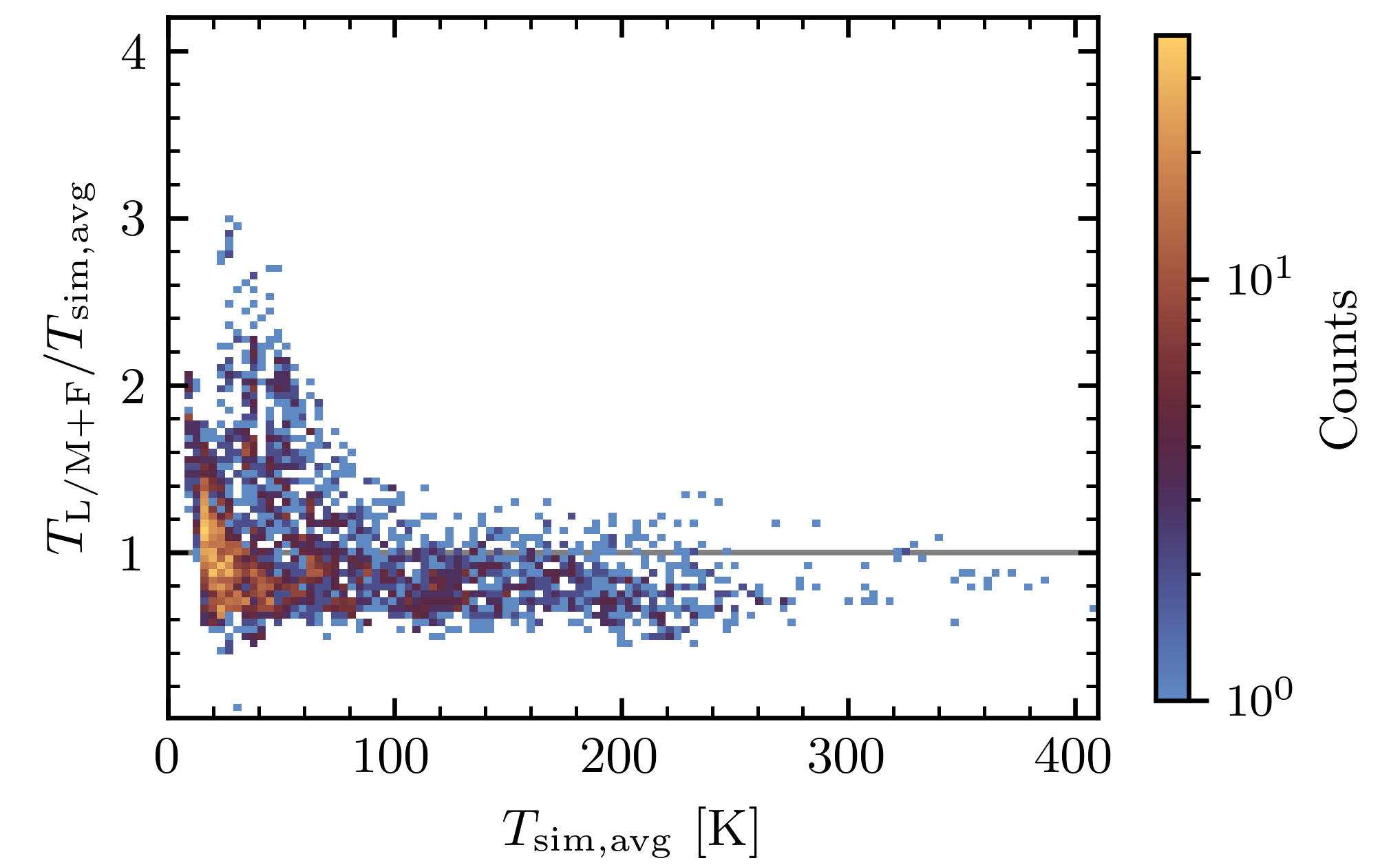}
     \end{subfigure}\\
     \begin{subfigure}[b]{0.45\textwidth}
         \centering
         \vspace{-0.9cm}
         \includegraphics[width=\columnwidth,trim=0 0 0 0.1cm, clip]{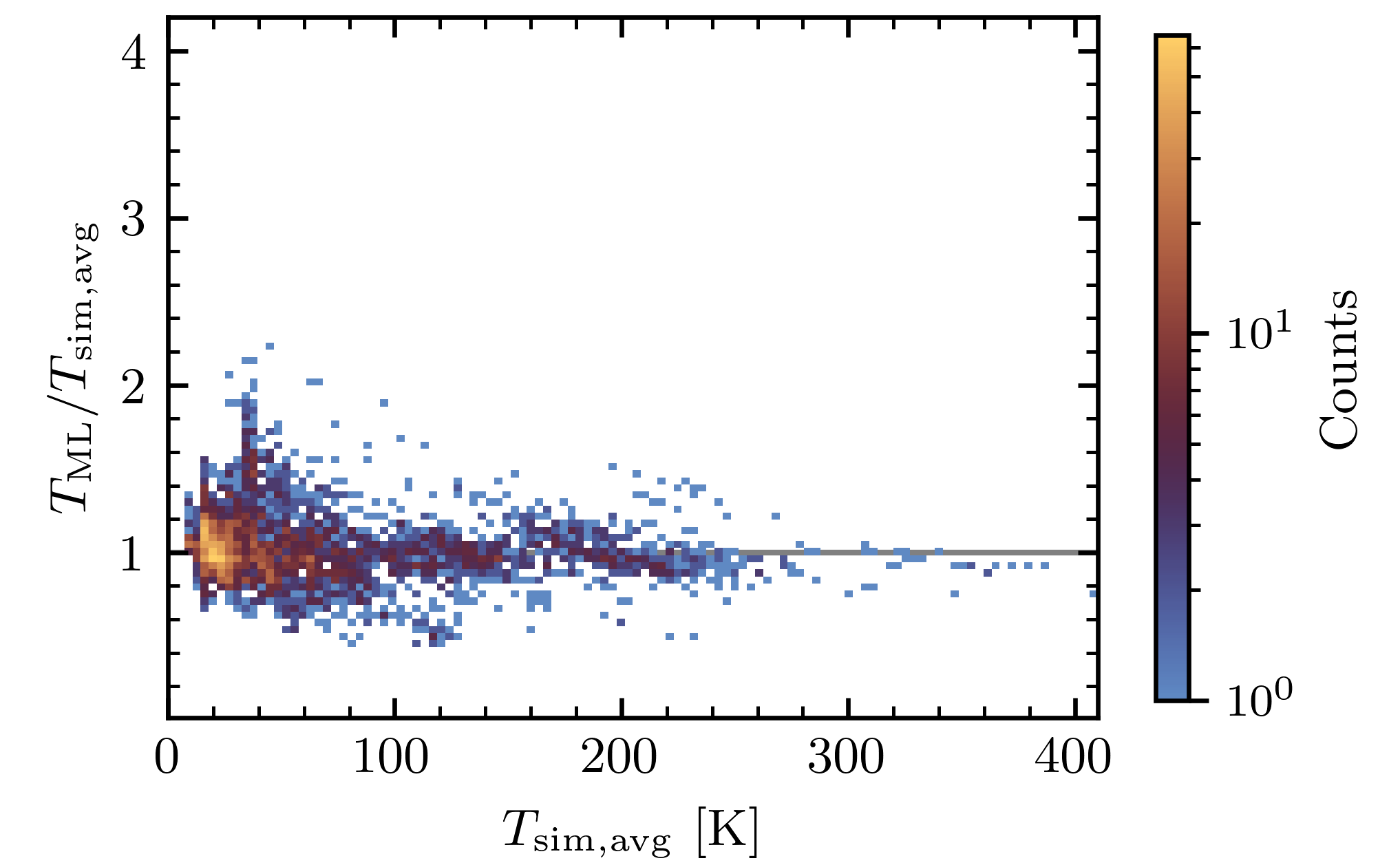}
     \end{subfigure}\\
     \caption{Ratio between estimated and simulated mass-weighted core temperatures as a function of $T_{\rm D}$.
     \emph{Top:} $T_{L/M}/T_{\rm sim,avg}$.  
     \emph{Middle:} $T_{L/M+F}/T_{\rm sim,avg}$ for the $L/M{+}$flux model.  
     \emph{Bottom:} $T_{\rm ML}/T_{\rm sim,avg}$ for the machine learning model.
     The horizontal line indicates equal temperatures. The empirical $L/M$ method systematically overestimates the temperatures of cold cores and underestimates those of warm cores. Including flux information reduces this bias and scatter, while the machine learning model recovers the simulated temperatures with minimal residual trends.}
     \label{fig:rfl9_update_hist2D_diff_tdus}
\end{figure}

Figure~\ref{fig:rfl9_update_hist2D_diff_tdus} shows the ratio of each estimated to simulated core temperature as a function of $T_{\rm sim,avg}$. The top panel demonstrates that the $L/M$ method systematically overestimates the temperatures of cold cores ($T_{\rm sim,avg} \lesssim 50$~K) and underestimates those of warm cores, yielding an overall $R^2 = 0.43$. This pronounced temperature-dependent bias highlights the limitations of using a single global $L/M$ indicator to characterise the thermal state of individual cores. The middle panel shows that the combined $L/M{+}$flux model substantially reduces the overall scatter and softens the strong temperature-dependent bias present in the pure $L/M$ prescription. The data cluster far more tightly around unity, increasing the overall $R^2$ to 0.83. Some residual structure remains, but the distribution is markedly more concentrated than for the simple $L/M$ method. The bottom panel displays the machine learning model, which provides the most accurate reconstruction of the simulated temperatures. The ratios remain tightly concentrated around unity across the entire temperature range, and the residual scatter is small. The resulting $R^2 = 0.96$ indicates excellent agreement between the estimates and the simulation values. 

Overall, our analysis shows that supervised machine learning offers an effective route towards accurate dust-temperature estimation, especially when trained on realistic simulations. Fully exploring its capabilities in observational contexts will be an important direction for future works.

\subsection{Consistency between flux and temperature estimates}
\label{sec:flux_consistency}

In addition to using accurate dust temperature estimates (see Sect.~\ref{sec:temperature_estimates}), the determination of core masses also requires good measurements of the dust continuum flux emission (see Eq.~\ref{eq:mass}). In the following, we examine the accuracy of the ALMA observations to recover the original fluxes as well as the consistency between the fluxes and temperatures used in the mass derivation. 

\begin{figure}
    \centering
    \includegraphics[width=1.0\linewidth]{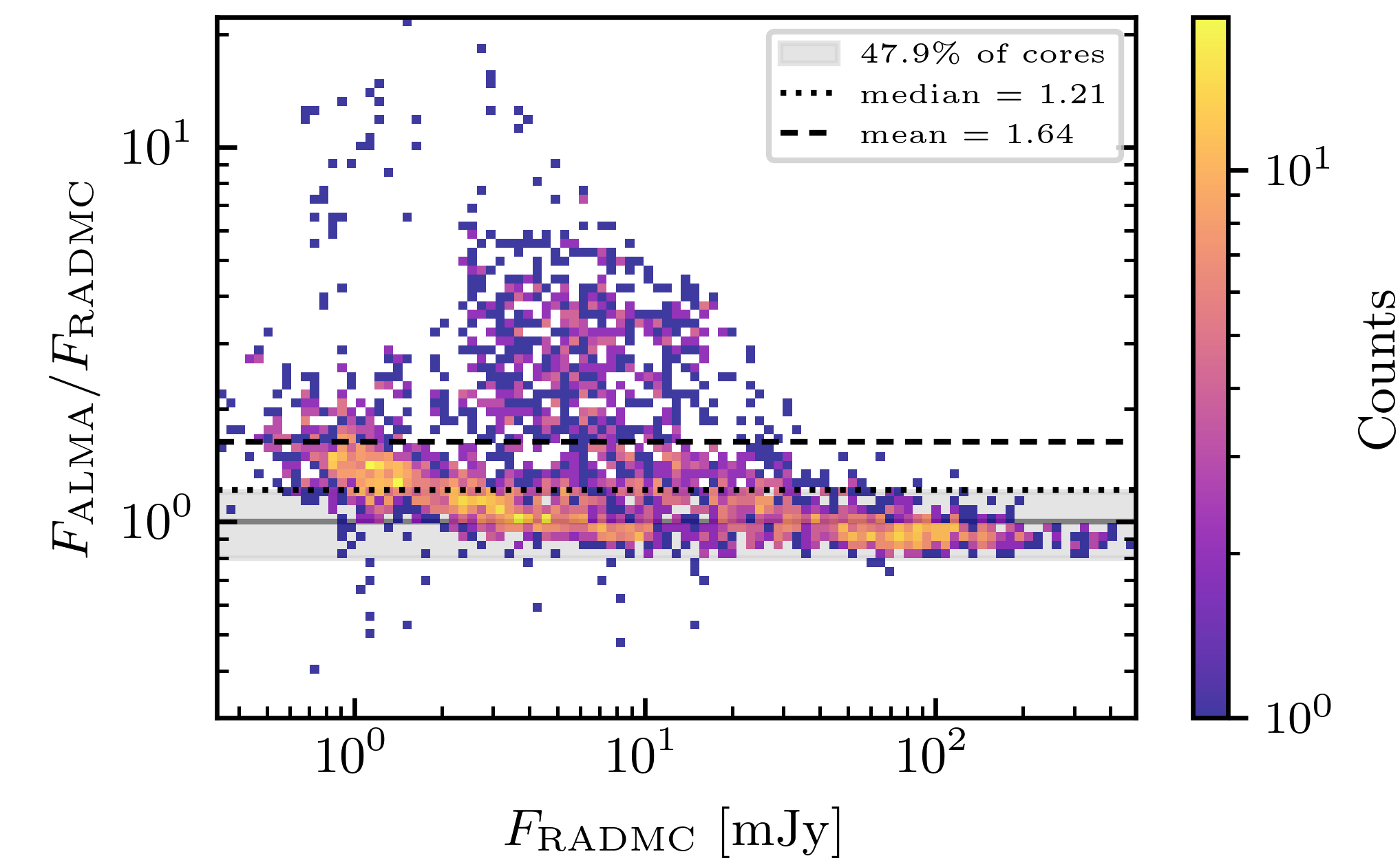}
    \caption{Total fluxes of the synthetic ALMA cores compared to those derived directly from the \textsc{RADMC-3D} images. The grey line indicates the one-to-one relation. Overall, the ALMA fluxes reproduce the \textsc{RADMC-3D} fluxes well. At low fluxes, ALMA cores tend to exhibit slightly higher values than their \textsc{RADMC-3D} counterparts, while at intermediate flux levels the scatter increases, likely due to interferometric filtering and noise.}
    \label{fig:rfl9_update_hist2D_flux_ALMA_vs_RADMC_RADMC}
\end{figure}

We start by exploring whether the fluxes of the cores in the \textsc{RADMC-3D} maps are accurately recovered in the synthetic ALMA observations. Figure~\ref{fig:rfl9_update_hist2D_flux_ALMA_vs_RADMC_RADMC} shows the ratio of the ALMA flux of each core to its corresponding \textsc{RADMC-3D} flux as a function of the \textsc{RADMC-3D} flux. Overall, most points cluster around the unity line, indicating that interferometric filtering preserves the global flux scaling for most cores. At low fluxes ($\lesssim 2$~mJy), ALMA tends to slightly overestimate the emission relative to \textsc{RADMC-3D}, whereas the scatter increases at intermediate fluxes ($\sim 1.5$–12~mJy), which may arise from a combination of noise, beam convolution, and image artefacts likely produced by partially resolved-out structures not properly recovered in the interferometric images.

\begin{figure}
     \centering
     \begin{subfigure}[b]{0.45\textwidth}
         \centering
         \includegraphics[width=\columnwidth,trim=0 0 0 0.056cm, clip]{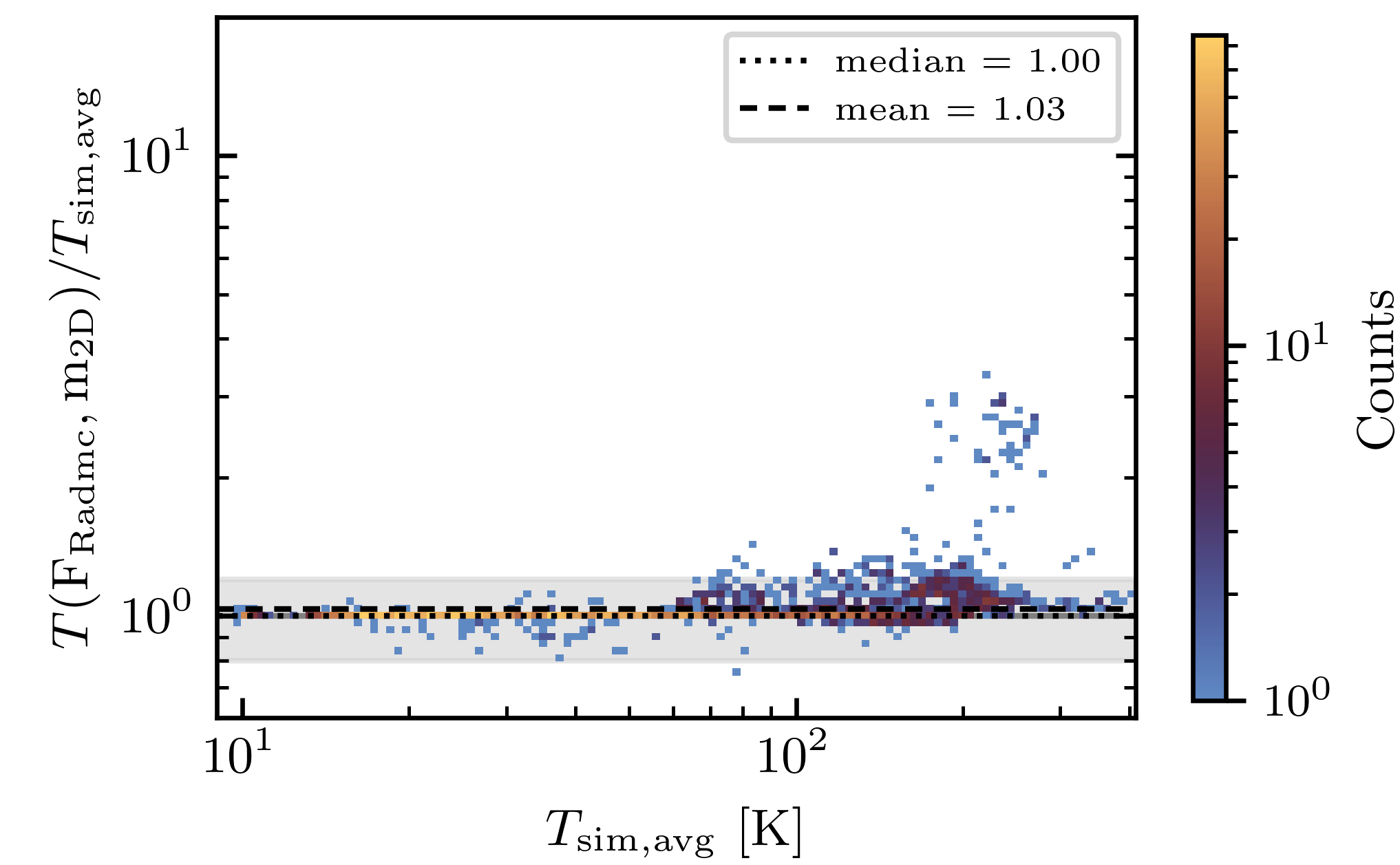}
     \end{subfigure}\\
     \begin{subfigure}[b]{0.45\textwidth}
         \centering
         \vspace{-0.9cm}
         \includegraphics[width=\columnwidth,trim=0 0 0 0.1cm, clip]{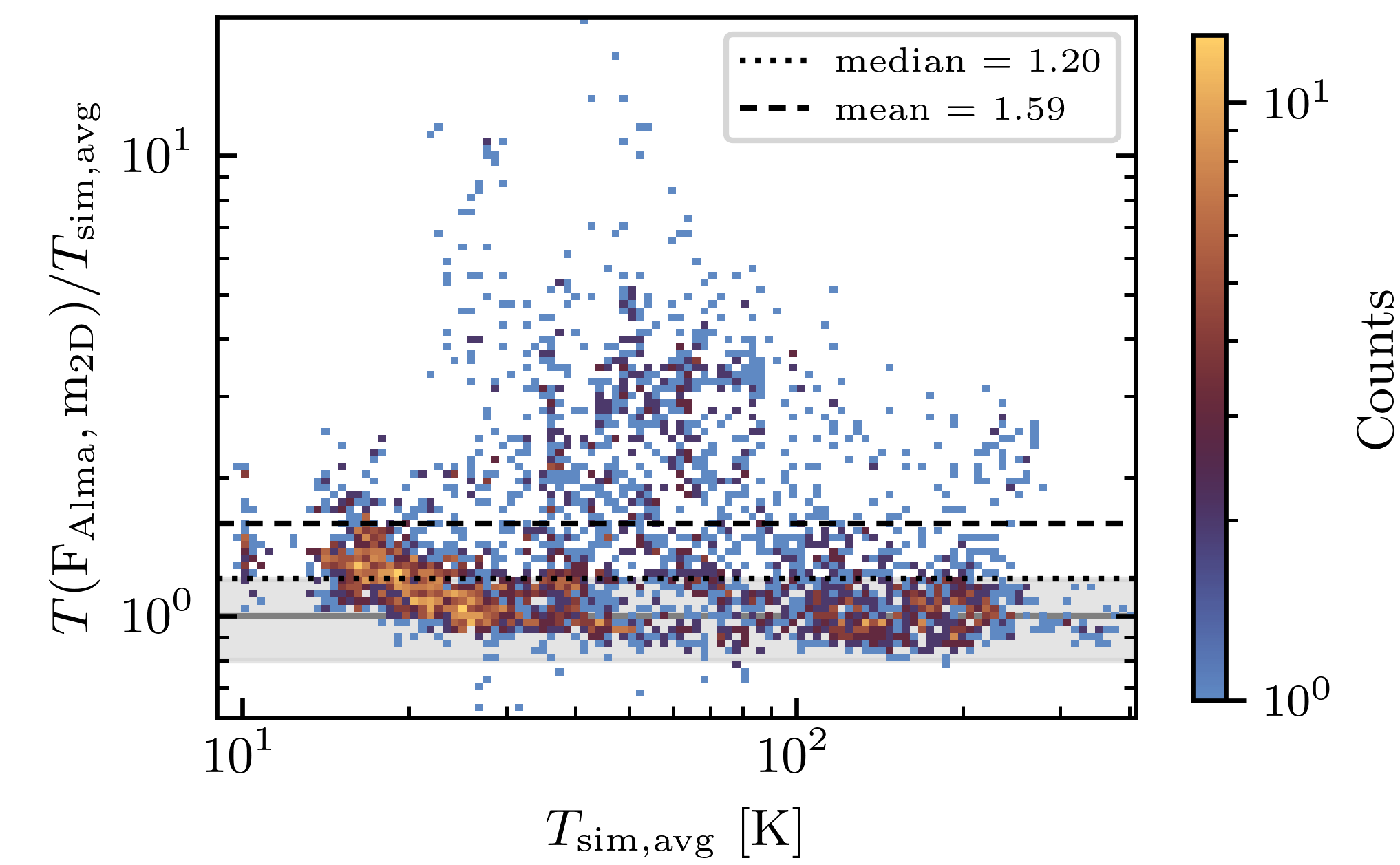}
     \end{subfigure}\\
     \caption{Ratio between the dust temperature inferred from the post-processed flux and simulation mass, $T_{\mathrm{F,m2D}}$, and the mass–weighted average core temperature from the simulation, $T_{\mathrm{sim,avg}}$, plotted as a function of $T_{\mathrm{sim,avg}}$ for \textsc{RADMC-3D} (\textit{top}) and \textsc{ALMA} cores (\textit{bottom}).
     The dotted and dashed horizontal lines indicate the median and mean of $T_{\mathrm{F,m2D}}/T_{\mathrm{sim,avg}}$, respectively. For the \textsc{RADMC-3D} cores, the derived temperatures agree very well with the simulation temperatures. Most ALMA cores lie near to the unity line, but a non-negligible fraction shows significantly higher ratios.}
     \label{fig:rfl9_update_hist2D_t_flux_mass_radmc}
\end{figure}

Since both the flux and the temperature are needed to compute the mass of a core, any change in the measured flux will automatically change the inferred mass for a given dust temperature. To isolate the effect of flux reconstruction, we invert the flux–mass relation and calculate the dust temperature $T_{\mathrm{flux,m2D}}$ that would be required in order to recover the true core mass from the simulation using the both the measured (noise-affected) ALMA fluxes and the (ideal) \textsc{RADMC-3D} fluxes. Figure~\ref{fig:rfl9_update_hist2D_t_flux_mass_radmc} compares $T_{\mathrm{flux,m2D}}$ to the mass-weighted temperatures from the simulation. The top panel shows the reference case in which the temperatures have been calculated using the (ideal) \textsc{RADMC-3D} fluxes. In this case, both the flux and the true core mass are known from the simulation, and the inferred temperatures match the simulation temperatures extremely well, with the distribution tightly clustered around unity (with median and mean ratios of 1.00 and 1.03, respectively). This confirms that when the flux is accurately recovered, the mass estimated from the flux and temperature is reliable. The bottom panel shows the corresponding result using the ALMA fluxes. Similar to the previous case, most cores lie close to the unity line, indicating that the temperatures implied by the ALMA fluxes are broadly consistent with the intrinsic core temperatures. However, the elevated median and mean ratios (1.2 and 1.6, respectively) indicate a systematic tendency for the ALMA fluxes to be somewhat higher than the \textsc{RADMC-3D} fluxes or to miss some extended emission, depending on the configuration. A non-negligible fraction of cores shows significantly higher ratios, corresponding to cases with large ALMA–to–\textsc{RADMC-3D} flux discrepancies.

\subsection{Core mass recovery using different temperature estimators}
\label{sec:obs_masses}

\begin{figure}
     \begin{subfigure}[b]{0.45\textwidth}
         \centering
         \includegraphics[width=\columnwidth,trim=0 0 0 0.1cm, clip]{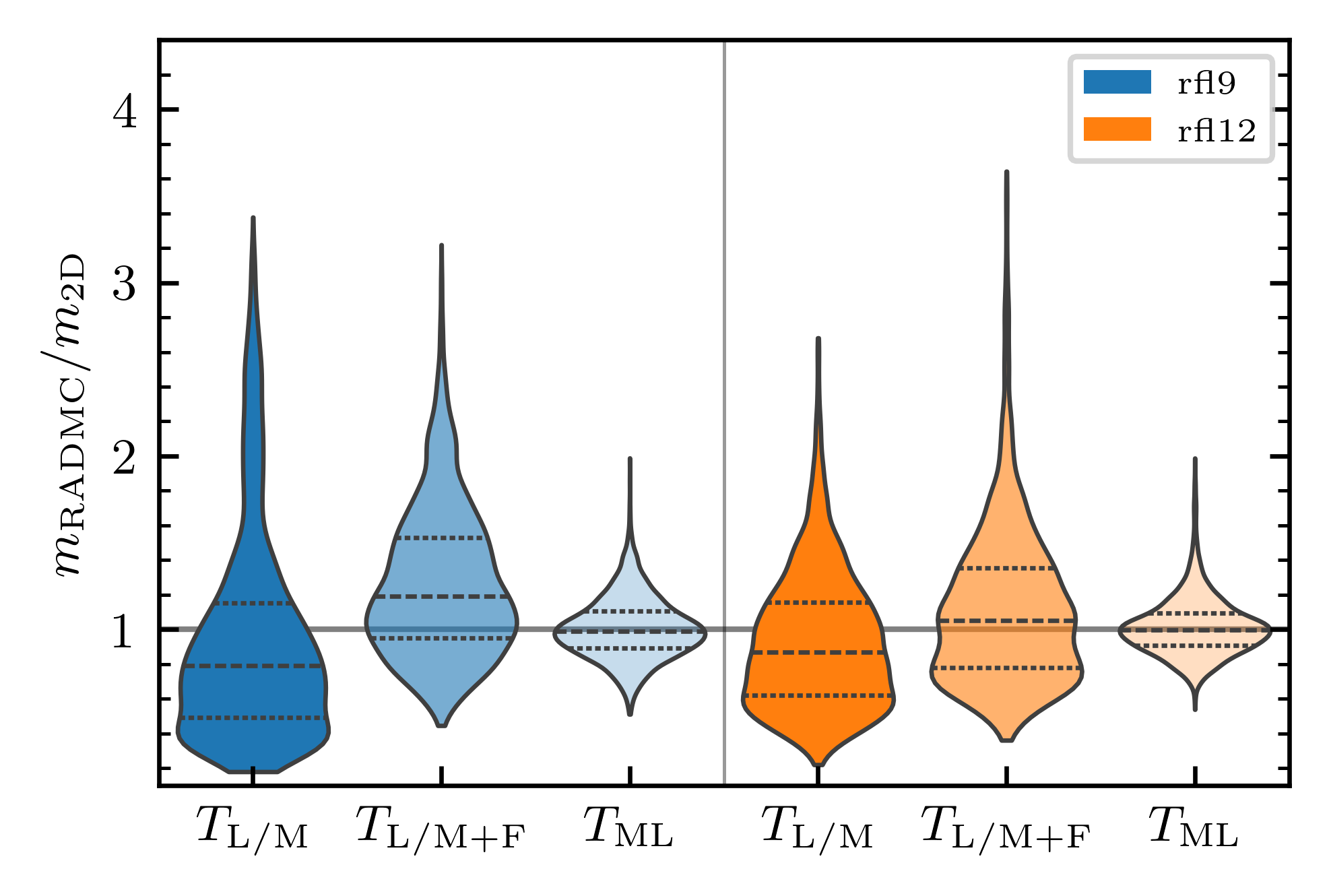}
     \end{subfigure}\\
     \begin{subfigure}[b]{0.45\textwidth}
         \centering
         \vspace{-0.74cm}
         \includegraphics[width=\columnwidth,trim=0 0 0 0.24cm, clip]{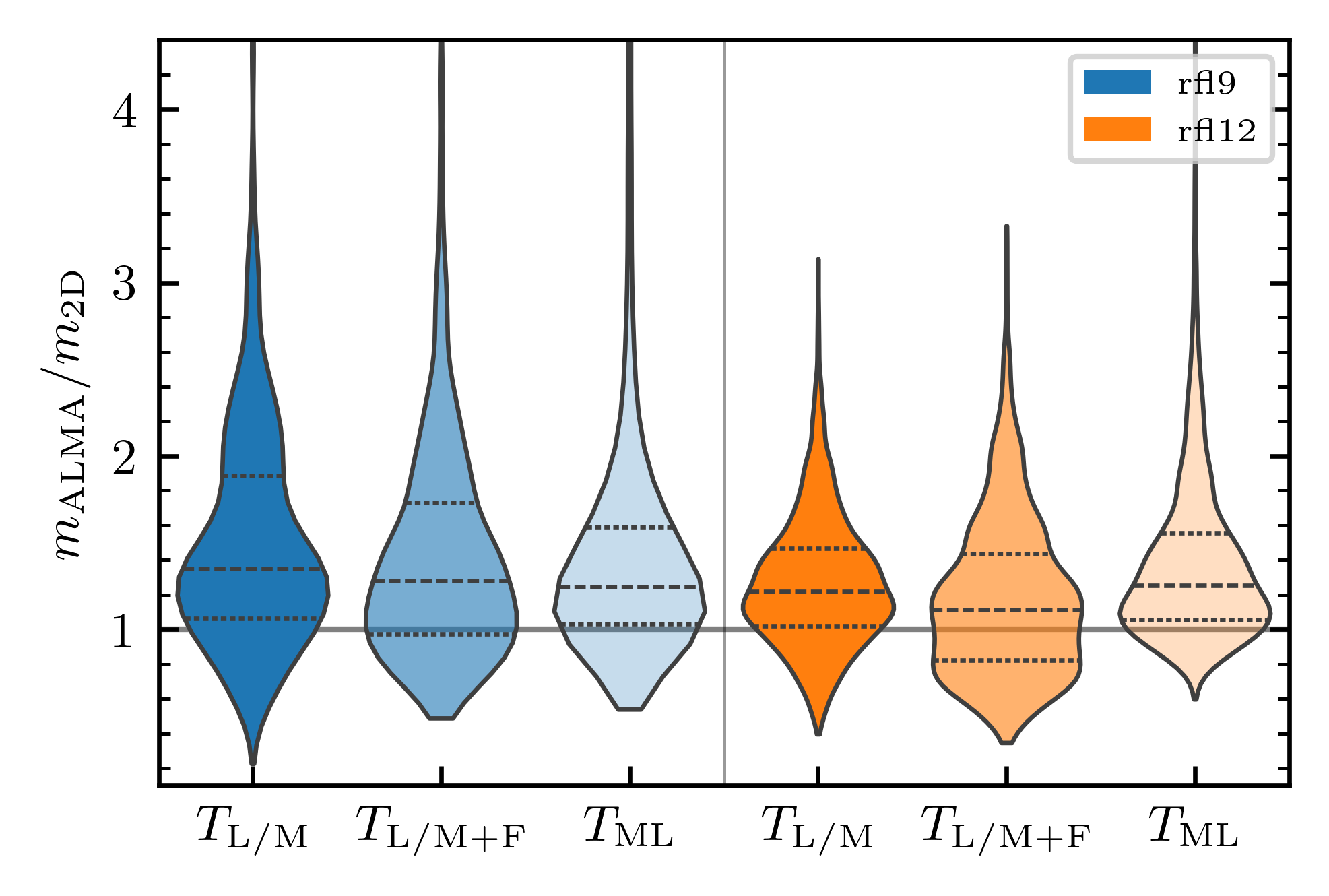}
     \end{subfigure}\\
     \caption{Ratio of reconstructed core masses to the true 2D masses for \textsc{rfl9} and \textsc{rfl12}. The panels show masses derived from the idealised \textsc{RADMC-3D} maps (\textit{top}) and from \textsc{ALMA} maps (\textit{bottom}). In both cases, core masses are reconstructed using the $L/M$ temperature estimate alone (dark colours), the combined $L/M$+flux temperature estimate (medium colours), or the machine learning approach, $T_{\rm ML}$ (light colours). For the \textsc{RADMC-3D} case (\textit{top}), the $L/M$ method underestimates the core masses, whereas the $L/M$+flux method yields more accurate results, and $T_{\rm ML}$ performs best. For the $m_\mathrm{ALMA}$ (\textit{bottom}), all three temperature estimates perform similarly in a statistical sense.}
     \label{fig:LM_vs_LM_flux}
\end{figure}

In this section we asses how accurately the different temperature prescriptions recover the true core masses. For this, we compare the reconstructed masses to the intrinsic 2D masses ($m_{\rm 2D}$) for both resolution runs: \textsc{rfl9} and \textsc{rfl12}. Figure~\ref{fig:LM_vs_LM_flux} presents these mass ratios for the identified cores in both the idealised \textsc{RADMC-3D} maps (top panels) and the synthetic ALMA observations (bottom panels). In both cases, we test the three temperature models presented in Sect.~\ref{sec:temperature_estimates}, corresponding to the standard $L/M$ approach, the improved $L/M{+}{\rm flux}$ method, and the machine learning-based model.

For the \textsc{RADMC-3D} emission maps, where the fluxes are ideal and noise-free, the standard $L/M$ method systematically underestimates the core masses, with median ratios $m_{\rm RADMC}/m_{\rm 2D}\simeq 0.8$ for \textsc{rfl9} and $0.85$ for \textsc{rfl12}. The $L/M+F$ approach, however, provides a notably more accurate reconstruction with median ratios of $1.2$ (\textsc{rfl9}) and $1.05$ (\textsc{rfl12}), demonstrating its ability to account for variations in dust temperature more effectively than the standard $L/M$ prescription. The machine learning approach yield the most accurate results for both runs, producing $m_{\rm RADMC}/m_{\rm 2D}$ values tightly clustered around unity. In the synthetic \textsc{ALMA} images, by contrast, the three temperature estimators yield broadly comparable $m_{\rm ALMA}/m_{\rm 2D}$ statistically across all cores, with modest differences in the distribution shapes and median values in the range $1.1-1.3$.

\begin{figure*}
     \centering
     \begin{subfigure}[b]{0.33\textwidth}
         \centering
         \includegraphics[width=1.0\columnwidth,trim=0 0 1.02cm 0.0cm, clip]{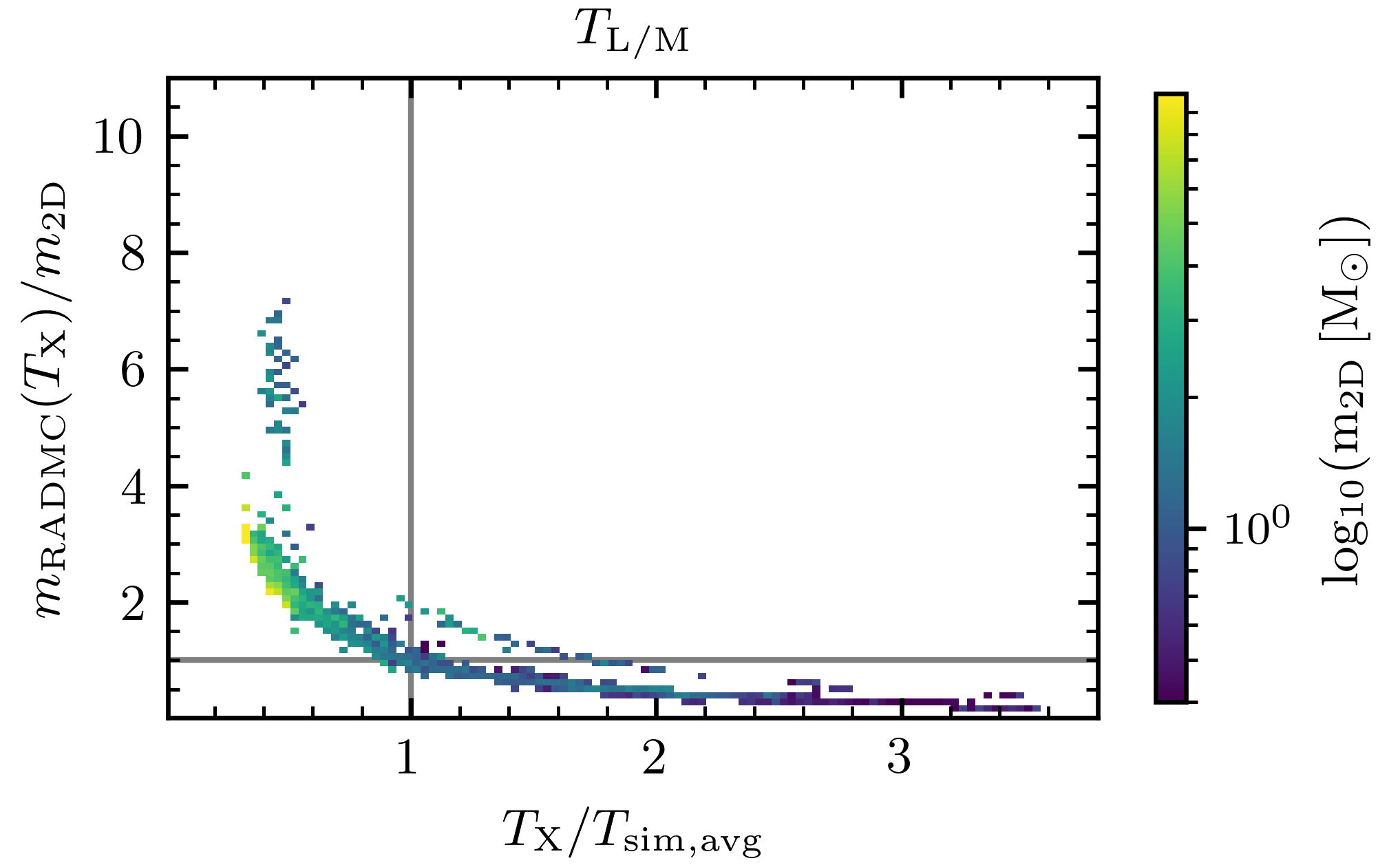}
     \end{subfigure}
     \hspace{-0.86cm}
     \begin{subfigure}[b]{0.33\textwidth}
         \centering
         \includegraphics[width=\columnwidth,trim=1.02cm 0 0 0.0cm, clip]{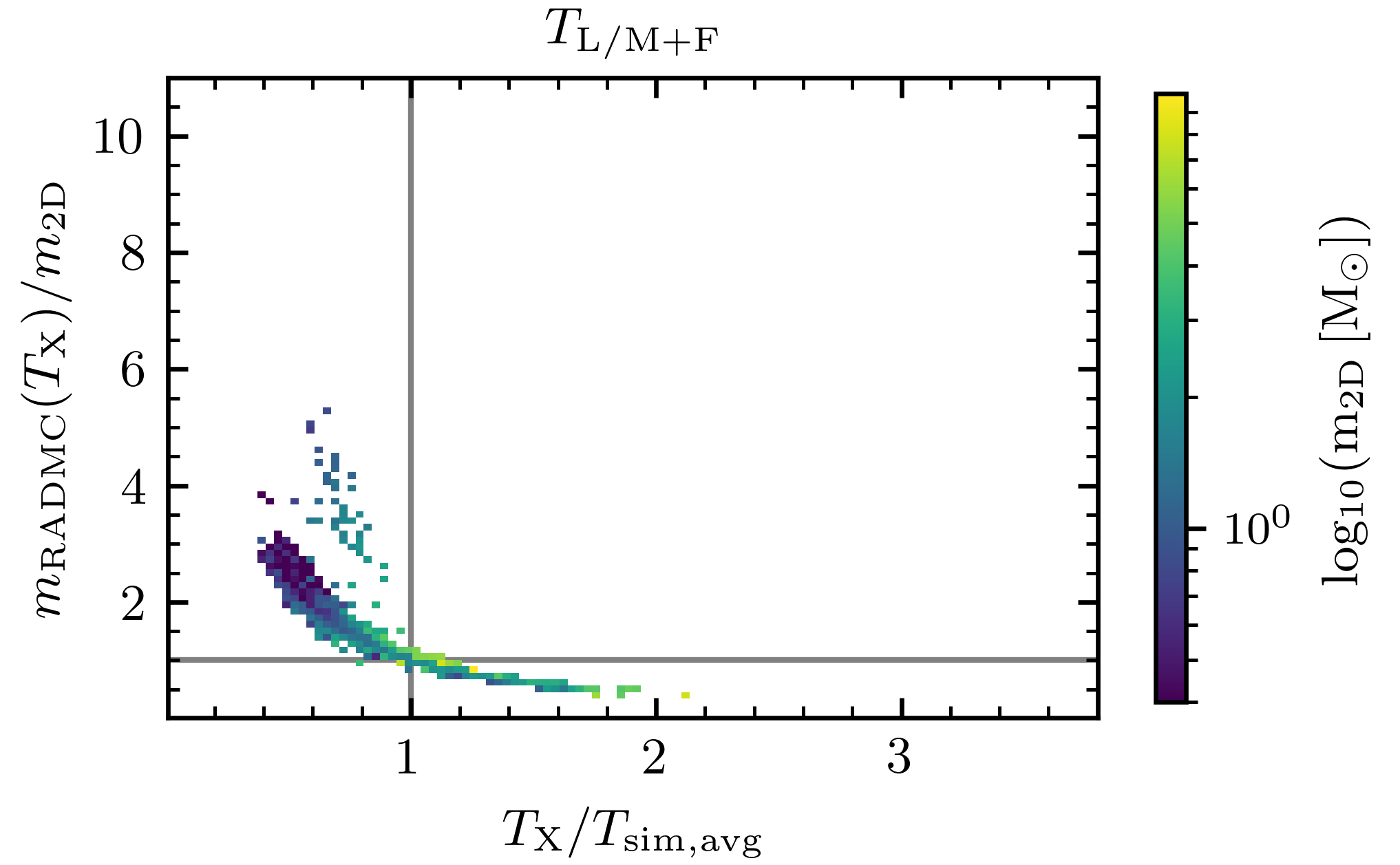}
     \end{subfigure}
     \hspace{-1.67cm}
        \begin{subfigure}[b]{0.33\textwidth}
         \centering
         \includegraphics[width=\columnwidth,trim=1.02cm 0 0 0.0cm, clip]{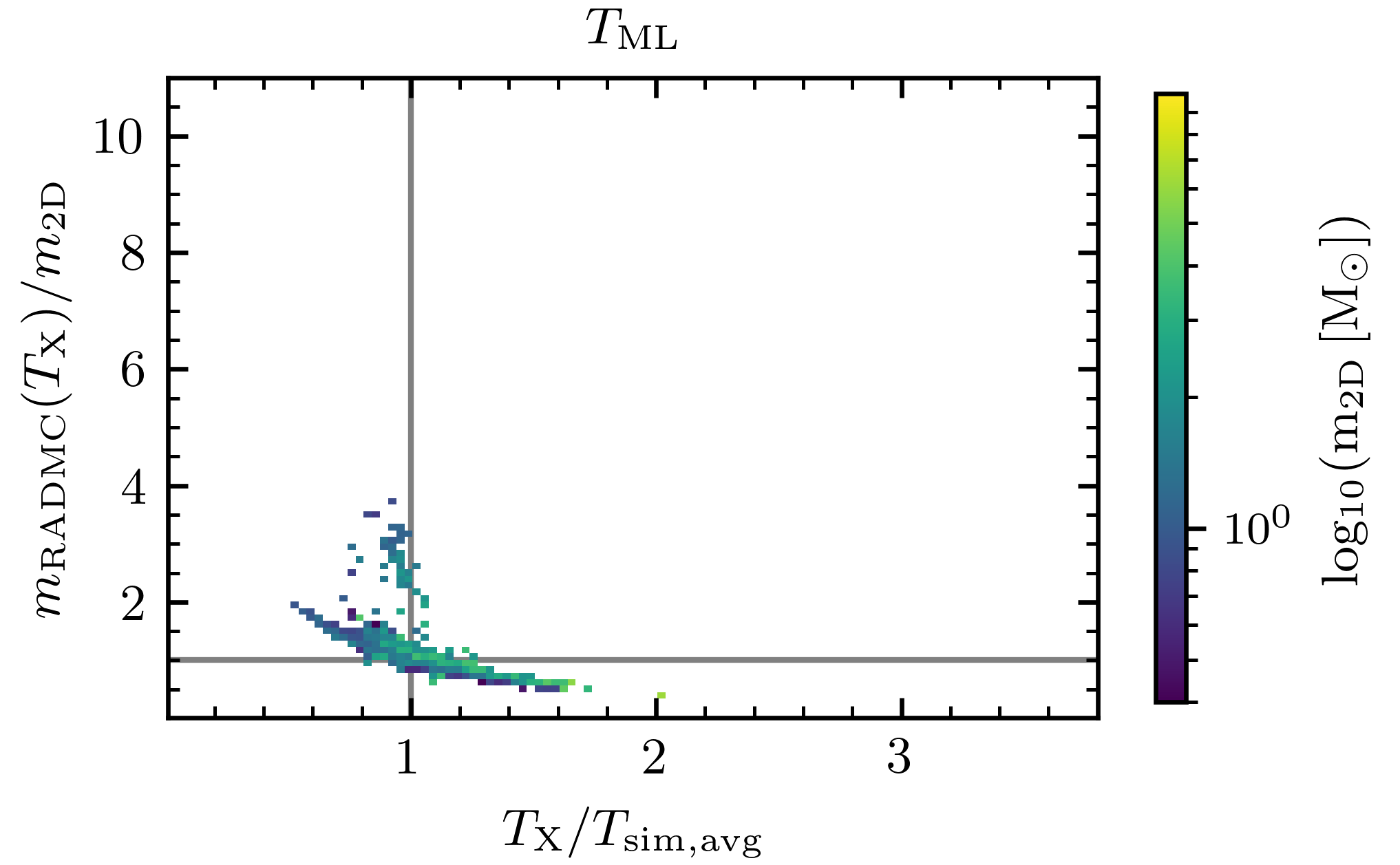}
     \end{subfigure}\\
     \vspace{-0.77cm}
     \begin{subfigure}[b]{0.33\textwidth}
         \centering
         \includegraphics[width=\columnwidth,trim=0 0 1.02cm 0.47cm, clip]{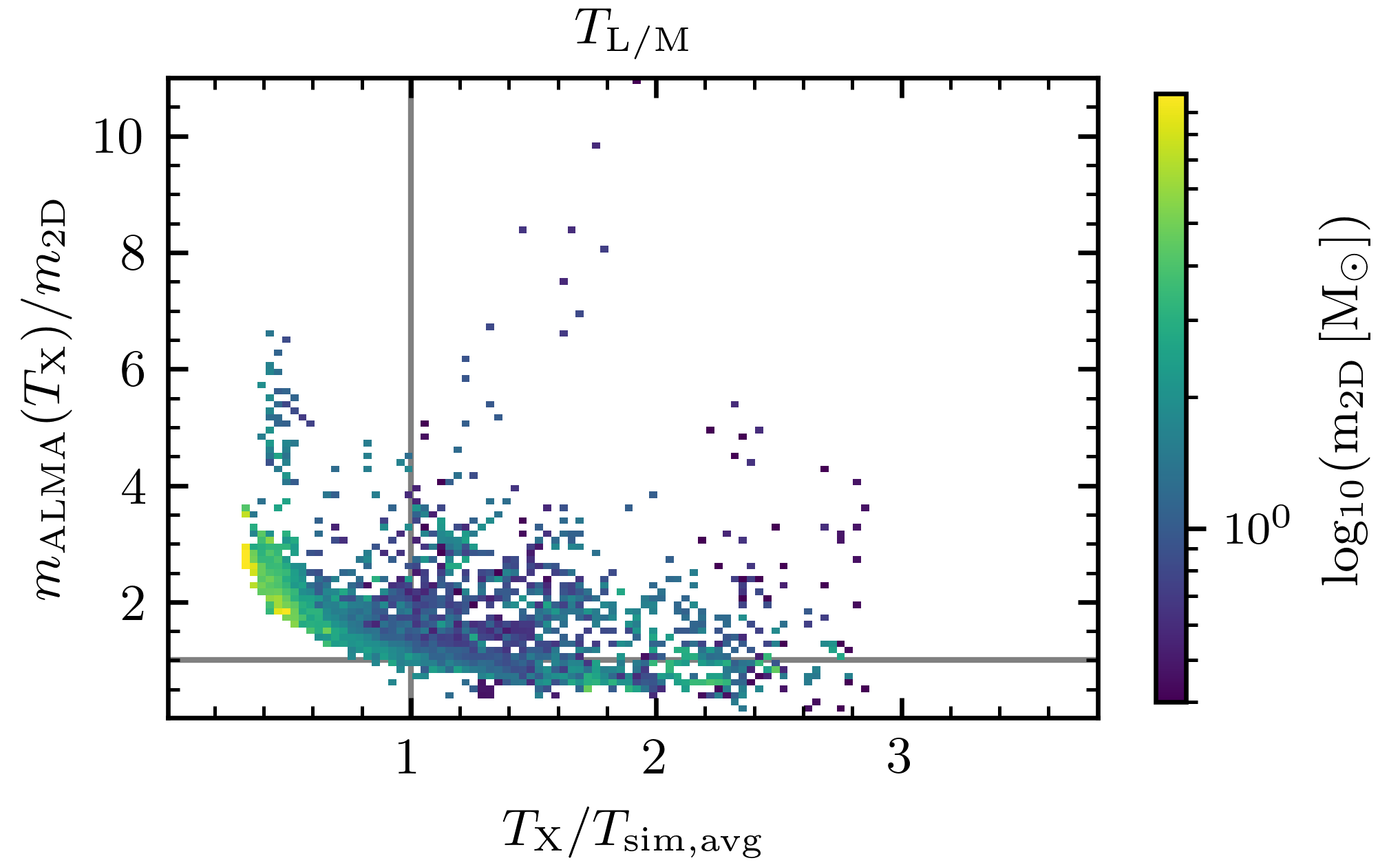}
     \end{subfigure}
     \hspace{-0.86cm}
     \begin{subfigure}[b]{0.33\textwidth}
         \centering
         \includegraphics[width=\columnwidth,trim=1.02cm 0 0 0.47cm, clip]{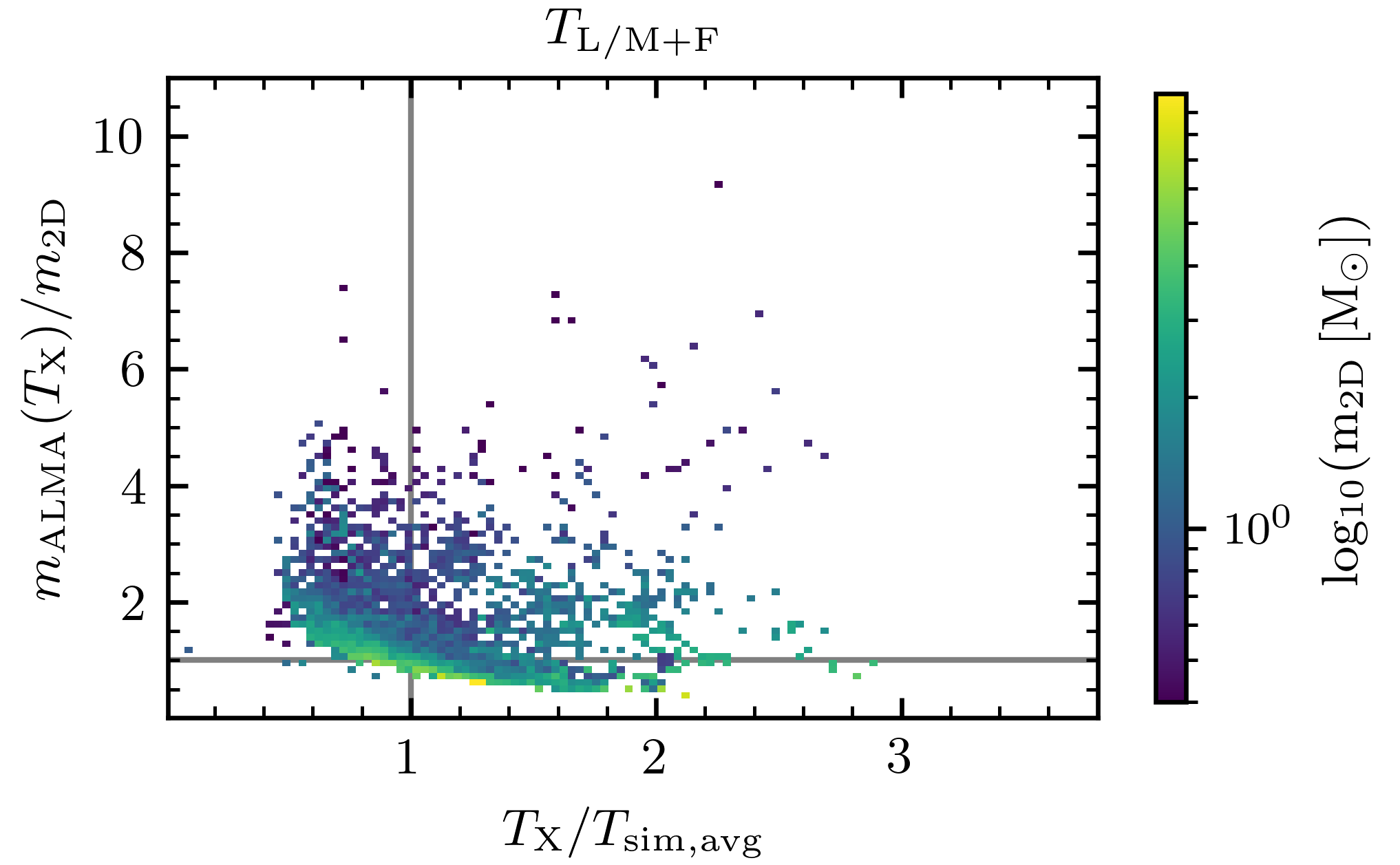}
     \end{subfigure}
     \hspace{-1.67cm}
     \begin{subfigure}[b]{0.33\textwidth}
         \centering
         \includegraphics[width=\columnwidth,trim=1.02cm 0 0 0.47cm, clip]{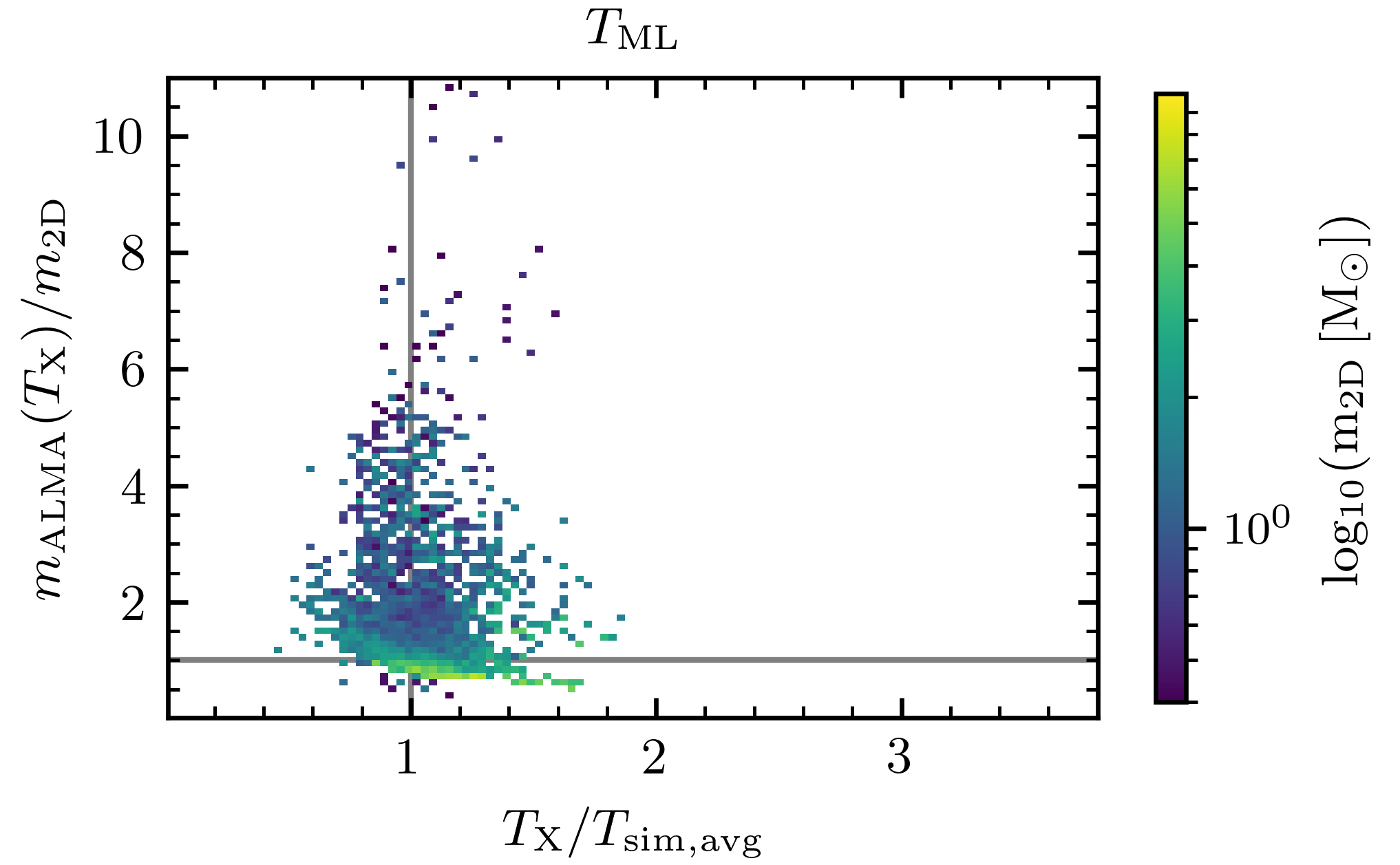}
     \end{subfigure}\\
     \caption{Two-dimensional histograms comparing the temperature and mass biases of core masses derived with \textsc{RADMC-3D} (\textit{top row}) and with the synthetic \textsc{ALMA} analysis (\textit{bottom row}). From left to right, we use temperatures estimated from $L/M$, from $L/M$ plus flux ($L/M{+}F$), and from the machine learning ($\mathrm{ML}$) model. The $x$-axis shows the temperature bias (estimated temperature relative to the simulation temperature), while the $y$-axis shows the corresponding mass bias (derived core mass relative to the projected simulation core mass). The colour scale indicates the projected simulation core mass (logarithmic). Grey lines mark equality. Overall, the \textsc{ALMA} cores show a larger scatter in mass bias than the \textsc{RADMC-3D} cores.
     For the $L/M$ method, \textsc{RADMC-3D} shows that an underestimated temperature relative to the simulation leads to an overestimated derived mass, and vice versa. \textsc{ALMA} cores show a similar trend. In both cases, the most massive cores are overpredicted in mass, while their temperatures tend to be underpredicted.
     Including the flux term ($L/M{+}F$) in \textsc{RADMC-3D} yields a more compact distribution around the one-to-one relation, but it also alters the mass dependence. Lower-mass cores now preferentially show underestimated temperatures with correspondingly overestimated masses, whereas higher-mass cores are recovered comparatively well. \textsc{ALMA} cores show a similar trend, though with noticeably more scatter.
     The $\mathrm{ML}$ model provides the best overall temperature agreement. However, in the \textsc{ALMA} case the lowest-mass cores remain systematically overestimated, even when the temperature bias is small.
     }
     \label{fig:rfl9_update_hist2D_mcasa_diff_vs_tdus_diff}
\end{figure*}

We now compare temperature and mass biases for core masses derived from the three temperature prescriptions (Fig.~\ref{fig:rfl9_update_hist2D_mcasa_diff_vs_tdus_diff}). The top row shows the idealised \textsc{RADMC-3D} case, while the bottom row shows the synthetic \textsc{ALMA} analysis. In each panel the temperature bias (estimated temperature relative to the simulation value) is plotted against the corresponding mass bias (derived core mass relative to the projected simulation core mass), with colour indicating the projected simulation core mass.

For temperatures estimated from the $L/M$ relation (left column), both \textsc{RADMC-3D} and \textsc{ALMA} exhibit the expected anti-correlation: underestimated temperatures lead to overestimated core masses, and vice versa. The most massive cores preferentially populate the regime of underestimated temperatures and overestimated masses. While the trend is tight in \textsc{RADMC-3D}, the \textsc{ALMA} results show substantially larger scatter in mass bias at fixed temperature bias.

Including the flux term ($L/M{+}F$; middle column) yields a more compact distribution around the one-to-one relations in the \textsc{RADMC-3D} case, but it also changes the mass dependence. Lower-mass cores tend to have underestimated temperatures with correspondingly overestimated masses, whereas higher-mass cores are recovered more accurately. The \textsc{ALMA} cores follow the same overall behaviour, again with noticeably increased scatter. However, the more massive cores now have a mass bias close to unity.

Finally, the machine learning model (right column) provides the best overall agreement in temperature. In \textsc{RADMC-3D} this translates into mass ratios clustered close to unity, whereas in the \textsc{ALMA} case the lowest-mass cores remain systematically overestimated even when the temperature bias is small.

The observational effects, such as noise, filtering, and imperfect flux recovery, dominate the mass uncertainties and reduce the sensitivity to the choice of temperature prescription. As shown above, inaccurate flux reconstruction directly propagates into mass errors, even when the temperature estimate is accurate. Consequently, a correct flux recovery is essential for a reliable mass determination in interferometric observations.

\section{The observer's guide to the study of cores}
\label{sec:warnings-guidelines}

In this section we summarise the main aspects and caveats to consider when interpreting interferometric dust-continuum observations of star-forming clumps and cores, similar to those performed in the ALMAGAL project \citep[see][]{2025Molinari, 2025Sanchez}, based on our comparison between simulations and synthetic observations. The main points are as follows.

\begin{itemize}

    \item Projection effects in the core mass determination: Our simulations show that 2D projected core masses systematically exceed the corresponding 3D spherical masses (see Sect.~\ref{sec:geometry} and Fig.~\ref{fig:violin_flash_2D_vs_3D}). The major contribution to this mass excess is the presence of additional material along the LOS (i.e. in the foreground or background of the core), rather than from geometric effects when projecting a 3D spherical core into a 2D circle (see Fig.~\ref{fig:resolution_mean_mass_vs_time_all_cnt_flash.png}).

    \vspace{0.2cm}
    \item Sink (stellar) masses and core masses: Our simulations show that the core mass has a general decrease with increasing (stellar) sink mass (see Sect.~\ref{sec:sink_core_mass} and Fig.~\ref{fig:rfl9_update_diffm2D_sink_vs_time}). At earlier stages, cores are dominated by massive envelopes with small sink masses, while at later stages, sinks have accreted most of the available material contained in the envelope. Though inferring stellar masses directly from core masses may be highly uncertain, we have defined an empirical relation between the sink (stellar) mass and the corresponding core mass (see Fig.~\ref{fig:rfl9_update_hist2D_m2D_vs_msink}, Eqs.~\ref{eq:msink_m2d} and \ref{eq:msink_malma}) based on our simulations.

    \vspace{0.2cm}
    \item Global clump properties do not uniquely determine the core population: Based on our simulations, before ionising feedback sets in, the CFE increases approximately linearly with time \citep[consistent with observational findings, see][]{2025Elia}, and the clump mass can be inferred from the total core mass. Once feedback becomes important, this relation breaks down and no clear correlation between clump-scale properties and the underlying core population remains (see Sect.~\ref{sec:clump} and Figs.~\ref{fig:rfl9_update_core_mass_vs_cloud_mass} and \ref{fig:rfl9_update_core_cloud_mass_vs_time}).

    \vspace{0.2cm}
    \item $L/M$ clump-based temperatures do not accurately describe core temperatures: Even though the clump $L/M$ traces the global thermal evolution and the onset of ionising feedback (see Fig.~\ref{fig:rfl9_update_L_over_M_in_time}), core temperatures span more than one order of magnitude at a fixed $L/M$. The empirical $L/M$ prescription \citep[see e.g.][]{2016Molinari, 2025Coletta} can underestimate or overestimate individual core temperatures by factors of around two to three (see Sect.~\ref{sec:LM_relation} and Fig.~\ref{fig:rfl9_update_hist2D_tdus_sim_LM_binned_mass}).
    
    \vspace{0.2cm}
    \item Core flux information improves $L/M$ clump-based temperature estimates: The $L/M$ clump-based temperature estimate for a core can be improved by incorporating the measured dust continuum flux of each core into a simple $L/M+F$ model (see Sect.~\ref{sec:LMflux}, Eq.~\ref{eq:model}). This substantially reduces the scatter in the recovered temperatures (see Fig.~\ref{fig:rfl9_update_hist2D_diff_tdus})
    and increases the coefficient of determination from $R^{2}=0.43$ for the $L/M$-only estimate to $R^{2}=0.83$ for $L/M+F$. Temperature prescriptions that combine clump-scale ($L/M$) and core-scale (flux) information are therefore preferred. The preliminary study of a machine learning model suggests further improvements in the core temperature determination with $R^{2}=0.96$ (see Sect.~\ref{sec:MLmodel} and Fig.~\ref{fig:rfl9_update_hist2D_diff_tdus}), but its application to real observations still requires further investigation.

    \vspace{0.2cm}
    \item Mass recovery also depends on flux reconstruction: Imperfect recovery of the core fluxes (e.g. due to artefacts or noise in interferometric observations) introduces additional errors in the mass determination for individual cores of up to a factor of about three, even when the dust temperature is known with perfect accuracy (see Sect.~\ref{sec:flux_consistency}, Figs.~\ref{fig:rfl9_update_hist2D_flux_ALMA_vs_RADMC_RADMC} and \ref{fig:rfl9_update_hist2D_t_flux_mass_radmc}).
    
\end{itemize}

\subsection{Expected uncertainties in core-mass estimates}
\label{sec:guideline_observers}

Based on the comparison between numerical simulations, \textsc{RADMC-3D} emission maps, and synthetic ALMA images, the following quantitative biases may be expected when deriving core masses from (sub-)millimetre continuum interferometric observations.
\begin{itemize}

    \item Projection effects: The (projected) core mass, $m_\mathrm{2D}$, exceeds the true spherical core mass, $m_\mathrm{3D}$, due to foreground and background contamination from the clump material. This results in core masses overestimates by a factor of two to three.

    \item Temperature uncertainty: True core temperatures span a broad range that is not captured by a single clump-averaged $L/M$ value. If the core temperature is underestimated by a factor of two (e.g.\ $20$~K instead of $40$~K), the mass is overestimated by a factor of three to seven. If the temperature is overestimated by a factor of two, the mass is underestimated by a factor of around two. Typical errors in the core mass estimate based only on temperature uncertainties are usually about a factor of two to five.

    \item Interferometric filtering: Instrumental effects from interferometers such as ALMA may typically overestimate the core mass by a factor of about one to two due to incorrect flux reconstruction.

\end{itemize}
Combining all dominant uncertainties, we find that individual ALMA continuum–derived core masses may typically be uncertain by a factor of about three to five and up to ten for extreme cases.

\section{Summary}
\label{sec:conclusions}

We have used radiation–hydrodynamic simulations, radiative-transfer post-processing, and synthetic ALMA imaging to quantify how reliably dense-core properties can be inferred from ALMAGAL-like continuum observations of massive star-forming clumps. We find that geometric effects alone introduce a substantial mass bias, with projected core masses systematically exceeding (at the level of a factor of roughly two to three) the corresponding 3D spherical masses due to foreground and background contamination. Correct flux reconstruction and dust temperature estimation are the other main sources of uncertainty in continuum-based core masses. Imperfect flux recovery due to interferometric filtering and atmospheric noise, for example, introduces additional mass biases of up to a factor of about two for individual cores, even when the temperature is known. For ALMAGAL-like observations, we found a minimum reliably detectable core mass of $\gtrsim 0.3$~$M_\odot$. Regarding core temperatures, the empirical clump-scale $L/M$-derived temperature prescription, used in some observational works, captures the global thermal evolution but fails to represent the large core-to-core temperature scatter. As a result, it biases the recovered projected core masses, especially for the most massive cores. Including the ALMA-measured core flux in a simple $L/M{+}$flux model significantly improves the temperature estimates, and a supervised machine learning model trained on the simulations performs best in our tests. We also explored the relation between the sink and core mass in the simulations and derived an empirical relation that provides the expected sink mass for a given projected core mass. Finally, at clump scales, the simulations reveal a tight, nearly linear growth of the CFE during the early pre-feedback phase, followed by strong fluctuations and a decline once ionising radiation disrupts the clump. Synthetic observations reproduce these qualitative evolutionary trends and the overall behaviour of the core mass budget, even though individual core masses remain uncertain at a level of a factor of a few.

\begin{acknowledgements}
We thank the anonymous referee for their valuable comments on our manuscript.
We gratefully acknowledge the Gauss Centre for Supercomputing e.V. (www.gauss-centre.eu) for funding this project by providing computing time through the John von Neumann Institute for Computing (NIC) on the GCS Supercomputer JUWELS at Jülich Supercomputing Centre (JSC).
B.Z., S.W.\  and G.A.F.\ gratefully acknowledge funding via the Collaborative Research Center 1601 (SFB 1601, sub-projects A5 and B1) funded by the German Science Foundation (DFG). S.W and G.A.F further acknowledges support by the Deutsche Forschungsgemeinschaft (DFG, German Research Foundation) under Germany’s Excellence Strategy via EXC 3037, the Cluster of Excellence “Our Dynamic Universe”, no. 533607693 and its Global Faculty programme. 
A.S.-M.\ acknowledges support from the RyC2021-032892-I grant funded by MCIN/AEI/10.13039/501100011033 and by the European Union `Next GenerationEU’/PRTR, as well as the program Unidad de Excelencia María de Maeztu CEX2020-001058-M, and support from the PID2023-146675NB (MCI-AEI-FEDER, UE). 
\end{acknowledgements}

\bibliographystyle{aa}
\bibliography{bibliography}

\begin{appendix}

\section{Effects of viewing angle}
\label{sec:viewing_angle}

\begin{figure}[ht!]
    \centering
    \includegraphics[width=1.0\linewidth,trim=0 0cm 0.6cm 0cm, clip]{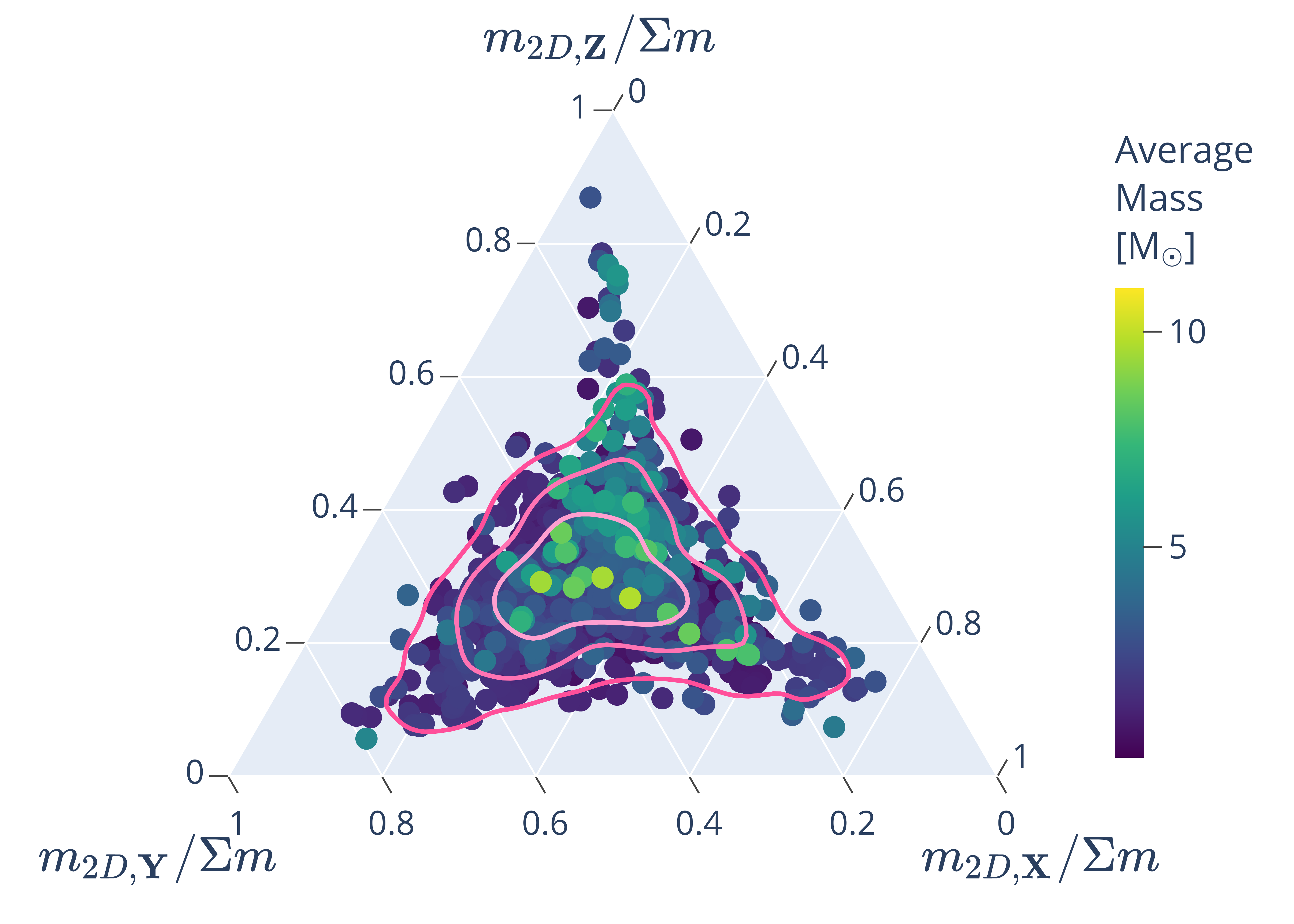}
    \caption{Ternary diagram showing the relative contributions of the 2D projected mass components derived from the simulation along the $X$, $Y$, and $Z$ coordinates. Each point represents a mass configuration, where the three coordinates correspond to $m_{2D,X}/\Sigma m$, $m_{2D,Y}/\Sigma m$, and $m_{2D,Z}/\Sigma m$, such that their sum is unity. Points located near a vertex indicate a projection dominated by a single spatial direction, whereas points near the centre of the triangle correspond to nearly isotropic mass distributions. The colour scale corresponds to the average mass associated with each configuration. Contours indicate the regions containing 50\,\%, 80\,\%, and 95\,\% of the cores.}
    \label{fig:rfl9_update_mass_ternary_new_flash}
\end{figure}

To characterise the geometry of the mass distribution, we investigate the fractional contributions of the projected mass components along the three coordinate axes. The ternary diagram in Fig.~\ref{fig:rfl9_update_mass_ternary_new_flash} shows the normalised projected masses $m_{2D,X}/\Sigma m$, $m_{2D,Y}/\Sigma m$, and $m_{2D,Z}/\Sigma m$, with $\Sigma m = m_{2D,X}+m_{2D,Y}+m_{2D,Z}$. Around 50\% of the cores cluster near the centre of the diagram, which corresponds to comparable contributions from all three projection directions. The remaining data are distributed in an extended structure in the diagram, indicating that the mass distribution around the other 50\% of the cores is not isotropic and/or that these cores themselves might have elongated structures, with a preferred direction. This is consistent with an elongated, filamentary morphology rather than a spherical one. Such anisotropic structures arise naturally in a turbulent medium, where material is funnelled along filamentary channels that can occur in any direction.

The higher-mass cores span a range of positions in the diagram. A small number appears closer to the $m_{2D,Z}/\Sigma m$ vertex, indicating cases where the projected mass is more strongly concentrated along the $Z$ direction. However, the most massive cores are located near the central region of the diagram, where the three projected components contribute in more comparable proportions. Lower-mass cores are more broadly scattered and show a correspondingly wider range of projection ratios. The distribution suggests a variety of geometric structures of the cores.

\end{appendix}

\end{document}